\documentclass[11pt]{article}
\usepackage{epsf}

\newcommand{ \centeron }[2]{{\setbox0=\hbox{#1}\setbox1=\hbox{#2}\ifdim
                             \wd1>\wd0\kern.5\wd1\kern-.5\wd0\fi \copy0
                             \kern-.5\wd0\kern-.5\wd1\copy1\ifdim\wd0>\wd1
                             \kern.5\wd0\kern-.5\wd1\fi}}
\newcommand{ \ltap }{\>\centeron{\raise.35ex\hbox{$<$}}
                     {\lower.65ex\hbox{$\sim$}}\>}
\newcommand{ \gtap }{\>\centeron{\raise.35ex\hbox{$>$}}
                     {\lower.65ex\hbox{$\sim$}}\>}
\newcommand{ \gsim }{\mathrel{\gtap}}
\newcommand{ \lsim }{\mathrel{\ltap}}

\newcommand{ \slashchar }[1]{\setbox0=\hbox{$#1$}   
   \dimen0=\wd0                                     
   \setbox1=\hbox{/} \dimen1=\wd1                   
   \ifdim\dimen0>\dimen1                            
      \rlap{\hbox to \dimen0{\hfil/\hfil}}          
      #1                                            
   \else                                            
      \rlap{\hbox to \dimen1{\hfil$#1$\hfil}}       
      /                                             
   \fi}                                             %

\newcommand{ \Munif    }{M_{\mathrm{unif}}}
\newcommand{ \MFN      }{M_{FN}}

\newcommand{ \ph       }{\gamma}

\newcommand{ \tr       }{\mathop{\mathrm{tr}}}
\newcommand{ \Str      }{\mathop{\mathrm{Str}}}

\newcommand{ \Li       }[1]{\mathop{\mathrm{Li}_{#1}}}

\newcommand{ \ra       }{\rightarrow}

\newcommand{ \LambdaDSB }{\Lambda_{\mathrm{DSB}}}

\newcommand{ \vev      }[1]{\langle {#1} \rangle}

\newcommand{ \one      }{\mathbf{1}}
\newcommand{ \two      }{\mathbf{2}}
\newcommand{ \three    }{\mathbf{3}}
\newcommand{ \threebar }{\mathbf{\overline{3}}}

\newcommand{ \textfrac }[2]{ {\textstyle\frac{#1}{#2}} }

\def\singleandabitspaced{\baselineskip=\normalbaselineskip\multiply
    \baselineskip by 105\divide\baselineskip by 100}
\def\abstractspacing{\baselineskip=\normalbaselineskip\multiply
    \baselineskip by 110\divide\baselineskip by 100}
\def\singlespaced{\baselineskip=\normalbaselineskip}

\newcommand{\Journal}[4]{{#1}\ \textbf{#2}, #3 (#4)}  
\newcommand{ \NPB    }[3]{\Journal{Nucl. Phys.}{B#1}{#2}{#3}}
\newcommand{ \PLB    }[3]{\Journal{Phys. Lett. B}{#1}{#2}{#3}}

\newcommand{ \PRD    }[3]{\Journal{Phys. Rev.}{D#1}{#2}{#3}}

\newcommand{ \PRL    }[3]{\Journal{Phys. Rev. Lett.}{#1}{#2}{#3}}
\newcommand{ \MPL    }[3]{\Journal{Mod. Phys. Lett.}{#1}{#2}{#3}}
\newcommand{ \PREP   }[3]{\Journal{Phys. Rept.}{#1}{#2}{#3}}
\newcommand{ \ZPC    }[3]{\Journal{Z. Phys. C}{#1}{#2}{#3}}

\newcommand{ \NPPS   }[3]{\Journal{Nucl. Phys. Proc. Suppl.}{#1}{#2}{#3}}
\newcommand{ \RMP    }[3]{\Journal{Rev. Mod. Phys.}{#1}{#2}{#3}}

\newcommand{ \xxx    }[1]{\texttt{[#1]}}

\begin{document}

\singlespaced

\begin{titlepage}

\begin{flushright}
UW/PT 99-12 \\
hep-ph/9906341 \\
June 1999
\end{flushright}

\vspace{0.5cm}

\begin{center}
\mbox{\Large \textbf{Phenomenology of flavor-mediated supersymmetry breaking}}

\vspace*{1.1cm}
{\large D.~Elazzar~Kaplan$^{(a)}$ and Graham~D.~Kribs$^{(b)}$} \\
\vspace*{0.8cm}
\textit{$^{(a)}$Department of Physics 1560, University of Washington, 
        Seattle, WA~~98195-1560} \\
\vspace*{0.3cm}
\textit{$^{(b)}$Department of Physics, Carnegie Mellon University, 
        Pittsburgh, PA~~15213-3890} \\
\vspace*{0.5cm}

\begin{abstract}
\indent

\abstractspacing

The phenomenology of a new economical supersymmetric model that utilizes 
dynamical supersymmetry breaking and gauge-mediation for the generation 
of the sparticle spectrum and the hierarchy of fermion masses is
discussed.  Similarities between the communication of supersymmetry
breaking through a messenger sector, and the generation of flavor 
using the Froggatt-Nielsen (FN) mechanism are exploited, leading to the 
identification of vector-like messenger fields with FN fields, 
and the messenger U(1) as a flavor symmetry.  An immediate 
consequence is that the first and second generation scalars acquire 
flavor-dependent masses, but do not violate flavor changing neutral
current bounds since their mass scale, consistent with ``effective 
supersymmetry'', is of order 10 TeV\@.  We define and advocate a 
``minimal flavor-mediated model'' (MFMM), recently introduced in the 
literature, that successfully accommodates the small flavor-breaking 
parameters of the standard model using order one couplings and ratios 
of flavon field vevs.  The mediation of supersymmetry breaking
occurs via two-loop log-enhanced gauge-mediated contributions,
as well as several one-loop and two-loop Yukawa-mediated contributions 
for which we provide analytical expressions.  The MFMM is parameterized 
by a small set of masses and couplings, with values restricted by 
several model constraints and experimental data.  Full two-loop 
renormalization group evolution is performed, correctly taking 
into account the negative two-loop gauge contributions from heavy 
first and second generations.  Electroweak symmetry is radiatively broken 
with the value of $\mu$ determined by matching to the $Z$ mass.  
The weak scale spectrum is generally rather heavy, except for the 
lightest Higgs, the lightest stau, the lightest chargino, the lightest 
two neutralinos, and of course a very light gravitino.  The 
next-to-lightest sparticle (NLSP) always has a decay length that is 
larger than the scale of a detector, and is either the lightest 
stau or the lightest neutralino.  Similar to ordinary gauge-mediated 
models, the best collider search strategies are, respectively, inclusive 
production of at least one highly ionizing track, or events with 
many taus plus missing energy.  In addition, $D^0 \leftrightarrow 
\overline{D}^0$ mixing is also a generic low energy signal.
Finally, the dynamical generation of the neutrino masses is 
briefly discussed.

\end{abstract}

\end{center}
\end{titlepage}

\newpage
\setcounter{page}{2}
\renewcommand{\thefootnote}{\arabic{footnote}}
\setcounter{footnote}{0}
\singleandabitspaced

\section{Introduction}
\label{introduction-sec}

Two central problems pervade the standard model (SM), namely 
understanding the how the scalar Higgs mass is stabilized against
radiative corrections and the origin of the disparate fermion masses.
Weak scale supersymmetry is well-known as the premiere solution to 
the first problem.  The most successful attempts on the second 
are made primarily through flavor symmetries.  It is natural to ask
if we can construct a theory that can explain the fermion mass hierarchy 
simultaneously with an explanation of the gauge hierarchy problem.

Moreover, there are reasons to suggest supersymmetry breaking and 
flavor symmetry breaking ought to be closely related in supersymmetric
models.  The main impetus arises from the supersymmetric 
nonrenormalization theorem that protects superpotential 
couplings from radiative corrections.  In particular, couplings
that vanish at one scale (unbroken flavor symmetry) remain zero
for all scales at all orders in perturbation theory, unless one of 
the following occurs:  (i) a mass scale, put into the theory ``by hand'',
explicitly or spontaneously breaks the flavor symmetry,
(ii) nonperturbative effects generate Yukawa couplings (i.e., the
flavor symmetry is anomalous), or (iii) supersymmetry is 
broken \cite{Nimaframework}.  Because we are interested in 
dynamically generating all scales below the Planck scale, we
discuss exclusively the latter possibility.

Building a model that is both supersymmetric and incorporates a flavor 
symmetry is strongly constrained, particularly due to processes
that can be enhanced by the presence of superpartners.
In general, any successful supersymmetric flavor model must have
a spectrum of scalar partners of quarks and leptons that do not 
contribute to low energy processes beyond experimental bounds.  
It is possible to respect limits on flavor changing neutral currents
(FCNC) and CP violation \cite{GGMS}, such as those from 
$K^0 - \overline{K}^0$ mixing, 
$\epsilon_K$, and lepton flavor violation, if the model contains one 
of the following spectra of squarks and sleptons:
\begin{itemize}
\item{Their masses are generation independent (``degeneracy'').}
\item{The scalar (mass)$^2$ matrices and the (mass)$^2$ matrices of 
      their fermionic partners are both diagonal in the same
      basis (``alignment'') \cite{NS}.}
\item{The masses of the first two generations are much larger than the
      weak scale (``decoupling'') \cite{DG, CKN}.}
\end{itemize}
To solve the flavor problem of the standard model, one must 
take care to respect the flavor problem of the minimal 
supersymmetric standard model as well.

One possibility is that the two solutions are decoupled from each other.
The flavor symmetry could be broken at a high scale $M_F \gg M_Z$, below
which the effective action contains the normal (MS)SM Yukawa couplings.  
Supersymmetry could then break in a hidden sector and the breaking 
be communicated via gravitational interactions \cite{nilles} or 
gauge interactions \cite{GM82, DNS, gmsbrev}.  Gauge-mediated supersymmetry
breaking (GMSB), in it's canonical form \cite{DNS}, leaves scalars of 
the same quantum numbers approximately degenerate, thus avoiding the flavor 
problem of the MSSM\@.  In supergravity-mediated models this problem 
can be circumvented by choosing boundary conditions for soft-breaking 
parameters that leave the sfermions approximately degenerate \cite{sugra}, 
or (as has been suggested recently) produce the decoupling 
solution \cite{BFP}.  In these cases, diagonal soft masses 
would be generated, thereby respecting any flavor symmetry,
and flavor breaking would only affect the form of supersymmetry
breaking via the renormalization group.

The other possibility is that the flavor symmetry dictates the form 
of the soft masses.  As in the models of Ref.~\cite{NS, LNS}, one or more
U(1) flavor symmetries could restrict the quark and squark mass matrices
to be approximately diagonal in the flavor basis.  This approximate 
alignment of quark and squark masses is able to suppress the additional 
sparticle contributions to FCNC (though to satisfy experimental bounds, 
the models are rather complex).  Similar efforts using a U(2)
flavor symmetry that forced the first two generations of scalars to be
degenerate at some scale was also able to suppress dangerous contributions 
\cite{hall}.  Alternatively, supersymmetry could be broken by an 
``anomalous'' U(1) gauge symmetry\footnote{The U(1) is rendered 
anomalous below the Planck scale by the Green-Schwarz mechanism \cite{GS}.}.  
In this scenario, the first two generations carry equal non-zero charges 
and their scalar components become heavy and degenerate \cite{BDDP},
and the U(1) could play the role of a flavor symmetry 
\cite{MRNW} in a Froggatt-Nielsen mechanism \cite{FN}.  
A completely different approach utilized a composite dynamical 
supersymmetry breaking, in which in some cases promising fermion 
textures and heavy first and second generations could result
\cite{preons}.  Finally, several groups have recently studied 
the possibility of generating the Yukawa couplings 
radiatively \cite{AHCH, BFPT}.

In this paper, gauge-mediation is utilized for the communication 
of supersymmetry breaking, which has several benefits for flavor physics.  
First, there naturally appears a hierarchy of scales in the simplest 
class of models, due to the separation of the dynamical supersymmetry 
breaking sector (DSB), the messenger sector, and the MSSM\@.  Furthermore,
additional gauge symmetries are naturally required for the DSB fields 
to communicate their supersymmetry breaking to the messenger fields.

The model presented here \cite{KLMNR} is supersymmetric, incorporates 
a gauged U(1)$_F$ flavor
symmetry, and utilizes a modified Froggatt-Nielsen mechanism to generate 
the necessary Yukawa coupling hierarchy.  The Froggatt-Nielsen 
mechanism \cite{FN} was invented two decades ago in the context of 
the SM as a way to reconcile the hierarchy of Yukawa couplings.  
The general principle is to forbid some tree-level Yukawa couplings, 
but introduce additional matter that when integrated out gives rise 
to the smaller Yukawa couplings.  In our interpretation the 
third generation is considered ``fundamental'', with nonzero Yukawa
couplings allowed even with the flavor symmetry unbroken, while fermion 
masses for the first and second generations arise after integrating out 
heavy FN fields.  We also insist that couplings are of order one,
by which we mean quantitatively that no couplings differ by more
than a factor of ten (i.e., all couplings are larger than about 
$0.1$).  In this way our quantitative improvement over an ordinary
supersymmetry model is to reduce the hierarchy of Yukawa
couplings $\sim 10^{-5}$--$1$ to $\sim 0.1$--$1$.
We utilize the two-Higgs doublet structure of the MSSM 
and demand the Yukawa couplings for the third generation also 
satisfy this requirement.  For third generation Yukawa couplings 
larger than $0.1$, it follows that $\tan\beta \gsim 10$, 
which is the lower bound that we use in this paper.\footnote{The
upper bound is determined by the model dynamics, discussed
in detail later in the paper.}  

Our analysis focuses almost entirely on one model \cite{KLMNR}, 
the minimal flavor-mediated model (MFMM), that has one set of fields 
and one parameterized superpotential.  We use the term ``flavor-mediated'' 
since the same gauge group that communicates supersymmetry breaking 
from the dynamical supersymmetry breaking sector also serves as the 
flavor symmetry.  In this way supersymmetry breaking and flavor
symmetry breaking are inseparably linked together, which is
the precise meaning of ``flavor-mediation''.

The main thrust of this paper is to derive the complete phenomenological 
spectrum of this model so that the most promising experimental signals 
can be extracted.  Unlike ordinary gauge-mediation, there are 
several technical subtleties in accurately calculating the
spectrum at the lowest scales.  We believe we have overcome 
the vast majority of these difficulties.  The model does have
several new couplings and masses that are not precisely known,
forcing us to parameterize this ignorance and show the results
as a function of these high(er) scale quantities.  

There are, however, several contortions that could be done to the model.
Some of these could potentially address problems that we do not solve 
or only have partial solutions in the MFMM\@.  For example,
the $\mu$ problem persists in the MFMM, and we implement 
no particular solution (instead fixing $\mu^2$ using electroweak
symmetry breaking (EWSB) constraints).  Nevertheless, there are various 
ways that a $\mu$ term could be generated from the dynamics of the
model.  Since the details of such $\mu$ term generation dynamics 
are essentially irrelevant to the low energy phenomenology, we can
rightly neglect this issue for all practical purposes.

The one aspect of the MFMM that may affect the weak scale 
phenomenology is the generation of neutrino masses.
Several experiments, in particular those measuring atmospheric 
and solar neutrinos, have now convincingly established that
some (if not all) generations of neutrinos are very likely mixing 
with one another.  This mixing presumably arises due to nonzero 
neutrino masses.  With certain assumptions, it is possible to
modify and augment the MFMM to incorporate neutrino masses 
of a size that can generate mixings to explain both the atmospheric 
neutrino results \cite{superK} and the solar neutrino results 
\cite{solar}.  However, no completely satisfactory solution is
presented since specific assumptions about the physics at 
scales beyond the DSB scale are required.  We will discuss this 
more fully in future work, but we are nevertheless encouraged 
by these preliminary results.  

The outline of the paper is as follows.  In Sec.~\ref{model-sec}
we describe in detail the sectors that make up the 
flavor-mediated model.  The role and function of the flavor
symmetry is given, along with the predictions for the fermion
masses and CKM elements.  The required and undetermined flavor charge 
assignments are given, as well as the resulting 
messenger--Froggatt-Nielsen superpotential.  
In Sec.~\ref{mediation-sec} the mediation of supersymmetry 
breaking is discussed, and the resulting holomorphic and
nonholomorphic contributions to the various fields are written.
In Sec.~\ref{RGE-sec} the renormalization group evolution
is discussed in detail, including several important effects 
that arise in our particular model.  These include
consequences of an anomalous U(1)$_F$ below the scale of
dynamical supersymmetry breaking, Yukawa-mediated
supersymmetry breaking contributions, and two-loop
gauge contributions induced by heavy first and second 
generations.  In Sec.~\ref{dynamics-sec} we present many of
the main results of our paper.  We discuss in detail
the relevant parameters of the model, and define a 
minimal parameter set associated with the MFMM\@.
Several phenomenological constraints restrict the
parameter space in highly nontrivial ways, and we illustrate
these constraints with a series of graphs and discussion.
In Sec.~\ref{signals-sec} the main signals of the model
are discussed in detail.  Our emphasis is on the collider
signals of the weak-scale spectrum, that depend ultimately
on several factors (MFMM parameters).  We have been able to apply
to our scenario several recent results from the literature 
that used event-level simulations, and this has allowed us to be rather 
precise in the predictions for the MFMM\@.  We also discuss
signals in low energy flavor processes (i.e., FCNC and CP
violation).  Sec.~\ref{modify-model-sec} is devoted to discussing 
some variations of the MFMM\@.  Our primary motivation is 
to briefly examine the impact of other choices of FN fields
and superpotential couplings that result if the SU(5) ansatz
is relaxed.  In addition, we discuss some directions toward
a solution of the $\mu$-problem, and perhaps most importantly
we discuss ways to augment and modify the MFMM to accommodate
nonzero neutrino masses.  Finally, our conclusions appear
in Sec.~\ref{conclusions-sec}.

\section{The model}
\label{model-sec}

There are essentially four sectors in this model.  At the highest
energy scales there is a DSB sector, of which we have certain
requirements but we leave largely unspecified.
Some DSB sector fields are charged under an additional nonanomalous
U(1), which serves both as the messenger U(1) responsible for
transmitting supersymmetry breaking to the messenger fields
(among others), and also acts as the flavor U(1) responsible for
specifying the necessary (or absent) couplings in the model.
Vector-like messenger fields also serve in a dual capacity 
by transmitting supersymmetry breaking to the MSSM via ordinary 
gauge interactions\footnote{We define ``MSSM gauge interactions''
to mean only the SM gauge and the supersymmetrized gaugino interactions 
between sectors of the theory.}, and also as Froggatt-Nielsen (FN) fields
that are needed to generate the effective Yukawa coupling hierarchy.
To emphasize the dual role of this sector, we call this the 
messenger--Froggatt-Nielsen sector, or simply ``FN sector''.
Similarly, messenger--Froggatt-Nielsen fields are called
simply ``FN fields''.
In order to generate the Yukawa hierarchy, several SM gauge singlets 
must also be present that are charged under the U(1)$_F$ and obtain 
vacuum expectation values.  These fields
are called flavons, and in this model there are two distinct
varieties.  Three $\chi^{(i)}$ fields are responsible for generating
the supersymmetric mass for the vector-like FN fields, 
and as such indirectly assist the generation of the Yukawa coupling
hierarchy.  Five $\phi^{(j)}$ fields are solely responsible to 
generate the effective Yukawa couplings, after the FN sector
and the flavon sector is integrated out.  Any given Yukawa
coupling (except those of the third generation) or CKM element 
is ultimately proportional to the ratio
$\vev{\phi^{(j)}}/\vev{\chi^{(i)}}$.
Thus, judicious U(1)$_F$ charge assignments with a properly
generated flavon vev hierarchy give the required fermion masses
and mixing angles.  Calculating the flavon vev hierarchy
with eight flavon fields amounts to a doing a complicated 
multi-dimensional effective potential minimization.  It was shown
in Ref.~\cite{KLMNR} that additional flavons are typically
necessary to generate the required vevs, but they do not
impact any other aspect of the model.  We therefore
simply assume flavons can be added such that the required local minimum 
is generated and is sufficient stable (lifetime of the universe).  
More concrete work detailing the generation of flavon vevs
with the desired minimum can be found in Ref.~\cite{Nimaframework}.
Finally, the MSSM scalar fields acquire supersymmetry breaking
masses from several sources.  It is on this aspect of the
model that we spend the most effort, since all signals of
the model are crucially dependent on precise calculations 
of these weak scale quantities.  

We have already remarked that the first and second generation
fields are charged under the U(1)$_F$.  There are several immediate 
consequences of this which we now discuss.  The motivating
reason is to forbid the first and second generation Yukawa
couplings through the U(1)$_F$ flavor symmetry.  These couplings
are regenerated in the effective theory through FN ladder 
diagrams with FN vector-like matter coupling to flavons 
and the first and second generations.  Since the first and
second generations must be charged in this scenario, the
scalar components acquire two-loop U(1)$_F$ gauge-mediated
masses directly induced after integrating out the DSB sector.  
These are much larger than the contributions induced by 
ordinary two-loop gauge interactions after integrating out 
the FN sector.  Thus, the MSSM fields also have a hierarchy:
the first and second generations are expected to be much
heavier than the other fields, due to the U(1)$_F$ coupling.

To generate the proper hierarchy of fermion masses without 
generating significant squark mixing, the first
and second generations should be assigned different U(1)$_F$ charge
for their component fields\footnote{U(1)$_F$ charge is assigned 
to respect an SU(5) ansatz.  This will be discussed below.}.
Hence, the model generically has large differences between the 
first and second generation scalar masses which can induce large 
FCNC\@.  This is precisely the supersymmetric flavor problem.
However, their masses are expected to be much heavier than 
the other MSSM fields, as explained above.  
It is well known that the decoupling solution to the FCNC problem in a 
generic MSSM requires the first and second generation scalar masses to be
so large they cause other phenomenological problems (e.g., unnatural
contributions to the Higgs masses, and third generation scalar masses 
running negative \cite{AHM, AG}).  Thus, the decoupling solution must be
augmented by some amount of degeneracy or alignment.
Such an ``effective supersymmetry'' spectrum has been discussed 
in different contexts in Refs.~\cite{DG, CKN, BDDP, CKLN, MRNW, 
BFP, other-effective-SUSY}
and indeed has been recently advocated as the most natural 
scenario \cite{baggertalk, BFP}.  In our model, it is remarkable
that the same mechanism that allows the generation of the
Yukawa hierarchy and generates generically large splittings
between the first and second generations, also solves the 
supersymmetric flavor problem since the first and second generations
receive large supersymmetry breaking masses and the squark masses
are approximately diagonal in the flavor basis.  This one of the 
central motivations for pursuing this approach to flavor.

The messenger U(1)$_F$ is anomalous after the DSB sector 
is integrated out.  This has several important consequences.
A Fayet-Iliopoulos (FI) term is automatically generated at one-loop, 
but is calculable only by specifying a DSB model.  Fields charged
under the U(1)$_F$ acquire supersymmetry breaking
masses near the DSB scale after the DSB fields are integrated out.
Furthermore, $\tr q_i m_i^2$ is necessarily nonzero, where 
$q_i$ and $m_i$ are respectively the U(1)$_F$ charge and
the mass of the scalar field $i$.  This generates significant 
one-loop corrections to the U(1)$_F$ charged scalar (mass)$^2$, 
analogous to when hypercharge $D$-term is nonzero in other 
supersymmetric models.

To successfully accommodate the fermion masses and CKM angles
with $\mathcal{O}(1)$ couplings, we must construct a (modified)
FN model.  In Ref.~\cite{KLMNR}, it was shown that the form of the
fermion mass matrices is tightly constrained due to the restrictions
on the first two generation masses and on the couplings of the up-type
Higgs\footnote{Allowed variations of these matrices are discussed in
Sec.~\ref{modify-model-sec}.}.
Although we will not repeat the detailed arguments, we do present
the up-type quark, down-type quark, and charged lepton mass matrices
in the notation used in this paper to show how the fermion masses
are generated.
 
After the U(1)$_F$ breaks and we are in the proper ground state,
several flavon fields acquire vevs.  These have the effect of giving
large (supersymmetric) masses to the FN fields and mixing FN fields 
with MSSM fields of the same quantum numbers, leading to the mass 
matrices we present below.  The generic form of the superpotential is 
\begin{eqnarray}
W &\supset& \chi F \overline{F} + \phi f \overline{F} + H f F \; ,
\end{eqnarray}
where the $F$ and $\overline{F}$ are FN fields, the $f$ are SM fields and 
$H$ is a Higgs field.  Before any FN fields are integrated out, the 
up-type fermion mass matrix is
\begin{eqnarray}
\mathbf{M}^u &=& \left( \begin{array}{ccccc} 
    0 & 0 & 0 & 0 & b_1^{(Q)} \vev{\phi^{(1)}} \\
    0 & 0 & 0 & 0 & b_2^{(Q)} \vev{\phi^{(2)}} \\
    0 & 0 & Y_t \vev{H^u} & \lambda^u_{2,1} \vev{H^u} & 0 \\
    0 & 0 & \lambda^u_{2,2} \vev{H^u} & 0 & a_1^{(Q)} \vev{\chi^{(1)}} \\
    b_1^{(u)} \vev{\phi^{(1)}} & b_2^{(u)} \vev{\phi^{(2)}} & 0
        & a_1^{(u)} \vev{\chi^{(1)}} & 0 \\
                        \end{array} \right)
\label{up-matrix-eq}
\end{eqnarray}
where the columns represent ($u^1$, $u^2$, $u^3$, $u^{FN}$, 
$\overline{Q}^{FN}$) and the rows represent 
($Q^1$, $Q^2$, $Q^3$, $Q^{FN}$, $\overline{u}^{FN}$).
We have inserted the component superpotential parameters that
are defined in Appendix~\ref{superpot-component-app}.  
The down-type fermion mass matrix is
\begin{eqnarray}
\mathbf{M}^d &=& \left( \begin{array}{cccccc} 
    0 & 0 & 0 & \lambda^d_{1,1} \vev{H^d} & 0 & b_1^{(Q)} \vev{\phi^{(1)}} \\
    0 & 0 & 0 & 0 & \lambda^d_{2,1} \vev{H^d} & b_2^{(Q)} \vev{\phi^{(2)}} \\
    0 & 0 & Y_b \vev{H^d} & 0 & 0 & 0 \\
    0 & 0 & \lambda^d_{4,1} \vev{H^d} & 0 & 0 & a_1^{(Q)} \vev{\chi^{(1)}} \\
    b_3^{(d)} \vev{\phi^{(3)}} & 0 & b_5^{(d)} \vev{\phi^{(5)}}
        & a_2^{(d)} \vev{\chi^{(2)}} & 0 & 0 \\
    0 & b_4^{(d)} \vev{\phi^{(4)}} & 0 & 0 
        & a_3^{(d)} \vev{\chi^{(3)}} & 0 \\
                        \end{array} \right)
\label{down-matrix-eq}
\end{eqnarray}
where the columns represent ($\overline{d}^1$, $\overline{d}^2$, 
$\overline{d}^3$, $\overline{d}^{FN(1)}$, $\overline{d}^{FN(2)}$, 
$\overline{Q}^{FN}$) and the rows represent 
($Q^1$, $Q^2$, $Q^3$, $Q^{FN}$, $d^{FN(1)}$, $d^{FN(2)}$).
Finally, for completeness, the charged lepton mass matrix is
\begin{eqnarray}
\mathbf{M}^e &=& \left( \begin{array}{cccccc} 
    0 & 0 & 0 & 0 & b_3^{(L)} \vev{\phi^{(3)}} & 0 \\
    0 & 0 & 0 & 0 & 0 & b_4^{(L)} \vev{\phi^{(4)}} \\
    0 & 0 & Y_\tau \vev{H^d} & \lambda^d_{4,2} \vev{H^d} & 
        b_5^{(L)} \vev{\phi^{(5)}} & 0 \\
    \lambda^d_{1,2} \vev{H^d} & 0 & 0 & 0 & a_2^{(L)} \vev{\chi^{(2)}} & 0 \\
    0 & \lambda^d_{2,2} \vev{H^d} & 0
        & 0 & 0 & a_3^{(L)} \vev{\chi^{(3)}} \\
    b_1^{(e)} \vev{\phi^{(1)}} & b_2^{(e)} \vev{\phi^{(2)}} & 0 & 
        a_1^{(e)} \vev{\chi^{(1)}} & 0 & 0 \\
                  \end{array} \right)
\label{lepton-matrix-eq}
\end{eqnarray}
where the columns represent 
($e^1$, $e^2$, $e^3$, $e^{FN}$, $L^{FN(1)}$, $L^{FN(2)}$)
and the rows represent ($\overline{L}^1$, $\overline{L}^2$, 
$\overline{L}^3$, $\overline{L}^{FN(1)}$, $\overline{L}^{FN(2)}$, 
$\overline{e}^{FN}$).
Notice that all MSSM Yukawa couplings are forbidden other than among the
third generation, just as the $U(1)_F$ flavor symmetry demands.

Assuming $a_i \vev{\chi^{(i)}} \gg \vev{\phi^{(j)}}, \vev{H^u}, \vev{H^d}$,
we can integrate out the FN fields giving the $3 \times 3$ fermion mass 
matrices:\footnote{Our results differ from that in Ref.~\cite{KLMNR} 
by expressing the mass matrices using superpotential couplings 
associated with the SU(3)$_c$ $\times$ SU(2)$_L$ $\times$ U(1)$_Y$ 
``component'' fields of the global SU(5) ansatz, which differentiate
couplings in the mass matrices.}
\begin{eqnarray}
\mathbf{M}^u_{\mathrm{eff}} &=& \vev{H^u} \left( \begin{array}{ccc}
    0 & 0 & \lambda^u_{2,2} \epsilon_{11}^{(Q)} \\
    0 & 0 & \lambda^u_{2,2} \epsilon_{21}^{(Q)} \\
    \lambda^u_{2,1} \epsilon_{11}^{(u)} 
        & \lambda^u_{2,1} \epsilon_{21}^{(u)} & Y_t
    \end{array} \right) \\
\mathbf{M}^d_{\mathrm{eff}} &=& \vev{H^d} \left( \begin{array}{ccc}
    \lambda^d_{1,1} \epsilon_{32}^{(d)} & 0 
        & \lambda^d_{4,1} \epsilon_{11}^{(Q)} 
          + \lambda^d_{1,1} \epsilon_{52}^{(d)} \\
    0 & \lambda^d_{2,1} \epsilon_{43}^{(d)} 
        & \lambda^d_{4,1} \epsilon_{21}^{(Q)} \\
    0 & 0 & Y_b
    \end{array} \right) \\
\mathbf{M}^e_{\mathrm{eff}} &=& \vev{H^d} \left( \begin{array}{ccc}
    \lambda^d_{1,2} \epsilon_{32}^{(L)} & 0 & 0 \\ 
    0 & \lambda^d_{2,2} \epsilon_{43}^{(L)} & 0 \\
    \lambda^d_{4,2} \epsilon_{11}^{(e)} + \lambda^d_{1,2} \epsilon_{52}^{(L)} 
        & \lambda^d_{4,2} \epsilon_{21}^{(e)} & Y_\tau \\
    \end{array} \right) \; ,
\end{eqnarray}
where
\begin{eqnarray}
\epsilon_{ij}^{(f)} 
&=& -\frac{b_i^{(f)}\vev{\phi^{(i)}}}{a_j^{(f)} \vev{\chi^{(j)}}}\; .
\end{eqnarray}
These matrices predict the following quark masses 
\begin{eqnarray}
\frac{m_d}{m_s} &=& \frac{\lambda^d_{1,1}
    \epsilon_{32}^{(d)}}{\lambda^d_{2,1}\epsilon_{43}^{(d)}} \\
\frac{m_s}{m_b} &=& \frac{\lambda^d_{2,1} 
    \epsilon_{43}^{(d)}}{Y_b} \\
\frac{m_c}{m_t} &=& \frac{\lambda^u_{1,2}
    \lambda^u_{2,2}\epsilon_{21}^{(Q)}\epsilon_{21}^{(u)}}{Y_t^2}
\end{eqnarray}
and CKM matrix
\begin{eqnarray}
\mathbf{V}_{\mathrm{CKM}} \simeq \left( \begin{array}{ccc}
    1 & - \frac{\epsilon_{11}^{(Q)}}{\epsilon_{21}^{(Q)}} 
        & \frac{\lambda^d_{1,1}}{Y_b} \epsilon_{52}^{(d)} \\
    \left(\frac{\epsilon_{11}^{(Q)}}{\epsilon_{21}^{(Q)}}\right)^* 
        & 1 & - \left(\frac{\lambda^u_{2,2}}{Y_t} - 
        \frac{\lambda^d_{4,1}}{Y_b} \right) \epsilon_{21}^{(Q)} \\
    \left(\frac{\lambda^u_{2,2}}{Y_t} - \frac{\lambda^d_{4,1}}{Y_b} \right)^* 
        \epsilon_{11}^{(Q)*} - 
        \left(\frac{\lambda^d_{1,1}}{Y_b}\right)^* \epsilon_{52}^{(d)*}
        & \left(\frac{\lambda^u_{2,2}}{Y_t} - 
        \frac{\lambda^d_{4,1}}{Y_b} \right)^* \epsilon_{21}^{(Q)*}
        & 1
    \end{array} \right) . \; & &
\end{eqnarray}

The one striking feature of these matrices 
is that $m_u$ is zero\footnote{This is an exact statement as the 
matrix in Eq.~(\ref{up-matrix-eq}) has a zero eigenvalue which is 
protected by a chiral symmetry.}.  This is a direct consequence of 
limiting ourselves to five $q, \overline{q}$ pairs of (color triplet)
FN fields.  Refs.~\cite{m-u-zero} suggest that 
$m_u = 0$ is not excluded by experiment, and furthermore has the virtue of 
solving the strong CP problem, though this solution is as of yet 
controversial \cite{Leut}.  Modifications to the FN sector which allow for 
$m_u \not= 0$ are discussed in Sec.~\ref{m-u-nonzero-sec}.

The charged lepton masses are equal to the down-type quark masses 
up to couplings of $\mathcal{O}(1)$.  For example,
\begin{equation}
\frac{m_e}{m_d}=\frac{\lambda_{1,2}^d \epsilon_{32}^{(L)}}{\lambda_{1,1}^d
        \epsilon_{32}^{(d)}} = \frac{\lambda_{1,2}^d b_3^{(L)} a_2^{(d)}}{
        \lambda_{1,1}^d b_3^{(d)} a_2^{(L)}}\sim\frac{1}{10} -\frac{1}{20}
        \; 
\end{equation}
is the smallest ratio between the two sectors.
Thus ratios of individual couplings of $\sim 2/5$ can explain these mass
differences.

To achieve this result the U(1)$_F$ charges must be judiciously
chosen, although several degrees of freedom remain even after
all the constraints are satisfied.  We require the first
and second generations are charged respecting an SU(5) ansatz, 
and all the other fields of the MSSM (third generation, Higgs, gauginos) 
are uncharged.  Although we do not attempt to embed the
model in an SU(5) GUT, there are four main motivations to
enforce the charge assignments commute with SU(5):
It is sufficient for gauge coupling unification to be at least naively 
preserved (to one-loop), U(1)$_F$ anomaly cancellation 
and $\tr Y_i m_i^2 \simeq 0$ are relatively easy to achieve,
and any assignment that did not compute with SU(5) (or 
SU(4)$\times$SU(2)$\times$SU(2), etc.)
would immediately preclude any hope of 
embedding the model in a GUT at the unification scale.  Having said this, 
we should also note that gauge coupling unification can be achieved with 
fields in representations that do not fit into SU(5) or any group
\cite{MartinGMSB, GDK-gaugino} (although such a messenger sector cannot
generate flavor) and there are numerous examples of 
free fermionic string constructions that break directly 
to the SM gauge group, never passing through a GUT in the first 
place.

Assuming the charge assignments to fields are distinct and that
the Higgs and third generation fields are uncharged, the above 
mass matrices imply the following constraints on the charges:
\begin{eqnarray*}
q_{\overline{5}^{FN(1)}} &=& - q_{10^1} \\
q_{\overline{5}^{FN(2)}} &=& - q_{10^2} \\
q_{5^{FN(1)}}            &=& - q_{\chi^{(2)}} + q_{10^1} \\
q_{5^{FN(2)}}            &=& - q_{\chi^{(3)}} + q_{10^2} \\
q_{10^{FN}}              &=& 0 \\ 
q_{\overline{10}^{FN}}   &=& - q_{\chi^{(1)}} \\ 
q_{\phi^{(1)}}           &=& - q_{10^1} + q_{\chi^{(1)}} \\
q_{\phi^{(2)}}           &=& - q_{10^2} + q_{\chi^{(1)}} \\
q_{\phi^{(3)}}         &=& - q_{10^1} - q_{\overline{5}^1} + q_{\chi^{(2)}} \\
q_{\phi^{(4)}}         &=& - q_{10^2} - q_{\overline{5}^2} + q_{\chi^{(3)}} \\
q_{\phi^{(5)}}           &=& - q_{5^{FN(1)}} \; . 
\end{eqnarray*}
Furthermore, requiring the mixed [SU(5)]$^2$U(1)$_F$ anomaly to vanish
imposes the additional constraint
\begin{eqnarray*}
0 &=& 3 (q_{10^1} + q_{10^2} - q_{\chi^{(1)}}) 
      + q_{\overline{5}^1} + q_{\overline{5}^2} 
      + q_{5^{FN(1)}} + q_{\overline{5}^{FN(1)}}
      + q_{5^{FN(2)}} + q_{\overline{5}^{FN(2)}} \; .
\end{eqnarray*}
We can rewrite the charges of $\chi^{(2)}$ and $\chi^{(3)}$ as
\begin{eqnarray*}
q_{\chi^{(2)}} &=& \frac{3 (q_{10^1} + q_{10^2} - q_{\chi^{(1)}}) 
                   + q_{\overline{5}^1} + q_{\overline{5}^2}}{q_r^{-1} + 1} \\
q_{\chi^{(3)}} &=& \frac{3 (q_{10^1} + q_{10^2} - q_{\chi^{(1)}}) 
                         + q_{\overline{5}^1} + q_{\overline{5}^2}}{q_r + 1}
\end{eqnarray*}
in terms of the ratio $q_r \equiv q_{\chi^{(2)}}/q_{\chi^{(3)}}$.
Thus, the set of undetermined charges can be taken to be
\begin{eqnarray*}
q_{10^1}, \; q_{10^2}, \; q_{\overline{5}^1}, \; q_{\overline{5}^2}, \; 
q_{\chi^{(1)}}, \; q_r \; .
\end{eqnarray*}

Naively the anomaly could be canceled with a particular choice
of the first and second generations (i.e., applying the anomaly constraint 
on the four charges).  However, this necessarily requires that some of 
the first and second generation charges have a different sign,
which is phenomenologically excluded.  (One of the (mass)$^2$
of the first or second generations would be negative.)
Thus, at least one of $q_{\chi^{(2)}}$ or $q_{\chi^{(3)}}$ must 
be nonzero (and negative).  It turns out that to construct
a viable effective potential that yields the desired hierarchy
of flavon vevs requires that either $q_{\chi^{(2)}}$ or
$q_{\chi^{(3)}}$ be larger (in absolute value) than
the first and second generation charges.  Hence, the remaining
charges for the $\chi$ flavons are the ratio $q_r$ and $q_{\chi^{(1)}}$.
We nominally take $q_{\chi^{(1)}}$ to vanish for our analysis
(as it reduces contact between $H^u$ and U(1)$_F$ charged fields),
however we discuss the consequences of a nonzero charge for
$\chi^{(1)}$ in Sec.~\ref{modify-model-sec}.

The full superpotential can be written using the SU(5)
ansatz as
\cite{KLMNR}\footnote{Our notation differs slightly from that
in Ref.~\cite{KLMNR}.}
\begin{eqnarray}
W &=&{} a_1 \chi^{(1)} 10^{FN} \overline{10}^{FN}
      + a_2 \chi^{(2)} 5^{FN(1)} \overline{5}^{FN(1)}
      + a_3 \chi^{(3)} 5^{FN(2)} \overline{5}^{FN(2)} \nonumber \\
& &{} + b_1 \phi^{(1)} 10^{1} \overline{10}^{FN}
      + b_2 \phi^{(2)} 10^{2} \overline{10}^{FN}
      + b_3 \phi^{(3)} \overline{5}^{1} 5^{FN(1)}
      + b_4 \phi^{(4)} \overline{5}^{2} 5^{FN(2)}
      + b_5 \phi^{(5)} \overline{5}^{3} 5^{FN(1)} \nonumber \\
& &{} + \lambda^{d}_1 H^{d} 10^{1} \overline{5}^{FN(1)}
      + \lambda^{d}_2 H^{d} 10^{2} \overline{5}^{FN(2)}
      + \lambda^{d}_3 H^{d} 10^{3} \overline{5}^{3}
      + \lambda^{d}_4 H^{d} 10^{FN} \overline{5}^{3} \nonumber \\
& &{} + \lambda^{u}_1/2 \> H^{u} 10^{3} 10^{3}
      + \lambda^{u}_2 H^{u} 10^{3} 10^{FN} \; .
\label{superpot-SU5-eq}
\end{eqnarray}
We have presented the complete superpotential in component form 
in Appendix~\ref{superpot-component-app}.

There are several comments that should be made regarding the form of 
Eq.~(\ref{superpot-SU5-eq}):
\begin{itemize}
\item We have implicitly assumed $R$-parity ($R \equiv (-1)^{3B + L + 2S}$)
with $B$, $L$, and $S$ representing baryon number, lepton number, and
spin respectively) is exactly conserved, and thus we have not 
written any baryon or lepton number violating terms.  
\item The messenger--Froggatt-Nielsen sector consists 
of two $5 + \overline{5}$ pairs and one $10 + \overline{10}$ pair.
For ordinary gauge-mediated models this implies that the gauge
couplings encounter a Landau pole prior to their apparent 
unification near $\Munif = 10^{16}$~GeV\@.  In this model,
however, partial unification is still possible with
semi-perturbative ($\alpha \lsim 0.5$) couplings, as we will see.
\item If $q_{\overline{10}^{FN}} = 0$, then following term is allowed:
\begin{eqnarray}
W &\subset& a' \chi^{(1)} H^u H^d
\end{eqnarray}
which generates an effective $\mu$ term when the flavon $\chi^{(1)}$
acquires a vev.  This term is too large to be consistent with
electroweak symmetry breaking.  Later we discuss to what extent 
it is a possible to generate the $\mu$ term in this model.
\end{itemize}

\section{Mediation of supersymmetry breaking}
\label{mediation-sec}

Supersymmetry is assumed to be dynamically broken in the DSB
sector.  This is communicated to the FN fields, flavons, and some
MSSM fields in two important ways.  First, the heavier $\chi$
flavons are assumed to acquire holomorphic ($F$-term) supersymmetry
breaking vevs.  Second, integrating out the DSB sector induces 
two-loop U(1)$_F$ gauge-mediated nonholomorphic contributions 
[soft (mass)$^2$] to the U(1)$_F$ charged scalar fields.

\subsection{Holomorphic contributions}
\label{holomorphic-sec}

In ordinary models of GMSB \cite{DNS, gmsbrev} there are one or 
more fields whose role is to communicate supersymmetry breaking 
to messenger quarks and leptons.  These fields acquire non-zero 
vacuum expectation values for their scalar and auxiliary components, 
which give the messengers supersymmetric and supersymmetry breaking 
masses respectively.  In our model, it is the flavons that play the 
role of these fields.  While their holomorphic contribution to 
supersymmetry breaking in the FN sector is no longer the only 
contribution, it remains significant (and essential to generate 
nonzero gaugino masses, as we shall see).

The auxiliary components of the flavons are in general unknown 
parameters unless the interactions between flavons are specified.  
However, their ``natural'' sizes can be estimated.  To produce the 
hierarchy of Yukawa couplings, the (flavon) superpotential must be 
such that only the $\chi$ flavons receive vevs at tree level.  
The $\phi$ fields would have non-zero vevs due to loop effects as 
discussed in Ref.~\cite{KLMNR}.  In constructing toy models, 
we have found that additional flavons are required to produce the
proper hierarchy, and that most $\chi$ and $\phi$ fields are not 
directly coupled.  Thus, $\left| F_{\chi} \right|^2$ can in general 
be approximated by the vacuum energy density 
$V\sim \tilde{m}^2 \langle\chi\rangle^2$.  The 
$F_{\phi}$ ``turn on'' at one or more loops.  A rough guess,
$F_{\phi^{(i)}} \sim \vev{\chi}\vev{\phi^{(i)}}$, agrees with the 
typical values produced in these toy models.

For these values, the $F_{\phi^{(i)}}$ do not contribute significantly
to the low-energy spectrum, with two exceptions.  One is the contribution
of $F_{\phi^{(5)}}$ to the mass of $\overline{5}^3$.  
A non-zero $F_{\phi^{(5)}}$ mixes $\overline{5}^3$ and 
$5^{FN(1)*}$ giving a contribution to the lightest
eigenvalue of $\mathcal{O}(\vev{\phi^{(5)}}^2)$.  As we shall see,
$\vev{\phi^{(5)}}$ may be relatively large ($\lsim$ a few TeV) thus giving 
a large negative contribution to these third generation scalar (mass)$^2$s.
This contribution is largely dependent on given superpotential couplings
and unspecified flavon interactions.  For simplicity, we assume 
$F_{\phi^{(5)}}$ is small enough such that we can ignore it, and take 
this as a mild constraint on the unspecified flavon superpotential.

The other exception is the contributions of $F$-terms to $CP$ violation.  
Non-zero $F_{\phi^{(1)}}$ and $F_{\phi^{(2)}}$ cause mixing between the
first two generations and thus generating contributions to both 
$\Delta m_K$ and $\epsilon_K$.  
For $F_{\phi}\sim \vev{\phi}\vev{\chi}$, the contribution 
to $\Delta m_K$ is within experimental bounds, however with an 
$\mathcal{O}(1)$ phase of $F_{\phi^{(1)}} F_{\phi^{(2)}}^*$,
the contribution to $\epsilon_K$ is two orders of magnitude too 
large.  This effectively adds a constraint to the the
superpotential interactions.  Either the phase of 
$F_{\phi^{(1)}} F_{\phi^{(2)}}^*$ is (unnaturally) small, or one 
(or both) of these $F$-terms happen to be small.\footnote{One could imagine 
that $CP$ is an approximate symmetry and that the dominant contribution 
to $\epsilon_K$ comes from supersymmetry effects only.  However, 
this option is ruled out by the recent measurement of 
$\epsilon'$ \cite{KTeV}.}  In what follows, 
we assume at least one of these criteria are satisfied.

\subsection{Gaugino masses}

MSSM gaugino masses are induced at one-loop through ordinary gauge 
interactions after the FN fields are integrated out.  The
expressions for the masses are nearly identical to minimal
messenger gauge-mediated models, except that the FN scalars
have nonholomorphic contributions.  These contributions can
be written as \cite{PT, GDK-gaugino}
\begin{eqnarray}
M_a(\MFN) &=& \frac{\alpha_a}{4 \pi} \sum_i S(i) m_{f_i} \sin 2\theta_i 
\times \left[ \frac{y_1^i}{1 - y_1^i} \ln y_1^i 
              - \frac{y_2^i}{1 - y_2^i} \ln y_2^i \right]
\label{gaugino-masses-eq}
\end{eqnarray}
where $S(i)$ is the Dynkin index in a normalization where 
$S(5 + \overline{5}) = 1$ and $S(10 + \overline{10}) = 3$.
The sum is over all the FN 
fields\footnote{Eq.~(\ref{gaugino-masses-eq}) expression can 
also be easily generalized using
component fields (and component superpotential couplings)
for non-degenerate SU(5) fields (see also 
Ref.~\cite{MartinGMSB}).  This is actually how we carry out 
the calculation of the gauginos in the following.};
for our model 
$i = 5^{FN(1)} + \overline{5}^{FN(1)}, 5^{FN(2)} + \overline{5}^{FN(2)}, 
10^{FN} + \overline{10}^{FN}$.  $m_{f_i} = a_i \vev{\chi^{(i)}}$ 
is the FN fermion mass and $y_{1,2}^i = (m_{1,2}^i)^2/m_{f_i}^2$
are ratios of the light and heavy FN scalar masses.  
The FN scalar mass matrix is
\begin{eqnarray}
\left( \begin{array}{cc} FN^\dagger & \overline{FN} \end{array} \right) 
    \left( \begin{array}{cc} 
    |a_i \vev{\chi^{(i)}}|^2 + m_{FN}^2 & a_i F_i \\
    a_i^* F_i^* & |a_i \vev{\chi^{(i)}}|^2 + m_{\overline{FN}}^2 
    \end{array} \right)
\left( \begin{array}{c} FN \\ \overline{FN}^\dagger \end{array} \right)
\end{eqnarray}
where $m_{FN}^2$ and $m_{\overline{FN}}^2$ are the nonholomorphic
(mass)$^2$, $FN = 5^{FN(1)}, 5^{FN(2)}, 10^{FN}$, and
$F_{\chi}$ is vev of the auxiliary component of 
$\chi$.\footnote{In this one instance we denote a vev without braces.}
The mass eigenstates are
\begin{eqnarray}
m_{f_i}^2 + \textfrac{1}{2} (m_{FN_i}^2 + m_{\overline{FN_i}}^2) \mp
   \textfrac{1}{2} \sqrt{(m_{FN_i}^2 - m_{\overline{FN}_i}^2)^2 
                         + 4 |a_i F_i|^2}
\label{FN-scalar-masses-eq}
\end{eqnarray}
with a mixing angle
\begin{eqnarray}
\tan 2\theta_i &=& \frac{2 a_i F_i}{m_{FN_i}^2 - m_{\overline{FN}_i}^2} \; .
\end{eqnarray}
In practice, $a \vev{\chi} \gg m_{FN}^2, m_{\overline{FN}}^2$
and $2 |a F| \gg m_{FN}^2 - m_{\overline{FN}}^2$, thus the 
nonholomorphic correction to the gaugino masses is rather small.

We define $\MFN$ as the scale where the FN fields are 
integrated out.  Roughly, this is an average of the FN fermion
masses\footnote{Large splittings between the different FN fields 
would generate corrections to using a naive average, but in
practice this turns out not to be a concern for the model.}.
This serves as the boundary condition for the renormalization 
scale dependent supersymmetry breaking masses for those contributions
induced by the MSSM gauge interactions with FN fields.

If we take $F = F_{\chi^{(1)}} = F_{\chi^{(2)}} = F_{\chi^{(3)}}$
and define $\MFN \equiv a_1 \vev{\chi^{(1)}} = a_2 \vev{\chi^{(2)}} 
= a_3 \vev{\chi^{(3)}}$, then we find the gaugino masses at the 
weak scale to be about
\begin{eqnarray}
M_1(M_1) &\simeq& (190 \; \mathrm{GeV}) \frac{F/\MFN}{28 \; \mathrm{TeV}} 
  \label{M1-approx-eq} \\
M_2(M_2) &\simeq& (380 \; \mathrm{GeV}) \frac{F/\MFN}{28 \; \mathrm{TeV}} 
  \label{M2-approx-eq} \\
M_3(M_3) &\simeq& (1000 \; \mathrm{GeV}) \frac{F/\MFN}{28 \; \mathrm{TeV}} 
  \; . \label{M3-approx-eq}
\end{eqnarray}

\subsection{U(1)$_F$ charged scalar masses and gaugino mass}

After the DSB sector is integrated out, U(1)$_F$ charged scalars
and the U(1)$_F$ gaugino will acquire two-loop and one-loop 
contributions to their soft supersymmetry breaking masses
respectively.  In general they will take the form
\begin{eqnarray}
m_i^2 &=& c \frac{q_i^2 g_F^4}{256 \pi^4} \LambdaDSB^2 \\
M_F &=& C \frac{g_F^2}{16 \pi^2} \LambdaDSB \; ,
\end{eqnarray}
where $\LambdaDSB$ is the scale where the DSB field are integrated
out, $g_F$ is the U(1)$_F$ gauge coupling, 
and $q_i$ is the U(1)$_F$ charge of the field $i$.
The overall coefficients $C, c$ (as well as their signs) are not known 
without specifying a 
DSB sector\footnote{We do, however, expect the coefficient $c$ to
be the same for all the scalars.}.  However, we can parameterize 
the effective mass scale $\tilde{m}^2$ induced by the DSB fields as 
\begin{eqnarray}
\tilde{m}^2 &=& c \frac{g_F^4}{256 \pi^4} \LambdaDSB^2 \; ,
     \label{mtilde2-eq}
\end{eqnarray}
giving\footnote{We neglect the overall coefficient in front of the
U(1)$_F$ gaugino mass, which does not affect our results.}
\begin{eqnarray}
m_i^2 &=& q_i^2 \tilde{m}^2 \\
M_F &\sim& |\tilde{m}| \label{U1F-gaugino-mass-eq} \; .
\end{eqnarray}
There is one additional subtlety.  Since the U(1)$_F$ is anomalous after
the DSB sector is integrated out, there is an additional contribution
to the scalar (mass)$^2$ from an effective Fayet-Iliopoulos term $\xi^2$
which appears at one-loop (and thus, in general, $\xi^2\gg\tilde{m}^2$).
This can be seen by examining the effective potential,
\begin{eqnarray}
V_{\mathrm{eff}} &\supset& \frac{g_F^2}{2}
    \left[ \xi^2 + \sum_i q_i |\phi_i|^2 \right]^2
  + \tilde{m}^2 \sum_i q_i^2 |\phi_i^2| \; .
\end{eqnarray}
When appropriate superpotential couplings are included, some flavons have 
non-zero expectation values which mostly cancel the FI term\footnote{This
in general occurs at a long-lived, local minimum.}.
After shifting these fields by their vevs, a constant,
$\xi'^2\sim \mathcal{O}(\tilde{m}^2)$, remains in the U(1)$_F$ $D$-term. 
The mass of the scalar fields at this minimum is
\begin{eqnarray}
m_i^2 &=& q_i (q_i + q_\xi) \tilde{m}^2
\label{scalar-mass-eq}
\end{eqnarray}
where $q_\xi \equiv g_F^2 \xi'^2/\tilde{m}^2$.  This simple result
is central to the phenomenological analyses that we carry out in the
remainder of the paper.

\subsection{Uncharged scalar masses}

The matter fields that are uncharged under the U(1)$_F$, namely 
the third generation and the Higgs, acquire supersymmetry breaking
masses from several sources.  Ordinary two-loop gauge-mediated
contributions arise after integrating out the FN sector, 
however now the FN scalars acquire both holomorphic and
nonholomorphic contributions after the DSB sector is integrated out.
As we have emphasized above, these nonholomorphic contributions
are soft masses that amount to shifts in the scalar masses away from 
the naive holomorphic calculation involving simply $\vev{\chi}$
and $F_{\chi}$.  [See Eq.~(\ref{FN-scalar-masses-eq}).]
Unlike the one-loop calculation of the MSSM gaugino masses, 
however, the uncharged scalars acquire logarithmically enhanced
contributions proportional to the supertrace over each
messenger multiplet \cite{PT}.  Explicitly, the result 
is \cite{PT, AG}\footnote{We employ the $\overline{\mathrm{DR}}'$ 
scheme \cite{JJMVY, AG}.}
\begin{eqnarray}
m_j^2 &=& \sum_{i} m_{f_i}^2 \left[ (2 y_1^i + 2 y_2^i - 4)
      \times \left( 2 + \gamma_E - \ln 4 \pi  
                    - \ln \frac{\LambdaDSB^2}{m_{f_i}^2} \right)
      + G(y_1^i, y_2^i, \theta_i)
      + G(y_2^i, y_1^i, \theta_i) \right] \nonumber \\
& &{} \times \sum_a C_a(j) S_a(i) \left( \frac{\alpha_a}{4 \pi} \right)^2
\label{uncharged-scalar-masses-eq}
\end{eqnarray}
where $a =$ (U(1)$_Y$, SU(2)$_L$, SU(3)$_c$), $C_a(j)$ is the quadratic 
Casimir of the group $a$ for the scalar field $j$, $S_a(i)$ is the 
Dynkin index of the group $a$ for the messenger field $i$, and
\begin{eqnarray}
G(y_1, y_2, \theta) &=&{} 
      + (\sin 2\theta)^2 \, \left[ - \textfrac{1}{2} \ln^2 y_1 
      + y_1 \ln y_1 \ln y_2 - y_1 \Li{2} \left(1 - \frac{y_1}{y_2}\right)
      \right] \nonumber \\
& &{} + \textfrac{1}{2} \ln^2 y_1 
      + y_1 \left[ 2 \ln y_1 
                   + \ln^2 y_1 + \Li{2} \left(1 - y_1\right)
                   - \Li{2} \left(1 - \frac{1}{y_1} \right) \right] \; .
\end{eqnarray}
where $\Li{2}$ is the dilogarithm function.  This result
is clearly a sum of two components: one proportional to the
$\Str \, m^2 = m_{f_i}^2 (2 y_1^i + 2 y_2^i - 4)$,
and the other a threshold function $G(y_1^i, y_2^i, \theta_i)
+ G(y_2^i, y_1^i, \theta_i)$.  It straightforward to show that
in the limit $\Str \, m^2$ vanishes (and thus
$(\sin 2\theta)^2 = 1$), we recover the analytic result
for the minimal messenger model calculated in 
Refs.~\cite{DGP, MartinGMSB}.  
Conversely, in the limit $m_{f_i} \ra 0$ but evaluating
the logarithm at the scale $Q$, we obtain a finite result proportional 
only to $\Str \, m^2$.  Indeed, the nonlogarithmic finite 
terms partially cancel, and thus to a good approximation the 
contribution to the (mass)$^2$ of the uncharged scalar fields
at the scale $Q$, in limit $m_{f_i} \ra 0$, is
\begin{eqnarray}
m_j^2(Q) &=& -8 \sum_a C_a(j) (\mathrm{tr}_i \; S_a(i) m_i^2) 
    \left( \frac{\alpha_a}{4 \pi} \right)^2 \ln \frac{\LambdaDSB}{Q}
    \label{PT-approx-eq}
\end{eqnarray}
which is immediately recognized as simply the two-loop renormalization
of the uncharged fields due to MSSM gauge interactions with charged
fields.  This term was first calculated in Refs.~\cite{MartinVaughn, 
Yamada}, and recognized to be of crucial importance to dynamical 
supersymmetry breaking models in Refs.~\cite{RandallStr, AHMRM, PT}.

This result is actually more general than we have presented.
Anytime gauge or Yukawa interactions connect apparently
separate sectors of the theory, we can expect to find 
logarithmically divergent contributions to the scalar (mass)$^2$
proportional to the supertrace of the other fields.

Uncharged scalars, therefore, receive contributions from
logarithmically-enhanced two-loop gauge-interactions with 
FN fields, and also from the U(1)$_F$ charged first and second
generations.  It is important to stress that since the 
logarithmically enhanced correction dominates, there is 
little distinction between FN fields and the first and 
second generations, despite the former being integrated out
at $\MFN$.

The above contributions are generic in that they contribute
to all scalar fields.  In addition, some uncharged scalars also 
have couplings with charged scalar fields through one-loop,
two-loop, or higher order Yukawa interactions.  These
interactions induce additional contributions to the
uncharged scalar (mass)$^2$ that will be calculated
using the renormalization group in Sec.~\ref{DSB-to-FN-sec}.

\subsection{Scalar trilinear couplings}

Once couplings between the FN fields and the MSSM fields are
introduced, scalar trilinear couplings are generated 
\cite{GiudiceRattazziwf}.  They arise as a result of
holomorphic supersymmetry breaking, and are proportional
to $F_{\chi}/\MFN$ multiplied by a one-loop factor
proportional to superpotential couplings.  Since the
scalar trilinear couplings enter the off-diagonal elements
of mass matrices, we need only consider those fields 
where the diagonal elements are not tremendous, and
the off-diagonal elements are not suppressed by a small
fermion mass.  Thus, only the scalar trilinear couplings associated 
with the top, bottom, or tau Yukawa coupling are relevant to this 
discussion.

The scalar trilinear couplings that arise after $\chi^{(1)}$ 
acquires a vev for both its scalar and auxiliary components are 
\cite{GiudiceRattazziwf}
\begin{eqnarray}
A_t &=& -\frac{1}{16 \pi^2} \Big( \lambda^u_{2,1} \Big)^2 
        \frac{F_{\chi^{(1)}}}{\vev{\chi^{(1)}}} \\
A_b &=& -\frac{1}{16 \pi^2} \left[ \Big( \lambda^u_{2,1} \Big)^2 
        + \Big( \lambda^d_{4,1} \Big)^2 \right] 
        \frac{F_{\chi^{(1)}}}{\vev{\chi^{(1)}}} \\
A_\tau &=& -\frac{1}{16 \pi^2} \left[ \Big( \lambda^u_{2,2} \Big)^2 
        + \Big( \lambda^d_{4,2} \Big)^2 \right] 
        \frac{F_{\chi^{(1)}}}{\vev{\chi^{(1)}}} \; .
\end{eqnarray}
These couplings enter the off-diagonal elements of the well-known 
sfermion mass matrices:
\begin{eqnarray}
\mathbf{M}_{\tilde{f}} &=& \left( \begin{array}{cc} 
    m_L^2 + m_f^2 + D_L & m_f (A_f - \mu a_\beta) \\
    m_f (A_f - \mu a_\beta) & m_R^2 + m_f^2 + D_R 
    \end{array} \right) \; ,
\end{eqnarray}
where $m_L^2$, $m_R^2$ are the SU(2)$_L$ doublet and singlet soft (mass)$^2$,
$m_f$ is the fermion mass, $D_L$ and $D_R$ are the MSSM $D$-terms, 
and $a_\beta = 1/\tan\beta$ for up-type sfermions, and 
$a_\beta = \tan\beta$ for down-type sfermions.
If $|\mu| \gsim A_f$, which is expected, $A_b$ and $A_\tau$ can be 
largely ignored since $\tan\beta$ is large in the model, and thus
the off-diagonal term is roughly $-m_f \mu \tan\beta$.
$A_t$ is largely irrelevant since the diagonal components are 
expected to be much larger than $m_t A_t$ (or $m_t \mu / \tan\beta$).
Numerical calculations confirm these expectations, as we will see later
in the paper.

\section{Renormalization group evolution}
\label{RGE-sec}

There are two separate mass scales in this model where soft 
supersymmetry breaking masses are induced in light fields after 
the heavy fields of some sector of the theory are integrated out.  
Soft masses are induced for U(1)$_F$ charged fields after the 
DSB sector is integrated out (we assume this occurs at $\LambdaDSB$), 
and soft masses are induced for MSSM fields after the FN sector is 
integrated out (we assume this occurs at $\MFN \simeq a_i \vev{\chi^{(i)}}$).  
The first and second generations receive contributions after 
both sectors are integrated out, 
although the U(1)$_F$ induced mass is always much larger 
than ordinary SU(3)$_c \times$ SU(2)$_L \times$ U(1)$_Y$ 
gauge-mediated contributions.

Once one sector of the model receives supersymmetry breaking
masses, there are several ways in which masses can be induced 
for sparticles in other sectors of the model.  We will be most 
concerned with the following effects:  Two-loop MSSM gauge interactions
between the heavy first and second generations with the lighter
third generation.  One-loop gauge interactions between
the U(1)$_F$ charged fields due to the Fayet-Iliopoulos 
$D$-term.  Yukawa interactions between U(1)$_F$ charged fields 
and uncharged fields leading to one-loop and two-loop 
contributions to scalar masses.  All of these contributions 
arise from renormalization group evolution.

\subsection{DSB scale to the FN scale}
\label{DSB-to-FN-sec}

After the DSB sector is integrated out, it is assumed that all 
of the charged fields acquire masses which depend on their
charge, i.e., Eq.~(\ref{scalar-mass-eq}) for the scalars and 
Eq.~(\ref{U1F-gaugino-mass-eq}) for the U(1)$_F$ gaugino mass,
at the scale $\LambdaDSB$.  All of the uncharged field masses are 
assumed to vanish at $\LambdaDSB$.  This serves as the boundary 
conditions for renormalization group evolution between the 
DSB scale and the FN scale.  Note that the only parameters that
enter the boundary conditions are the charges of the fields, 
the universal (mass)$^2$ term $\tilde{m}^2$, and the ratio $q_\xi$ 
of the FI $D$-term to $\tilde{m}^2$.

It is in the evolution to lower scales that a proliferation 
of parameters enter the calculations of the renormalization 
group evolution, and hence the masses.  
These include the gauge couplings $g_1$, $g_2$, 
$g_3$, and $g_F$, and numerous superpotential parameters 
[see Eq.~(\ref{superpot-component-eq})].  
We have calculated the important renormalization group 
equations that are expected to evolve significantly.  In particular,
we can safely neglect the evolution of 
the superpotential parameters, gauge couplings, and the U(1)$_F$ 
gaugino mass between $\LambdaDSB$ and $\MFN$, 
although the scalar (mass)$^2$ for all the matter fields (FN and MSSM) 
must be evolved.  The one-loop RG evolution equations for the 
scalar (mass)$^2$ of all relevant fields are given in 
Appendix~\ref{DSB-FN-RGE-app}.

We numerically integrated the coupled RG evolution for the 
calculations below.  However, for several fields we provide 
approximate expressions for the dominant RG contributions for the 
scalar (mass)$^2$ applicable at the scale $\MFN$.  
The uncharged fields acquire masses from two-loop gauge-mediation,
Eq.~(\ref{uncharged-scalar-masses-eq}), as well as contributions from 
superpotential couplings at one-loop and two-loops to charged fields.
The third generation component fields in the $\overline{5}^3$
have the direct superpotential coupling $b_5$ to U(1)$_F$ charged
fields, giving the one-loop contributions
\begin{eqnarray}
\Delta m^2_{\overline{d}^3}(\MFN) &=& -\frac{\tilde{m}^2}{8\pi^2} 
    \Big( b_5^{(d)} \Big)^2 q_{\phi^{(5)}}^2
    \ln \frac{\LambdaDSB}{\MFN} \label{dbar3-approx-eq} \\
\Delta m^2_{\overline{L}^3}(\MFN) &=& -\frac{\tilde{m}^2}{8\pi^2} 
    \Big( b_5^{(L)} \Big)^2 q_{\phi^{(5)}}^2 
    \ln \frac{\LambdaDSB}{\MFN} \; . \label{Lbar3-approx-eq}
\end{eqnarray}
The down-type Higgs field also has direct superpotential
couplings to U(1)$_F$ charged fields, giving
\begin{equation}
\Delta m^2_{H^d}(\MFN) = -\frac{\tilde{m}^2}{8\pi^2} \left[ \, 
      3 \Big( \lambda^d_{1,1} \Big)^2 q_{10^1}^2 
    + \Big( \lambda^d_{1,2} \Big)^2 q_{10^1}^2 
    + 3 \Big( \lambda^d_{2,1} \Big)^2 q_{10^2}^2 
    + \Big( \lambda^d_{2,2} \Big)^2 q_{10^2}^2 
    \right] \ln \frac{\LambdaDSB}{\MFN} \; . \label{Hd-approx-eq}
\end{equation}
Notice that these one-loop contributions are positive if 
$\tilde{m}^2 < 0$.  The component fields $Q^3$ and $e^3$ do not have 
direct superpotential couplings to U(1)$_F$ charged fields.  However, 
they do have couplings to uncharged fields that acquire masses at 
one-loop, namely $\overline{d}^3$, $\overline{L}^3$, and $H^d$.
Thus, $Q^3$ and $e^3$ acquire effectively two-loop contributions,
\begin{eqnarray}
\Delta m^2_{Q^3}(\MFN) &=& \frac{Y_b^2 \tilde{m}^2}{128\pi^4} \bigg[ \,
      3 \Big( \lambda^d_{1,1} \Big)^2 q_{10^1}^2 
    + \Big( \lambda^d_{1,2} \Big)^2 q_{10^1}^2 
    + 3 \Big( \lambda^d_{2,1} \Big)^2 q_{10^2}^2 
    + \Big( \lambda^d_{2,2} \Big)^2 q_{10^2}^2 \nonumber \\
& &{} \qquad \quad + \Big( b_5^{(d)} \Big)^2 q_{\phi^{(5)}}^2 \bigg]
    \times \left( \ln \frac{\LambdaDSB}{\MFN} \right)^2 
    \label{Q3-approx-eq} \\
\Delta m^2_{e^3}(\MFN) &=& \frac{Y_\tau^2 \tilde{m}^2}{64\pi^2} \bigg[ \,
      3 \Big( \lambda^d_{1,1} \Big)^2 q_{10^1}^2 
    + \Big( \lambda^d_{1,2} \Big)^2 q_{10^1}^2 
    + 3 \Big( \lambda^d_{2,1} \Big)^2 q_{10^2}^2 
    + \Big( \lambda^d_{2,2} \Big)^2 q_{10^2}^2 \nonumber \\
& &{} \qquad \quad + \Big( b_5^{(L)} \Big)^2 q_{\phi^{(5)}}^2 \bigg]
    \times \left( \ln \frac{\LambdaDSB}{\MFN} \right)^2 \; .
    \label{e3-approx-eq}
\end{eqnarray}
Notice that these two-loop superpotential-induced contributions 
are \emph{negative} if $\tilde{m}^2 < 0$.  Finally, $u^3$ and 
$H^u$ acquire superpotential-induced contributions only at higher
orders, which we do not present here.  (These small contributions can be 
ignored since the two-loop gauge-mediated contributions dominate 
for these fields.)

There is also significant evolution among the charged scalar 
fields due to U(1)$_F$ gaugino interactions and the FI $D$-term.  
In both cases the RG contributions depend on the separation
between $\MFN$ and $\LambdaDSB$ and the U(1)$_F$ coupling strength $g_F$.  
The gaugino contribution can be estimated
\begin{eqnarray}
\Delta m_i^2 &=& \frac{q_i^2 g_F^2}{2\pi^2} |\tilde{m}^2| 
    \ln \frac{\LambdaDSB}{\MFN} \; .
    \label{gaugino-mass-approx-eq}
\end{eqnarray}
The contribution to the RG evolution from the FI $D$-term is shown
in Eq.~(\ref{weighted-sum-eq}), and is really a sum of two distinct
components.  At $\LambdaDSB$ the masses of all charged fields are 
equivalent to $q_i (q_i + q_\xi) \tilde{m}^2$, and so we can 
write this term at $Q = \LambdaDSB$ quite simply:
\begin{eqnarray}
S'(\LambdaDSB) &=& \tilde{m}^2 \sum_j S(j) q_j^3 
    + \tilde{m}^2 q_\xi \sum_j S(j) q_j^2 \; .
\end{eqnarray}
The first term is proportional to the [U(1)$_F$]$^3$ anomaly,
while the second component is proportional to the FI $D$-term
that results after the DSB fields are integrated out.
A term proportional to the anomaly should not be surprising since the 
U(1)$_F$ is both anomalous and has a FI $D$-term after the DSB sector 
is integrated out.
Since the anomaly contribution is a weighted sum over (charge)$^3$, 
several cancellations reduce this contribution to be typically 
much smaller than the second term proportional to $q_\xi$.  
We emphasize that the second term is generated as a
result of DSB sector dynamics, and in particular does not depend on 
whether, for example, the U(1)$_F$ anomaly is canceled among the light
(non-DSB) fields (see Ref.~\cite{DNS}).  We can approximate
this contribution to U(1)$_F$ charged scalar fields as
\begin{eqnarray}
\Delta m_i^2 &=& -\frac{q_i g_F^2}{8\pi^2} S' \ln \frac{\LambdaDSB}{\MFN} \; ,
    \label{FI-D-term-approx-eq}
\end{eqnarray}
which is negative if $q_i \tilde{m}^2 > 0$, as is the case 
for the first and second generations.

The size of these terms is not known precisely, since we 
expect, for example, additional flavon fields charged under the
U(1)$_F$ but uncharged under the SM gauge group to be present.  
If we assume that the anomaly term nearly vanishes,
then the term proportional to $q_\xi$ is really
a lower bound on the size of the RG contribution.  In the sum 
we include only the fields charged under the U(1)$_F$ that
we have stated in this paper, and thus it should be remembered that
the contribution we include is really a lower bound on the FI $D$-term 
induced evolution.

\subsection{FN scale to the weak scale}

After the FN sector is integrated out, the remaining fields
in the model are the MSSM fields and the light flavons.
Since the light (and heavy) flavons are uncharged under
the SM gauge group, they do not play any significant dynamical
role in the low energy phenomenology.  (They are, however,
essential to generating the Yukawa coupling hierarchy.)
The MSSM fields, in contrast, are precisely the ones that
we expect to find in upcoming collider experiments if 
weak scale supersymmetry is indeed manifest in Nature.

For the evolution between the FN scale and the weak scale, we have 
used the full two-loop RG equations for all superpotential couplings 
and soft mass parameters in the MSSM \cite{MartinVaughn, Yamada}.
These include the dominant two-loop SM gauge interactions between
the first and second generations with the third generation,
as well as several subdominant terms.

One technical subtlety arises as to how (or at what scale) 
to precisely decouple the first and second generations from
the third generation evolution.  Unlike the FN sector (or any other 
heavy vector-like sector), it is more subtle to integrate out these 
contributions due to the light fermion masses.  High precision 
demands two-loop matching conditions near the weak scale for the 
third generation sparticle masses, which could be comparable to 
three-loop log-enhanced renormalization group contributions.  
An alternative was performed in Ref.~\cite{AG} by taking $m_f \ra 0$
limit of Eq.~(\ref{uncharged-scalar-masses-eq}) and finding
that a nontrivial finite term proportional to the messenger 
supertrace is present.  In particular, they found that
Eq.~(\ref{PT-approx-eq}) should be modified by letting
\begin{eqnarray}
\ln \frac{\LambdaDSB}{Q} &\ra& 
    \left[ \ln \frac{\LambdaDSB}{Q} + \frac{\pi^2}{6} \right] \; ,
\end{eqnarray}
which we can rewrite as a shift of the scale 
\begin{eqnarray}
Q &\ra& \frac{Q}{\exp \pi^2/6} \; \sim \; \textfrac{1}{5} Q \; .
\end{eqnarray}
In their approach, the scalars were decoupled at the first and 
second generation mass scale $Q = m_{1st,2nd}$, which is evidently
approximately equivalent to performing log evolution \emph{only} via 
Eq.~(\ref{PT-approx-eq}), but down to a scale 
$Q \sim \textfrac{1}{5} m_{1st,2nd}$.  Hence, a good approximation
is to evolve the third generation \emph{including} the two-loop 
effects of the first and second generations down to the weak scale,
which is expected to be about a factor of $10$ below $m_{1st,2nd}$.
This is what we do in this paper.  

Another issue that arises in the evolution from the FN scale
to the weak scale is possible presence of the one-loop hypercharge 
$D$-term that is proportional to $\tr Y_i m_i^2$.  Ordinarily this
term is forbidden in supergravity-mediated models with common 
scalar masses since $m^2 \tr Y_i$ is simply proportional to the 
gravitational anomaly that is required to vanish.  In gauge-mediated 
models, each scalar $m_i^2$ is a sum over several contributions 
proportional to its gauge charges, but again the hypercharge $D$-term 
cancels due to requiring the mixed and triangle anomalies to vanish.
However, this problem would seem to be particularly acute for 
models with heavy 
first and second generations.  Interestingly, $\tr Y_i m_i^2$ vanishes
for matter in (for example) complete SU(5) multiplets,
which is a strong motivation in our scenario to enforce that
\emph{all} of the first and second generations are heavy
(see Sec.~\ref{modify-model-sec} for further discussion).
Furthermore, the gauge anomaly argument also applies to our model
for the two-loop gauge-mediated contributions, but does \emph{not} 
apply to the Yukawa-induced contributions in 
Eqs.~(\ref{dbar3-approx-eq})--(\ref{e3-approx-eq}).
This is obvious to see with, e.g., the contributions to $\overline{L}^3$
and $H^d$ that have identical gauge quantum numbers but 
distinct superpotential couplings.\footnote{Is is easy to show that 
as the superpotential parameters are taken to zero, the
hypercharge $D$-term generated at the FN scale goes to zero.}
The effect of these contributions on the size of the hypercharge 
$D$-term can be enhanced if the SU(5) component superpotential 
couplings are widely split apart.  In all cases the $\tr Y_i m_i^2$ 
contribution to the scalar (mass)$^2$ RGE for a given sparticle is 
suppressed by $g_1^2$, and thus the $D$-term term imparts its 
greatest relative one-loop effect on the stau mass.  Depending on the 
sign of $\tr Y_i m_i^2$ (summed over the MSSM fields), the contribution
can be either positive or negative, although a significant positive
contribution is excluded by requiring proper EWSB, since one would also 
have an undesired significant positive contribution to $m_{H^u}^2$.

\section{Dynamics of the model}
\label{dynamics-sec}

The dynamics of the model that is of interest to us includes the 
full spectrum of both the FN fields and the MSSM fields.  
It should now be obvious that this is a highly interconnected model, 
and thus to extract e.g.\ weak scale observables with any
accuracy requires a complete implementation of the model
with as few approximations as possible.  To this end, 
we have implemented the full set of RG evolution equations 
and the relevant boundary conditions in a computer program for 
numerical evaluation.  The program uses a fourth order Runge-Kutta 
algorithm to evaluate the the coupled differential equations.  
The general procedure is to:
\begin{itemize}
\item[1.] evolve the SM gauge couplings to the FN scale
\item[2.] set the DSB boundary conditions
\item[3.] evolve all of the scalar (mass)$^2$ from
          the DSB scale to the FN scale (neglecting evolution of the
          gauge couplings and the superpotential parameters)
\item[4.] integrate out the FN sector, adding the appropriate 
          contributions to the soft mass parameters
\item[5.] evolve to the weak scale, decoupling the heavy states
          as appropriate
\item[6.] evaluate the one-loop effective potential to extract
          $\mu^2$ and $B_\mu$
\item[7.] compute the mass eigenstates and appropriate mixing
          matrices of the weak scale sparticles
\item[8.] repeat steps $1$ through $7$ until one achieves convergence
          (about three or four repetitions is always sufficient)
\end{itemize}
Each one of these steps involves numerous technical subtleties,
and we discuss several important ones in the following sections.

\subsection{A minimal set of parameters}

Applying no constraints, the minimal model has:
\begin{itemize}
\item four gauge couplings
\item $30$ superpotential parameters (plus phases\footnote{When considering
the tree-level potential below the DSB scale with flavons replaced with
their vevs we find the following physical phases:  two in each of the full 
down-quark and charged lepton mass matrices (with the same phases appearing
in the sparticle matrices), one phase associated with the $\mu$-term and 
a phase for each flavon auxiliary component vev.  One combination of these
phases plays the role of the CKM phase while the remainder either contribute
to $B^0-\overline{B}^0$ mixing significantly or are unimportant 
for phenomenology
(except for one from the $F$-components of the $\phi$'s which must be
constrained).  This is true for the most part because all squarks in 
this model are $\gsim 1$ TeV\@.  We shall leave the phases out of the
parameter counting and discuss their generic effects in 
Sec.~\ref{CP-violation-sec}.})
\item two Higgs scalars that acquire vevs
\item eight flavons with scalar component vevs
\item three flavons with auxiliary component vevs 
      (see Sec.~\ref{holomorphic-sec})
\item $\tilde{m}^2$: universally induced scalar (mass)$^2$
      (we take $M_{\mathrm{U(1)}_F} \sim \sqrt{|\tilde{m}^2|}$)
\item $q_\xi$: ratio of FI $D$-term to $\tilde{m}^2$
\item six undetermined U(1)$_F$ charges
\end{itemize}
This may seem overwhelming, however consider that all\footnote{Note 
that the phase in the CKM matrix will be discussed in 
Sec.~\ref{CP-violation-sec}, and $\theta_{QCD}$ is not physical 
since the model has $m_u = 0$ (see also Sec.~\ref{m-u-nonzero-sec}).}
of the parameters of the SM, namely the the gauge couplings, quark
and charged lepton Yukawa couplings, CKM matrix elements, Higgs vev, 
and Higgs mass are included in the above parameter set.
In particular, the following is fixed by experiment:
\begin{itemize}
\item three gauge couplings $g_1$, $g_2$, and $g_3$
\item Higgs vevs determined by effective potential minimization
      up to $\mathrm{sign}(\mu)$ and the ratio 
      $\tan\beta \equiv \vev{H^u}/\vev{H^d}$
\item three superpotential parameters, $Y_t$, $Y_b$, and $Y_\tau$
\item six fermion masses and three CKM elements that are determined 
      by several superpotential parameters and ratios of light to
      heavy flavon vevs
\end{itemize}
Generally the superpotential parameters associated with 
the $\chi$ flavons can be absorbed into the vevs (although one 
may still need to distinguish between quark and lepton couplings).  
Most of the other superpotential couplings feed into the quark 
and lepton mass matrices, but are otherwise unimportant.  
This is fortunate in that the number of parameters that 
influence the weak scale phenomenology is a much smaller set.
The exceptions to this rule are the superpotential couplings
that enter the higher order calculations of the uncharged scalar 
masses, namely $b_5^{(d)}$, $b_5^{(L)}$, $\lambda^d_{1,1}$, 
$\lambda^d_{1,2}$, $\lambda^d_{2,1}$, and $\lambda^d_{2,2}$.

We can simplify the model by assuming
\begin{itemize}
\item $\MFN = a_i \vev{\chi^{(i)}}$, the latter assumed to be 
      roughly equal
\item $F_{\chi} = a_i F_{\chi^{(i)}}$, the latter assumed to be 
      roughly equal
\item $q_{\chi^{(1)}} = 0$; $q_{\chi^{(2)}}/q_{\chi^{(3)}} \equiv q_r \sim 1$
\item $q = q_{10^1} \sim q_{10^2} \sim q_{\overline{5}^1} 
        \sim q_{\overline{5}^2}$
\item $b_5 = b_5^{(d)} \sim b_5^{(L)}$
\item $\lambda^d_1 = \lambda^d_{1,1} \sim \lambda^d_{1,2}$
\item $\lambda^d_2 = \lambda^d_{2,1} \sim \lambda^d_{2,2}$
\end{itemize}
An overall U(1)$_F$ charge can be absorbed in $\tilde{m}$ (or equivalently 
in $g_F$), as is usual for an Abelian group.  Thus if we assume
all of the first and second generation charges are closely 
comparable (but not exactly equal), the only undetermined charge
is the ratio $q_\xi/q$.

Thus we arrive at the ``minimal set of parameters'' capable of
adequately describing the model:
\begin{equation}
\MFN, \; F_{\chi}, \; \LambdaDSB, \; \tilde{m}^2, \; b_5, \; \lambda^d_1, \; 
\lambda^d_2, \; q_\xi/q, \; \mathrm{sgn}(\mu), \; \tan\beta
\end{equation}
We call this the ``minimal flavor-mediated model'' (MFMM), 
that is hereafter the set of free parameters we consider in the 
model.\footnote{Note that in the limit $\tilde{m} \ra 0$ the MFMM
becomes an ordinary gauge-mediated model with $n_{10 + \overline{10}} = 1$ 
and $n_{5 + \overline{5}} = 2$.  We explicitly verified that our 
calculations do indeed reproduce the spectrum of an ordinary
gauge-mediated model \cite{DTW, BMPZ, AKM1} in this limit.}
Some variations from this minimal parameter set are explored 
in more detail in Sec.~\ref{modify-model-sec}.

\subsection{General discussion}

The MSSM gaugino masses are proportional to $F_{\chi}/\MFN$, and are
well approximated by Eqs.~(\ref{M1-approx-eq})-(\ref{M3-approx-eq}).
A rough lower bound is $F_{\chi}/\MFN \gsim 7$~TeV, since charginos 
lighter than about $90$~GeV have not been seen at LEP\@.
Imposing some degree of naturalness suggests the gluino ought
not to be too heavy (for a recent discussion, see Ref.~\cite{KaneKing}).
Requiring the gluino be lighter than $1 \, (1.5)$~TeV implies
$F_{\chi}/\MFN \lsim 28 \, (35)$~TeV\@.  

The universal (mass scale)$^2$ induced by the DSB sector, $\tilde{m}^2$,
must be negative.  This is evident from 
Eqs.~(\ref{dbar3-approx-eq}),(\ref{Lbar3-approx-eq}) that
imply the one-loop Yukawa-induced (mass)$^2$ for $\overline{d}^3$
and $\overline{L}^3$ is positive only if $\tilde{m}^2 < 0$.
One might speculate that such contributions could be smaller
than positive log-enhanced two-loop gauge-mediated contributions
from the FN scalars.  However, $\tilde{m}^2 < 0$ (i.e., a negative
messenger field supertrace) is required to generate positive 
contributions to the third generation masses via two-loop log-enhanced
gauge-mediation.  Self-consistency of the above
condition requires $q < 0 < q_\xi$ and $|q| < q_\xi$,\footnote{This
generalizes to 
$q_{10^1},q_{10^2},q_{\overline{5}^1},q_{\overline{5}^2} < 0 < q_\xi$ and
$|q_{10^1}|,|q_{10^2}|,|q_{\overline{5}^1}|,|q_{\overline{5}^2}| < q_\xi$ 
for first and second generation charges that are not approximately 
degenerate.} such that the (mass)$^2$ for the first and second 
generations is positive.  Finally, preliminary work regarding 
the U(1)$_F$ charge relations that would generate the required
flavon vev hierarchy suggested \cite{KLMNR} that
$|q_\chi| > |q|$, which we note is satisfied by the above relations.

The FN scale $\MFN$ is generically expected to be of order $100$~TeV 
\cite{KLMNR}, which is roughly the vev of the heavier $\chi$ flavons.
The ratios $\vev{\phi}/\vev{\chi}$ are determined by the fermion
mass hierarchy, and are thus largely fixed by experiment.  

$\tilde{m}^2$ and the DSB scale $\LambdaDSB$ are in principle 
related up to a group theory constant as shown in Eq.~(\ref{mtilde2-eq}).  
However, the group theory factor is not known without specifying
a DSB sector.  In ordinary gauge-mediation the group theory factor
that enters the two-loop expression for the MSSM scalar masses is 
$6 n_{10 + \overline{10}} + 2 n_{5 + \overline{5}}$ (i.e., 
determined by the content of the messenger sector),
so a reasonable estimate in our case is probably $1 \lsim |c| \lsim 10$.
We can rewrite Eq.~(\ref{mtilde2-eq}) as 
\begin{eqnarray}
\tilde{m}^2 &\simeq& (10 \; \mathrm{TeV})^2 \,\, \frac{c}{3} 
    \left( \frac{g_F}{0.3} \right)^4
    \left( \frac{\LambdaDSB}{10^7 \; \mathrm{GeV}} \right)^2 \; ,
    \label{mtilde2-approx-eq}
\end{eqnarray}
with some foresight into the next section where constraints on $g_F$
are discussed.

\subsection{The first and second generation mass scale}

\begin{figure}[!t]
\centering
\epsfxsize=5.0in
\hspace*{0in}
\epsffile{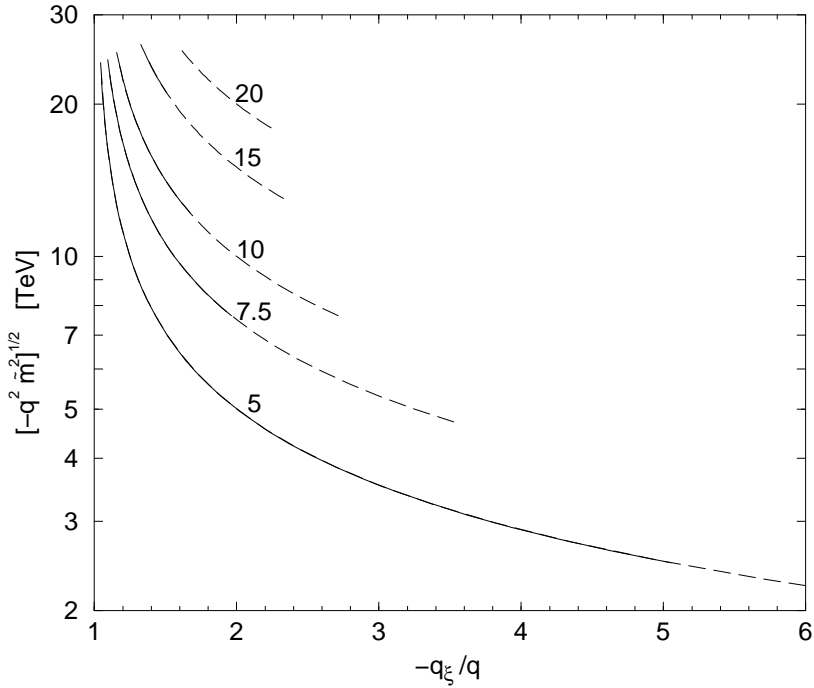}
\caption{Contours of the first and second generation masses
in units of TeV induced at the DSB scale as a function of the 
parameters $q^2 \tilde{m}^2$ and $q_\xi/q$.  We have chosen
$\tan\beta = 10$ to fully illustrate the dependence on $m_{1st,2nd}$, 
however it should be noted that the upper limit on $m_{1st,2nd}$
is significantly lowered as $\tan\beta$ is increased.  The solid lines 
correspond to the parameter space allowed for $\tan\beta = 30$,
while the solid $+$ dashed lines correspond to the parameter space 
allowed for $\tan\beta = 10$.}
\label{firstsecond-fig}
\end{figure}

The mass of the first and second generation scalars induced by the
DSB sector at the scale $Q = \LambdaDSB$ is given by 
Eq.~(\ref{scalar-mass-eq}).  This is the physical mass 
under the approximation that RG evolution for the first and 
second generation scalar masses can be ignored.  
In Fig.~\ref{firstsecond-fig}
we show contour lines for various first and second generation
masses within the allowed region of parameter space.  Upper bounds
on $|q^2 \tilde{m}^2|$ (corresponding to the contours' far left
endpoint) arise from requiring that the flavon scalar (mass)$^2$ 
are positive.  Upper bounds on $|q_\xi/q|$ (corresponding to the
contours' far right endpoint) arise from requiring the lightest 
stau (mass)$^2$ be positive.  Notice that lowering $\tan\beta$
admits larger $|q_\xi/q|$, since lowering $\tan\beta$ also increases
the stau mass.  Thus, the internal dynamics of the model restrict 
the allowed range of input parameters, although the precise cutoffs 
do depend on other parameters including $\MFN$ and $F_\chi$.

\begin{figure}[!t]
\centering
\epsfxsize=5.0in
\hspace*{0in}
\epsffile{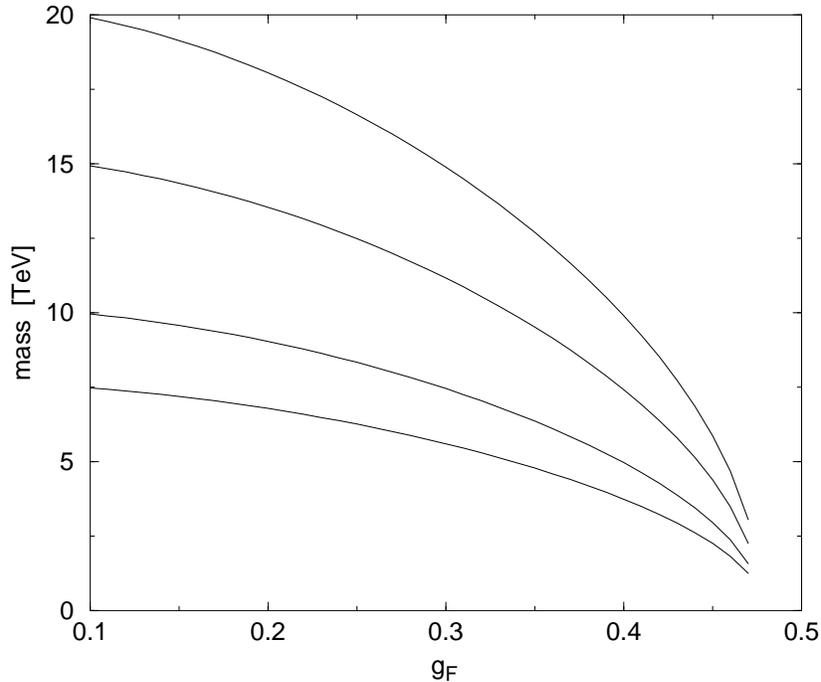}
\caption{RG evolution of U(1)$_F$ charged scalar masses resulting
from interactions with the U(1)$_F$ gaugino and the FI $D$-term.
The four lines correspond to $m_{1st,2nd}$ for a DSB model with
$(-q^2 \tilde{m}^2)^{1/2} = 20, 15, 10, 7.5$ TeV from top to
bottom.}
\label{gf-fig}
\end{figure}

There is, however, RG evolution of the first and second generation
masses.  Between the FN scale and the weak scale, there are
ordinary gauge interactions coupling to the MSSM gauginos,
but such contributions are very small since $M_a \ll m_{1st,2nd}$.
There are also the effective first and second generation Yukawa 
interactions that result after the flavons acquire vevs, but these 
contributions are also tiny.  It is the U(1)$_F$ interactions
induced by the U(1)$_F$ gaugino mass, and even more importantly, the
FI $D$-term that induce evolution among the U(1)$_F$ charged
scalar masses [see Eq.~(\ref{DSB-FN-RGE-eq})].  In 
Fig.~\ref{gf-fig} we illustrate the evolution
of the first and second generation scalar masses as a function of 
the undetermined gauge coupling $g_F$, for a particular 
choice of other parameters (key among them $\LambdaDSB/\MFN = 10^2$ and 
$\tan\beta = 10$).  Note that if the content of the DSB sector 
is fixed, then the scale $\LambdaDSB$ must also change inversely
proportional to $g_F^2$.
Other choices of the ratio $\LambdaDSB/\MFN$ can be 
obtained by rescaling
\begin{eqnarray*}
g_F^2 &\longrightarrow& \frac{g_F^2}{\ln 10^2} \ln \frac{\LambdaDSB}{\MFN} \; .
\end{eqnarray*}
Obviously the first and second generation scalar masses are 
dramatically reduced as the gauge coupling is increased, and this
provides another phenomenological constraint on the model.

The reason to emphasize the precise first and second generation mass
scale that is generated from the model is threefold.  First, the first and 
second generation mass scale determines the size of supersymmetric 
FCNC contributions to e.g.\ $K^0 \leftrightarrow \overline{K}^0$ mixing,
and so has indirect phenomenological relevance in FCNC processes.
Second, increasing the first and second generation mass scale reduces
the gauge couplings at a given high scale 
(i.e.\ $\Munif \sim 10^{16}$ GeV), avoiding the Landau 
pole that would be naively expected with the FN sector in the model.  
This will be discussed in greater detail in 
Sec.~\ref{gauge-coupling-unif-sec}.  Finally, and perhaps most
important, the first and second generation mass scale determines
the size of the two-loop negative gauge contributions to the
third generation.  Susinctly, increasing the first and second generation
mass scale implies the third generation is reduced.  This is
important because it suggests a no-lose principle for this model:  
Either supersymmetric contributions to FCNC should be seen 
or the third generation (in particular, the lightest stau) 
should be seen in experiment.  Simultaneously tightening these
experimental bounds, even by relatively small factors, rapidly 
reduces the allowed parameter space for this model.

\subsection{Superpotential couplings}

It is clear from Eqs.~(\ref{dbar3-approx-eq})-(\ref{e3-approx-eq})
that the superpotential couplings $b_5$, $\lambda^d_1$, and $\lambda^d_2$ 
determine both the positive one-loop and negative two-loop 
Yukawa-induced supersymmetry breaking contributions to the third generation.
Of greatest concern is the negative two-loop contribution to the
SU(2)$_L$ singlet stau (mass)$^2$, since the positive two-loop 
gauge-mediated log-enhanced contribution is proportional to the
relatively small coupling $g_1^4$ multipled by
\emph{one} power of $\ln (\LambdaDSB/\MFN)$.
The negative two-loop Yukawa-mediated contributions are proportional 
to $\tilde{m}^2$ multiplied by a third generation down-type Yukawa 
coupling that can be well approximated by 
$Y_{b, \tau} = \sqrt{2} m_{b, \tau} \tan\beta/v$ 
(i.e., proportional to $\tan\beta$).  The positive
two-loop gauge-mediated log-enhanced contributions are also 
proportional to $\tilde{m}^2$, and thus one can always find
an upper limit on $\tan\beta$ for fixed $b_5$, $\lambda^d_1$, $\lambda^d_2$ 
couplings that is essentially independent of $\tilde{m}^2$.

\begin{figure}[!t]
\centering
\epsfxsize=5.0in
\hspace*{0in}
\epsffile{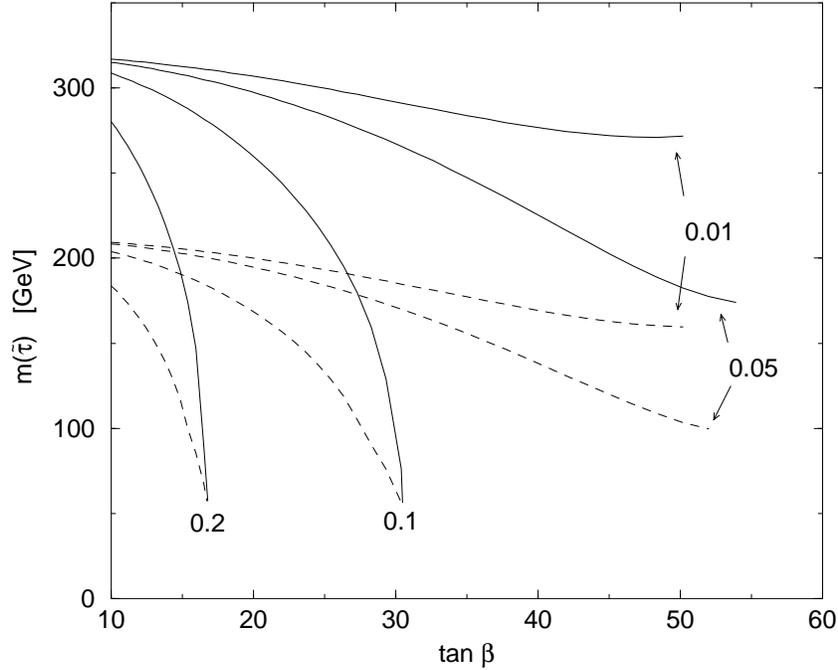}
\caption{The stau mass is shown as a function
of $\tan\beta$ with the contours corresponding to particular
choices of for the the superpotential parameters $b_5$, $\lambda^d_1$, 
and $\lambda^d_2$ (taken to be equal for illustration).  
The (solid, dashed) lines correspond to taking the parameters 
$F_\chi/\MFN = (30, 20)$ TeV and $\tilde{m}^2 = (-30^2, -20^2)$ 
TeV$^2$.}
\label{run6c-fig}
\end{figure}

In Fig.~\ref{run6c-fig} we show
the stau mass as a function of $\tan\beta$, for several
choices of the superpotential parameters $b_5$, $\lambda^d_1$, 
and $\lambda^d_2$ (taken to be equal for illustration).  
Requiring $\mathcal{O}(1)$ couplings, i.e. couplings 
larger than about $0.1$ (as explained in the introduction),
implies $\tan\beta \gsim 10$ such that $Y_{b,\tau} \gsim 0.1$.
If we impose the same restriction on the superpotential 
couplings, then $\tan\beta \lsim 30$.  This conclusion
is not significantly modified by changing other parameters
in the model.  Hence, for the model to survive experimental 
constraints we are faced with couplings 
$\sim 0.1 \ra 0.3$ in either the third generation Yukawa couplings
or the $b_5$, $\lambda^d_1$, $\lambda^d_2$ superpotential couplings.

In addition, the sizes of these couplings are interesting from a 
model building perspective, since their size (coupled with ratios of 
flavon vevs) impacts the predictions for the fermion masses.  
For example, if these superpotential couplings
(and no others) were small, then $\epsilon_{52}^{(d)}$, which
determines the CKM element $V_{ub}$, would be suppressed relative
to $\phi^{(5)}/\chi^{(2)}$.  Thus, while $\epsilon_{52}^{(d)}$
is required to be one of the smallest ratios, $\vev{\phi^{(5)}}$
would be comparable to the other flavon vevs.  It is therefore
possible to identify $\phi^{(5)} \equiv \phi^{(4)}$, giving 
one less flavon and an additional restriction on the U(1)$_F$ charges.
In general, several other variations with additional couplings
and other flavon identifications are possible, thereby reducing
the number of flavon fields required for the model (however, one must 
verify the stability of the non color-breaking vacua, etc.).  
Since there is no immediate phenomenological impact of a reduction
of the number of flavons, we do not pursue this model building exercise 
further in this paper.

\subsection{Electroweak symmetry breaking}

In ordinary supergravity-mediated models and minimal gauge-mediated
models, electroweak symmetry breaking (EWSB) is radiatively
induced due to the evolution of the up-type soft Higgs (mass)$^2$ into 
negative values near the weak scale.  The negative one-loop contributions 
to $H^u$ arise from the top Yukawa coupling, although we note that 
there are analogous negative one-loop contributions to $H^d$ from 
the bottom and tau Yukawa couplings (that are large for large $\tan\beta$).
In this model, EWSB occurs exactly as above, and we utilize the minimized 
Higgs effective potential at the weak scale to match to the $Z$ mass.  
The expressions are very well known, namely
\begin{eqnarray}
\mu^2 &=& -\textfrac{1}{2} M_Z^2 
    + \frac{m_{H^d}^2 - m_{H^u}^2 \tan^2\beta}{\tan^2\beta - 1} 
    + \cdots \label{mu2-eq} \\
m_{A}^2 &=& - \frac{2 B_\mu}{\sin 2\beta} \; = \; 
    \frac{1}{\cos 2\beta} \left( m_{H^u}^2 - m_{H^d}^2 \right) - M_Z^2
    + \cdots \label{A2-eq}
\end{eqnarray}
where $m_{A}$ is the CP-odd Higgs mass, $B_\mu$ is the soft (mass)$^2$
parameter for $H^u H^d$, and the ellipsis
stands for higher order corrections that we do
not write here.  We include one-loop corrections from all squarks 
and sleptons using the results of Ref.~\cite{PBMZ}, but neglect the 
much smaller contributions from gauginos and Higgs scalars.
We found that despite the large mass scale of the first
and second generations in this model, their summed contribution
to the effective potential is small compared with the third generation
(which are dominated by the contributions from the stops).\footnote{In 
particular, the electroweak $D$-term contributions that are naively 
proportional to $M_Z^2 \ln (m_{1st,2nd}/Q)$ approximately cancel upon 
summing over the matter of each generation.}  

Applying the constraint $m_{A}^2 > 0$ to Eq.~(\ref{A2-eq})
implies $m_{H^d}^2 > m_{H^u}^2$ is necessary to satisfy the EWSB
conditions at tree-level (and generally also at one-loop).
At large $\tan\beta$, the negative one-loop contributions to
$H^d$ and $H^u$ are comparable, and thus the requirement
$m_{A}^2 > 0$ establishes an upper bound on $\tan\beta$ 
in the model (for example, the endpoints of the lines on the far right
of Fig.~\ref{run6c-fig} arise from this constraint).   Since we have 
already argued that intermediate values of $\tan\beta$ lead to the 
most natural couplings, it follows that in general EWSB does not 
impose a severe constraint on the model.  

\begin{figure}[!t]
\centering
\epsfxsize=5.0in
\hspace*{0in}
\epsffile{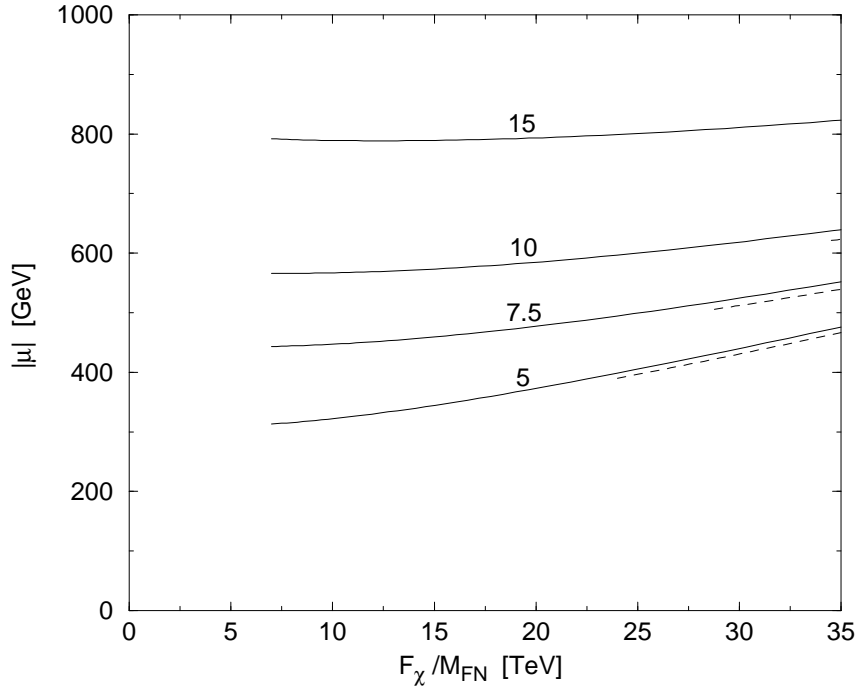}
\caption{The extracted value of $\mu$ using EWSB constraints
is shown as a function of the holomorphic supersymmetry breaking 
vev $F_\chi$.  The four solid (two dashed) contours correspond 
to $m_{1st,2nd}$ in TeV for $\tan\beta = 10$ ($\tan\beta = 30$).}
\label{mu-Fchi-fig}
\end{figure}

Nevertheless, matching the Higgs vevs to the $Z$ mass using the EWSB 
relation Eq.~(\ref{mu2-eq}) does allow us to determine the bilinear 
superpotential mass parameter $\mu$ (up to a sign) at the weak scale.
In Fig.~\ref{mu-Fchi-fig} we show the size of the $\mu$-term
that results in the model as a function of the holomorphic supersymmetry
breaking vev $F_\chi$.  We stress that $\mu$ is within
$1$~TeV for typical parameters of the model, although a low 
value, i.e.\ $|\mu| \lsim 500$ GeV, arises only for
$m_{1st,2nd} \lsim 5$ TeV which is difficult to reconcile with 
FCNC constraints.  The left-hand side endpoints of the
contours arise from the experimental bounds on the lightest 
chargino mass and the lightest stau mass.  Notice that at larger $m_{1st,2nd}$
there is essentially no dependence on the holomorphic supersymmetry 
breaking vev $F_\chi$.  Thus fine-tuning arguments that restrict the size 
of $|\mu|$ apply primarily to the DSB induced mass parameter $\tilde{m}^2$.

Finally, we note that although the EWSB conditions determine the 
value of $\mu$ at the weak scale, we have not specified a mechanism 
for generating $\mu$ from higher scale dynamics.  We defer 
a discussion of this to Sec.~\ref{q-chi1-nonzero-sec}.

\subsection{Gauge coupling unification}
\label{gauge-coupling-unif-sec}

One of the benefits of FN fields that fill complete SU(5) reps is that 
gauge coupling unification is preserved to one-loop.  
Unfortunately, this does not hold to two-loops (or higher), 
with the approximation increasingly poor as the gauge couplings become
larger.  It is well known that while the unification scale is unchanged
to one-loop as the number of messenger fields in complete SU(5) reps
is increased, the value of the unified gauge coupling increases.
The upper bound on the number of messenger fields is usually taken to be 
$3 n_{10 + \overline{10}} + n_{5 + \overline{5}} \le 4$ for 
a messenger scale not too far removed from the weak 
scale\footnote{Increasing the messenger scale allows for more 
messenger fields in the messenger sector while avoiding a Landau pole
at the unification scale, see e.g.\ Ref~\cite{KMR}.}. 
The model we have presented requires one $10 + \overline{10}$ pair 
and two $5 + \overline{5}$ pairs in order to generate the SM fermion 
masses, which nominally implies the gauge couplings encounter a Landau 
pole prior to unification.

In our model, however, the first and second generations are heavy.  
Using a decoupling scheme, the $\beta$-function coefficients 
of the gauge couplings are shifted from the MSSM values for 
scales between $M_Z$ and $m_{1st,2nd}$.
The gauge $\beta$-function coefficient at one-loop $B_a^{(1)}$ is shifted by
\begin{eqnarray}
\Delta B_a^{(1)} &=& -\frac{4}{3} 
\end{eqnarray}
assuming all of the first and second generation scalars are decoupled.
This gives a correction to the gauge couplings that is roughly
\begin{eqnarray}
\Delta g_a &\approx& \frac{g_a^3}{16 \pi^2} \Delta B_a^{(1)} 
   \ln \frac{m_{1st,2nd}}{M_Z} \; .
\end{eqnarray}
The relevance of this correction is apparent if we absorb this 
one-loop correction into a shift of the effective FN scale
(where the gauge $\beta$-function coefficients are again changed
due to the matter content of the FN fields)
\begin{eqnarray}
\frac{{\MFN}_{\mathrm{eff}}}{\MFN} &=& \left( \frac{m_{1st,2nd}}{M_Z} 
    \right)^{4/15} \; ,
\end{eqnarray}
which is nearly an increase of a factor of $4$ for 
$m_{1st,2nd} \sim 10$ TeV\@.
Hence, this one-loop analysis suggests that a model with heavy
first and second generations allows the FN scale to be lowered by 
a nontrivial factor, relative to a model without heavy first and 
second generations.

\begin{figure}[!t]
\centering
\epsfxsize=5.0in
\hspace*{0in}
\epsffile{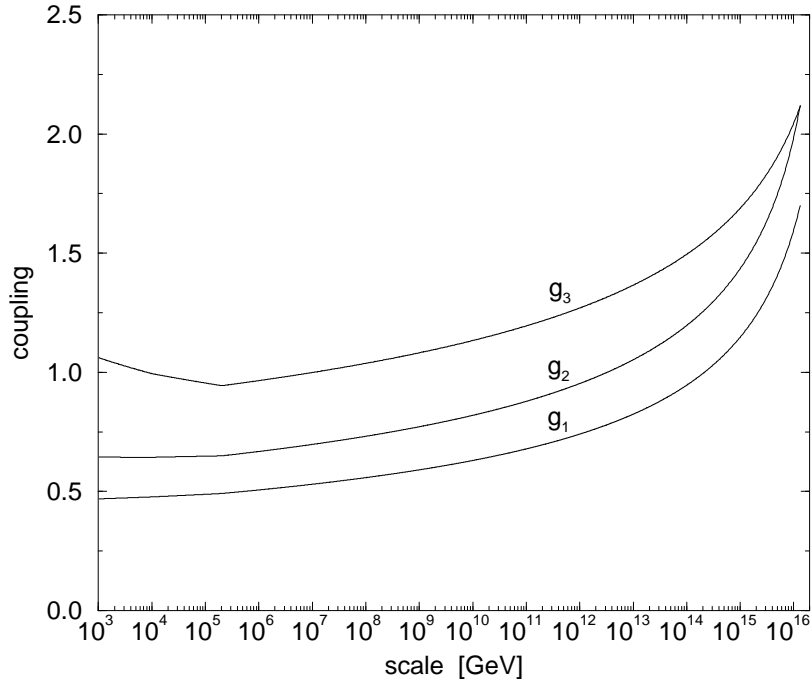}
\caption{The evolution of the gauge couplings from the weak scale 
to $\Munif \simeq 10^{16}$ GeV where $g_2(\Munif) = g_3(\Munif)$
is illustrated for a model with $m_{1st,2nd} \sim 10$ TeV
and $\MFN = 2 \times 10^5$ GeV\@.}
\label{gaugecouplings1-fig}
\end{figure}

\begin{figure}[!t]
\centering
\epsfxsize=5.0in
\hspace*{0in}
\epsffile{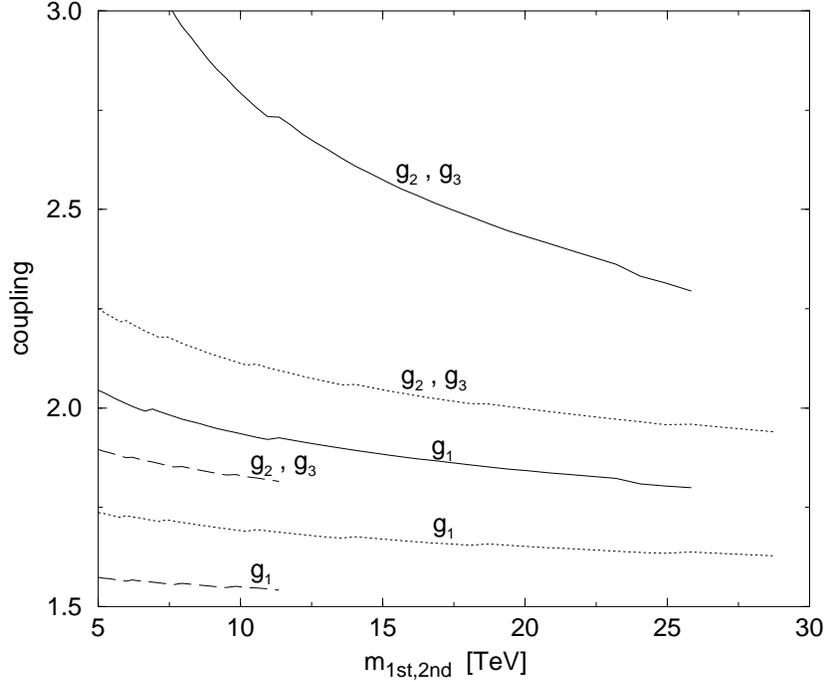}
\caption{The variation of the values of the gauge couplings
at $Q = \Munif$ as a function of the first and second generation
mass scale $m_{1st,2nd}$ and $\MFN$.  The solid, dotted, and dashed
lines correspond to $\MFN = 10^5$, $2 \times 10^5$, and 
$4 \times 10^5$ GeV respectively.}
\label{gaugecouplings2-fig}
\end{figure}

In practice, one must use the full two-loop RG equations to
calculate the evolution and determine if indeed the gauge
couplings encounter a Landau pole prior to unification.
In Fig.~\ref{gaugecouplings1-fig} we illustrate the evolution
of the three SM gauge couplings with the parameter choices
$m_{1st,2nd} \simeq 10$ TeV and $M_{FN} = 2 \times 10^{5}$ GeV\@.
It is obvious that the gauge couplings are diverging near
the unification scale, although they remain semi-perturbative 
($g_a \lsim 2.5$) up to $\Munif$ where $g_2(\Munif) = g_3(\Munif)$.
Notice that $g_1(\Munif)$ is also large but somewhat below the non-Abelian 
couplings, and therefore naive triple gauge coupling unification 
does not occur in the model.  
This conclusion appears to be robust in the model; 
other choices of parameters such as $\MFN$ or $m_{1st,2nd}$ 
do not affect this conclusion, as shown in Fig.~\ref{gaugecouplings2-fig}.
That $g_1$ is not unified is an intriguing and unexpected result.  
We explicitly checked that this result is \emph{not} significantly 
affected by large splittings between e.g., the colored and 
uncolored FN fields.  However, unlike some hypothetical model with 
$g_2(\Munif) \not= g_3(\Munif)$ or 
$g_1(\Munif) \gg g_2(\Munif) \sim g_3(\Munif)$, several remedies can
be applied in this case to restore full gauge coupling unification.
As usual, modifications come at the expense of either an ad hoc 
assumption (such as postulating the U(1)$_Y$ normalization is not 
the usual $\sqrt{5/3}$ that results from the SU(5) GUT; although
see Ref.~\cite{DFMR}) or adding fields with special properties
(in particular, fields that do not fill an SU(5) rep but instead
are charged only under U(1)$_Y$).  Since these modifications are
unlikely to have any significant impact on the weak scale 
phenomenology of the model, we will not pursue this further
in this paper.

\section{Signals of the Model}
\label{signals-sec}

There are numerous ways to search for signals of the flavor model, 
including through rare FCNC processes, collider phenomenology
at the threshold of sparticle production, etc.  
We focus on several aspects of phenomenology that will be relevant 
at current or upcoming collider experiments.
In particular, we provide a general overview of the model's spectrum,
the identity of the next-to-lightest sparticle (NLSP), the associated 
macroscopic decay length into a particle and a light gravitino, 
the prospects for stau production and the importance of taus in collider 
signals, and finally we examine $D^0 \leftrightarrow \overline{D}^0$ mixing 
as a window on a promising FCNC process that is expected in the model.

\subsection{General comments}

The weak scale spectrum of the model consists of the gauginos, 
the third generation scalars, and the Higgs scalars, with several 
of these fields expected to have $\mathcal{O}(1 \; \mathrm{TeV})$ masses.  
The chargino and neutralino
mass eigenstates arise from mixing between the gaugino and Higgsino
fields, as usual in the MSSM\@.  In our model, $\mu$ is always 
significantly larger than the Bino mass $M_1$, and is also larger 
than the Wino mass $M_2$ in all but the most highly contrived scenarios. 
This implies that the two lightest neutralinos and the lightest 
chargino are dominantly gaugino-like,
\begin{eqnarray}
\tilde{N}_1 \; \simeq \; \tilde{B} \quad & , & \quad 
\tilde{N}_2 \; \simeq \; \tilde{W}^0 \\
\tilde{C}_1^\pm & \simeq & \tilde{W}^\pm
\end{eqnarray}
with a mass spectrum that can be approximated by
\begin{eqnarray}
m_{\tilde{N}_1} & \simeq & M_1 \\
m_{\tilde{N}_2} \; \simeq \; m_{\tilde{C}_1} & \simeq & M_2 \\
m_{\tilde{N}_3} \; \simeq \; m_{\tilde{N}_4} \; \simeq \; 
    m_{\tilde{C}_2} & \simeq & |\mu| \; .
\end{eqnarray}
Some combination of the lightest chargino and the lightest two
neutralinos are among the most likely sparticles to be produced 
with a large rate at a collider.
Search strategies for these sparticles depend
heavily on three factors:  the identity of the NLSP, the decay
length of the NLSP, and the mass of the lightest stau in relative 
comparison with the lighter gauginos.  In virtually all of
our model's parameter space, there are only two possible
mass hierarchies:
\begin{eqnarray}
m_{\tilde{N}_1} &< \;  m_{\tilde{\tau}_1} \; <& m_{\tilde{N}_2, \tilde{C}_1} 
    \qquad \mbox{(neutralino-NLSP scenario)} \label{N1-NLSP-eq} \\
m_{\tilde{\tau}_1} &< \; m_{\tilde{N}_1} \; <& m_{\tilde{N}_2, \tilde{C}_1}
    \qquad \mbox{(stau-NLSP scenario)} \label{stau-NLSP-eq}
\end{eqnarray}
Thus, we expect that $\tilde{N}_2 \ra \tilde{\tau}_1^\pm \tau^\mp$ and 
$\tilde{C}_1^\pm \ra \tilde{\tau}_1^\pm \nu_\tau$ 
to dominate throughout parameter space\footnote{These decays 
are expected to be at least comparable to (or possibly dominate over) 
$\tilde{C}_1^\pm \ra W^\pm \tilde{N}_1$ and $\tilde{N}_2 \ra Z \tilde{N}_1$
even when the latter are kinematically open \cite{Wells-taus}.},
regardless of the identity of the NLSP\@.  Although a pure
Wino-like chargino does not couple to $\tilde{\tau}_R$, there is 
sufficient mixing within 
$\tilde{\tau}_1 = \cos\theta_{\tilde{\tau}} \tilde{\tau}_L - 
\sin\theta_{\tilde{\tau}} \tilde{\tau}_R$ at moderate to large $\tan\beta$ 
such that $\tilde{C}_1$ can proceed through the 2-body decay into 
$\tilde{\tau}_1 \nu_\tau$, albeit suppressed by the mixing angle 
$\cos^2\theta_\tau$.  If the NLSP is
the lightest stau, then there are potentially additional 
sources of taus in collider events, as we discuss below.  
In any case, the importance of searching for taus as signals of
flavor-mediated supersymmetric models cannot be overemphasized.

The third generation scalar spectrum is roughly split into
several heavy fields $\tilde{\tau}_2$, $\tilde{\nu}_\tau$,
$\tilde{t}_1$, $\tilde{t}_2$, $\tilde{b}_1$, and $\tilde{b}_2$,
and one light field $\tilde{\tau}_1$.  The heavier sleptons
acquire their mass dominantly from $m_{\overline{L}^3}$, hence
$\tilde{\tau}_2 \simeq \tilde{\tau}_L$, 
$m_{\tilde{\tau}_2} \simeq m_{\tilde{\nu}_{\tau}}$, and thus
$\tilde{\tau}_1 \simeq \tilde{\tau}_R$.  Since $\overline{d}^3$
acquires a positive one-loop Yukawa-induced contribution,  
Eq.~(\ref{dbar3-approx-eq}), while $Q^3$ acquires a negative 
two-loop Yukawa-induced contribution, Eq.~(\ref{Q3-approx-eq}), 
we find that $\tilde{b}_2 \simeq \tilde{b}_R$ is
the heaviest third generation scalar field.  The other third 
generation squarks are roughly $\tilde{b}_1 \simeq \tilde{b}_L$, 
$\tilde{t}_2 \sim \tilde{t}_L$, $\tilde{t}_1 \sim \tilde{t}_R$, 
and $m_{\tilde{t}_1} \lsim m_{\tilde{t}_2} \simeq m_{\tilde{b}_1}$.  
The stop mass
eigenstates are not as well defined by their left- or right-handed
squark eigenstates due to larger off-diagonal LR mixing
(as compared with the sbottoms or the staus).

\begin{figure}[!t]
\centering
\epsfxsize=5.0in
\hspace*{0in}
\epsffile{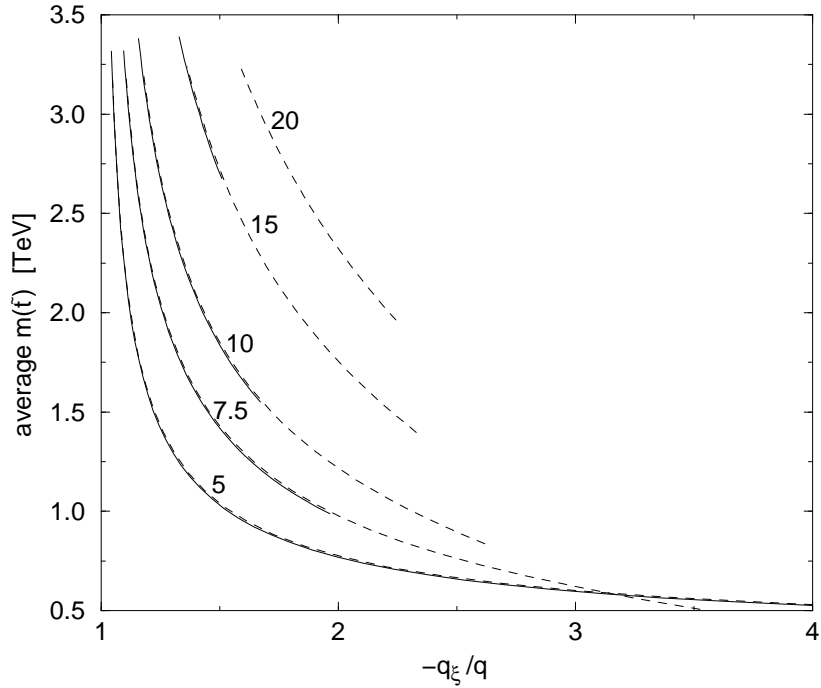}
\caption{The average stop mass $(m_{\tilde{t}_1} + m_{\tilde{t}_2})/2$
is shown as a function of the ratio of the normalized FI $D$-term $q_\xi$.
The solid (dashed) lines are contours of the first and second
generation mass scale $m_{1st,2nd}$ in TeV, for $\tan\beta = 10$
($\tan\beta = 30$).}
\label{stop-FI-fig}
\end{figure}

The mass scale of the third generation squarks is highly dependent 
on $\tilde{m}^2$ and the FI-term $q_\xi$, as shown in 
Fig.~\ref{stop-FI-fig}.
The strong correlation between the mass scale of the first and 
second generations and the third generation arises due to the 
log-enhanced gauge- and Yukawa-mediated contributions proportional 
to $\tilde{m}^2$.
For example, restricting the third generation to be less than 
about $1.5$ TeV implies the first and second generation masses must 
be less than about $10$ $(15)$ TeV for $\tan\beta \lsim 30$ $(10)$.

The difference between the heavy third generation sleptons and the
heavy third generation squarks is determined largely by the 
positive one-loop RG evolution induced by gauginos, just as
in ordinary minimal gauge-mediated models.  In particular,
the colored sparticles receive one-loop gauge contributions
proportional to $g_3^2 M_3^2$, as opposed to the left-handed sleptons 
which receive one-loop gauge contributions proportional to 
$g_2^2 M_2^2$.  Thus the colored sparticles are significantly boosted 
in mass relative to their slepton counterparts by an amount that is 
ultimately proportional to the holomorphic supersymmetry breaking 
vev $F_\chi$.  In typical cases, the heavier slepton masses for 
$\tilde{\tau}_2$ and $\tilde{\nu}_\tau$ are between about $0.6$ to $0.9$ 
times an average squark mass $(m_{\tilde{t}_1} + m_{\tilde{t}_2})/2$.

The Higgs scalars acquire masses that are dependent on 
the up-type and down-type Higgs soft masses $m_{H^u}^2$,
$m_{H^d}^2$ and the conditions for EWSB that determine $\mu^2$ 
[see Eqs.~(\ref{mu2-eq}), (\ref{A2-eq})].  At moderate 
$\tan\beta$, and assuming the radiative corrections to the EWSB 
conditions are small\footnote{This is a reasonable approximation to the 
accuracy we are considering in the following.}, we can make the approximations
\begin{eqnarray}
\mu^2   &\simeq& -m_{H^u}^2 \label{mu2-approx-eq} \\
m_{A}^2 \; \simeq \; m_{H}^2 \; \simeq \; m_{H^\pm}^2 \; &\simeq& 
    m_{H^d}^2 + \mu^2
\end{eqnarray} 
up to corrections of order $M_Z^2/m_{H^u}^2$, where the Higgs soft 
masses are evaluated at the ``best minimization'' scale that
we take to be $(m_{\tilde{t}_1} + m_{\tilde{t}_2})/2$.
Since $H^d$ and $\overline{L}^3$ have the same (local) gauge quantum 
numbers, they acquire identical gauge-mediated contributions
to their masses at $\MFN$, and have similar RG evolution to
the weak scale.  The one-loop Yukawa-mediated contributions
are typically only slightly different [compare Eq.~(\ref{Hd-approx-eq}) 
with Eq.~(\ref{Lbar3-approx-eq})], however in practice we find that
subdominant negative two-loop Yukawa-mediated corrections
reduce the mass of $H^d$ by typically 10--25\% relative to 
the mass of $\overline{L}^3$.\footnote{Since the relative size of 
$m_{H^d}$ and $m_{\overline{L}^3}$ at the messenger scale (and at
the weak scale) is determined by the Yukawa coupling contributions,
these quantities are sensitive to the U(1)$_F$ charge assignments 
for the first and second generations, as well as the relative size
of the superpotential couplings $b_5$, $\lambda^d_1$, and $\lambda^d_2$.}
The up-type Higgs acquires
a very small mass at the FN scale, and is driven to moderate-sized
negative values via the top Yukawa coupling by one-loop RG evolution.
For illustration, one can easily translate the values
of $\mu$ shown in Fig.~\ref{mu-Fchi-fig} into good approximations
of $(-m_{H^u}^2)^{1/2}$ using Eq.~(\ref{mu2-approx-eq}).
For $10 \lsim \tan\beta \lsim 30$, $m_{H^d}$ and $\mu$ are comparable,
and so a rough approximation of the heavier Higgs scalar masses
is $\sqrt{2} |\mu|$.  Thus, the heavier Higgs scalar fields 
have masses that are much larger than the weak scale (close to, 
for example, the heavier third generation sleptons).

The lightest Higgs mass is expected to fall within the bounds
that have been well established for the MSSM \cite{CarenaWagnerhiggs,
higgsbounds}.
We performed one-loop \cite{PBMZ} (and in some cases approximate 
two-loop \cite{CarenaWagnerhiggs}) 
calculations to extract its mass for several choices of parameters.  
We checked that at one-loop the lightest Higgs mass is not 
sensitive to the heavy first and second generation masses (considered
in this paper), and for moderate $\tan\beta$ depends largely on 
just the mass of the stops.  Since the stops are relatively heavy
in our model (see Fig.~\ref{stop-FI-fig}), the associated radiative 
corrections to the Higgs mass are significant.  A precise calculation
of the lightest Higgs mass requires careful treatment of the scale to 
evaluate the radiative corrections, and one must include the dominant 
two-loop corrections (which tend to reduce the Higgs mass relative to a naive 
one-loop calculation).  Our estimates suggest that the 
lightest Higgs mass is near the upper end of the MSSM bounds, 
$m_{h} = 110$--$125$ GeV, throughout the parameter space 
of the model.  We have not attempted a high precision ($\lsim$ few GeV) 
calculation, but do not expect our estimate to be in error by more
than about 5\%.  It is interesting to note that this mass safely 
evades the current bounds from LEP \cite{LEPhiggsbound} due to the heavy 
stops that are an inevitable consequence of the model.
 
In Appendix~\ref{example-app} we present an example set of
input parameters and some of the calculated sparticle masses 
to illustrate the generic spectra that are expected in the MFMM\@.
The evolution of some of the weak scale soft masses is also shown
as a function of scale.

\subsection{Gravitino mass}

In models of low energy supersymmetry breaking, the perhaps best-known
generic prediction is that the gravitino mass should be rather small.
Recall that after global supersymmetry is dynamically broken in the 
DSB sector, a massless spin-1/2 Goldstino is present.  Ordinarily this
would be phenomenological disaster, but in fact this would-be
Goldstino becomes the longitudinal components of the spin-3/2
gravitino, after the model is embedded in supergravity at
$M_{Pl} = 2.4 \times 10^{18}$~GeV\@.  The gravitino therefore
acquires mass by this super-Higgs mechanism, of order
\begin{equation}
m_{\tilde{G}} \; = \; \frac{\LambdaDSB^2}{\sqrt{3} M_{Pl}} \; = \; 
    (24 \; \mathrm{keV}) 
    \left( \frac{\LambdaDSB}{10^7 \; \mathrm{GeV}} \right)^2 \; .
\end{equation}
One of the central phenomenological consequences of a such a light
gravitino is that every sparticle must ultimately decay into 
it.\footnote{Assuming $R$-parity is (at least nearly) exact.}
The characteristic scale associated with the two-body interaction
of a gravitino and either a matter or vector supermultiplet
scales as $1/\LambdaDSB^2$ for the longitudinal components 
(i.e.\ the Goldstino).  Hence, the coupling is small in comparison
to the gauge interactions but not suppressed by the Planck scale.
All sparticles are therefore expected to decay into the next-to-lightest
supersymmetric particle (NLSP) with very short decay lengths, and
then the NLSP decays into the gravitino and a SM particle with
a decay length that is inversely proportional to the coupling
strength, or directly proportional to the gravitino mass.
This leads to an enormously rich phenomenology (for a review,
see Ref.~\cite{gmsbrev}.)

As we discussed in the previous section, the lightest neutralino
mass eigenstates are dominantly ``gaugino-like'', with the
lightest neutralino being composed dominantly of the Bino state.  
If the lightest neutralino is the NLSP, the dominant decay will always be 
$\tilde{N}_1 \ra \ph\tilde{G}$.  Otherwise, if the lightest
stau is the NLSP, the dominant decay will always be 
$\tilde{\tau}_1 \ra \tau\tilde{G}$.  The NLSP decay width
$\Gamma(\mathrm{NLSP} \ra \mathrm{particle} + \tilde{G})$
is virtually identical\footnote{$\Gamma(\tilde{N}_1 \ra \ph\tilde{G})/
\Gamma(\tilde{\tau}_1 \ra \tau\tilde{G}) = \cos^2 \theta_W \simeq 0.78$
for a purely Bino-like lightest neutralino.} regardless of which sparticle 
is actually the NLSP.  If the fundamental scale of supersymmetry
breaking is indeed $\LambdaDSB$, we can write the characteristic 
decay length $L = \Gamma^{-1}$ of the NLSP as \cite{AKM1}
\begin{eqnarray}
L(\mathrm{NLSP} \ra \mathrm{particle} + \tilde{G}) &=&
   (1.7 \; \mathrm{km}) \left( \frac{m_{\mathrm{NLSP}}}{150 \; \mathrm{GeV}} 
   \right)^{-5} \left( \frac{\LambdaDSB}{10^7 \; \mathrm{GeV}} \right)^4
   \left( \frac{E^2_{\mathrm{NLSP}}}{m^2_{\mathrm{NLSP}}} - 1 \right)^{1/2}
   \label{L-NLSP-eq}
\end{eqnarray}
where $E_{\mathrm{NLSP}}$ is the energy of the NLSP that is 
produced.\footnote{The decay length of any given NLSP is of course
probabilistic.  Eq.~(\ref{L-NLSP-eq}) is correctly interpreted
as the length within which a fraction $1 - 1/e$ of an ensemble of
NLSPs produced at the same energy would have decayed \cite{AKM1}.}

It is obvious that if $m_{\mathrm{NLSP}}$ and $\LambdaDSB$ are of order
the normalized values given in Eq.~(\ref{L-NLSP-eq}), the 
characteristic decay length is much larger than the scale 
of the detector, and thus virtually all produced NLSPs
decay \emph{outside} the detector.  This is a well-known 
expectation of nearly all (multi-sector) DSB models.
It has profound implications for the strategy to search for
signals of our flavor model, depending on the identity of the NLSP, 
as we discuss in the next two sections.

Due to the very strong dependence on the NLSP mass and the 
DSB scale, it is natural to ask if there are regions of 
parameter space that might give characteristic decay lengths
of order the scale of the detector.  In such a case there are a
host of fascinating, near background-free signals that could be
exploited to probe the supersymmetric parameter space 
(see e.g.\ \cite{AKM1}).  One trivial possibility is an $e^+e^-$ 
collider experiment operating at an energy slightly above the pair 
production threshold for NLSPs, thereby imparting very small 
energies to these sparticles.  More nontrivial possibilities
amount to shifts upward $m_{\mathrm{NLSP}}$ and/or shifts downward of 
$\LambdaDSB$.  These have a dramatic impact on the characteristic
decay length due to the large power dependence in Eq.~(\ref{L-NLSP-eq}).
Unfortunately it is difficult to lower $\LambdaDSB$ by any
significant factor, due to the direct connection to $\tilde{m}^2$
as shown in Eq.~(\ref{mtilde2-approx-eq}).  The mass of the NLSP
is simply $\mathrm{min}[m_{\tilde{N}_1}, m_{\tilde{\tau}_1}]$.
If we restrict $M_3 \lsim 1 \; (1.5)$ TeV, then 
$m_{\tilde{N}_1} \lsim 190 \; (285)$ GeV\@.  
Although the stau mass is expected to be light, it is strongly
dependent on the parameters of the model due to the several
sources of supersymmetry breaking that contribute to its mass.
In Fig.~\ref{run6c-fig} we showed the mass of the stau as
function of several parameters, with the result that
$m_{\tilde{\tau}_1} \lsim 300$ GeV is a conservative upper
bound.\footnote{Since both the $F_\chi$ vev and $\tilde{m}^2$
are ultimately related to $\LambdaDSB$ up to some factors,
it is inappropriate to consider simultaneously 
increasing $m_{\mathrm{NLSP}}$ while decreasing $\LambdaDSB$.}
Thus, unless the true parameters of the model are
highly unnatural (couplings significantly less than $0.1$
and/or a large $F_\chi$ vev giving a gluino mass significantly 
beyond $1.5$ TeV), the characteristic decay length of the NLSP
is almost certainly much larger than the scale of a collider
detector.

\subsection{Consequences of a neutralino NLSP}
\label{neutralino-NLSP-sec}

If the NLSP is the lightest neutralino, then the collider signals
are expected to be very similar to the original supergravity-mediated
models since virtually every $\tilde{N}_1$ escapes the detector carrying 
away missing energy.  As we discussed above, $\tilde{C}_1$ and/or
$\tilde{N}_2$ production is expected to dominantly produce taus in 
every event, due to Eq.~(\ref{N1-NLSP-eq}).  This is likely
to make detection more difficult due to the prompt decay of
the tau.  A tau decays either leptonically into $\ell\nu_\tau\nu_\ell$
or hadronically into several channels that include a $\nu_\tau$.
The neutrino(s) from tau decay provide an additional source of 
missing energy, suggesting a larger missing energy cut may assist 
discerning signal from background without rejecting a tremendous amount 
of signal.  Leptonic tau decays do yield ``clean'' events with only 
leptons and little hadronic activity, but the resulting leptons have 
significantly less energy than the parent tau due to the three-body decay,
and one faces a branching ratio suppression of $0.35$ for every
leptonic tau decay.  

Searching for taus a signal of supersymmetry has recently been
extensively studied, both in the context of supergravity-mediated
models as well as gauge-mediated models \cite{DDRT, DTW, BMPZ, 
BaerGMSB, AKM1, Nandipapers, AKM2, Baerandcompany, Wells-taus, 
MatchevLykken, MatchevPierce}.
In the absence of any signal at LEP, attention has been turned
to the prospects at the upgraded Tevatron, in several high
luminosity versions.  One of the most interesting signals
of gaugino production is the ``trilepton'' signal arising from 
$\tilde{C}_1^\pm \tilde{N}_2$ production.  If these gauginos
do indeed decay dominantly into a stau, then the signal
of this process in our model (with a neutralino NLSP) is
three $\tau$'s plus missing energy.  Matchev and Lykken 
\cite{MatchevLykken} have analyzed several ways to detect 
this particular signal, and emphasized searching for modes
with both some leptonic tau decays as well as some hadronic tau decays.
They performed an event-level analysis folding in detector cuts 
and a semi-realistic detector simulation, and found that
the best signature of tri-tau production is $\ell\ell\tau_h$
where two taus decay leptonically and one tau decays hadronically.
In this process the Tevatron had the greatest reach, although
at least $1$--$2$~fb$^{-1}$ of data is required for a (weak) signal
with low mass charginos ($m_{\tilde{C}_1} \lsim 120$~GeV), and tens 
of fb$^{-1}$ of data is needed to exclude any mass chargino beyond 
the LEP bound of about $90$~GeV\@.  

Chargino production ($\tilde{C}_1^\pm \tilde{C}_1^\mp$) and 
stau production are also processes that could have a large rate at 
the Tevatron.  The signature for both is expected to be two 
opposite-sign taus plus missing energy.  Unfortunately, this
signature suffers from a large Standard Model background forcing one
to be clever about the event-level cuts (for a recent analysis,
see e.g.~\cite{Baerandcompany, MatchevPierce}).

Neutralino production ($\tilde{N}_2 \tilde{N}_2$) has much
more interesting signals comprising four taus plus missing
energy.  One interesting signal is a pair of like-sign dileptons
plus two tau jets, where each neutralino decays ultimately
into two taus, one of which decays hadronically.
However, the pair production of the second-lightest neutralino
is somewhat suppressed compared with chargino production or
chargino-neutralino production, and requires a full event-level
simulation to determine the prospects for discovery 
(or exclusion) at the Tevatron.  If the results of Matchev
and Lykken \cite{MatchevLykken} can be roughly applied to this case, 
it is unlikely neutralino production will be detectable before
the LHC becomes operational.

\subsection{Consequences of a stau NLSP}
\label{stau-NLSP-sec}

If the NLSP is the stau, then every sparticle produced 
in a collider results in a charged stau track.  Searches
for heavy stable charged tracks have been performed at LEP
and Tevatron (run I) without success, suggesting that the 
upgraded Tevatron has, once again, the most promising future 
prospects for detection.  

An ionizing charged track resulting from a stau is a dramatic 
signal in a detector \cite{DDRT, DTW, AKM1, FengMoroi, MartinWells}.
However, the only relevant distinguishing characteristic for a detector 
between a muon and a quasi-stable stau is the stau mass.
If a produced stau is sufficiently relativistic (in the lab frame),
it will behave as a ``minimum-ionizing'' particle and thus 
appear virtually identical to a muon.  If a produced stau is slower, 
in particular \cite{FengMoroi, MartinWells}
\begin{eqnarray}
\beta\gamma \; \equiv \; 
    \left( \frac{E^2_{\tilde{\tau}_1}}{m^2_{\tilde{\tau}_1}} 
    - 1 \right)^{1/2} \lsim 0.85 \; ,
\end{eqnarray}
then the stau has a greater-than-minimum ionization rate 
in the detector that can be distinguished from a muon \cite{CDFslepton}.
The size of $\beta\gamma$ depends on the mass of the produced
sparticle, the energy of the collider, and whether the stau was 
directly produced or indirectly through heavier sparticle decay.
Feng, Moroi \cite{FengMoroi}, and Martin, Wells \cite{MartinWells}
analyzed in detail the detectability of quasi-stable staus at the Tevatron.
They found that by searching for at least one highly-ionizing track 
(HIT) per event, stau production could be probed for stau masses
up to roughly $150$--$200$ GeV with between a few to tens of 
fb$^{-1}$ of data.  Similarly gaugino production is expected
to yield several taus in the event (as described above) 
in addition to charged stau tracks.  These authors found
that chargino masses of up to $300$--$400$ GeV could be 
probed for similar amounts of data.  Here the process 
$\tilde{N_1} \tilde{N_1}$ gives the interesting signal
$\tilde{\tau}\tilde{\tau}\tau\tau$, half of the time
with same-sign staus and same-sign taus.  We should note, however,
that in some cases these authors' analyses included contributions 
from (and discussed particular signatures associated with) first 
and second generation slepton production, that is of course absent 
in our model.  In addition, we expect the lightest chargino
and the second lightest neutralino to decay to a stau, 
generating plenty of taus in the final state, just as discussed
in Sec.~\ref{neutralino-NLSP-sec}.

\subsection{FCNC and CP violation: signals}
\label{CP-violation-sec}

It has been shown \cite{KLMNR} that the additional contributions to 
$\Delta m_K$ and $\epsilon_K$ in the model under discussion do not exceed
those of the SM and therefore agree with current experimental measurements.  
This result is not universal for generic weak-scale squark masses.  The 
suppression of contributions to $K^0-\overline{K}^0$ mixing has two 
sources.\footnote{As discussed earlier, to avoid large contributions to 
$\epsilon_K$ we require $F_{\phi^{(1)}} F_{\phi^{(2)}}^*$ to have 
a small imaginary part 
($\sim \mathcal{O}(10^{-2}) \times \phi^{(1)}\phi^{(2)}(\chi^{(1)})^2$).}
One is that the first two generations of scalars are heavy 
($\sim 7$--$20$~TeV) 
and that their (mass)$^2$ matrices are approximately diagonal in the 
flavor basis.  The other source of suppression arises due to the
Cabibbo angle coming from the up sector.  Thus the gluino box diagrams 
which contribute to $K^0-\overline{K}^0$ mixing are small because 
of both large scalar masses and small mixing.\footnote{Box diagrams 
with wino exchange \emph{do} contain couplings with Cabibbo-like mixing.  
However, the suppression $(\alpha_w/\alpha_s)^2$ is enough to render 
these contributions sub-dominant.} 

A definite signal from this model is enhanced $D^0-\overline{D}^0$ mixing 
\cite{CKLN} since first and second generation mixing at the 
quark-squark-gluino vertex is of order the Cabibbo angle 
(as expected since the Cabibbo angle comes from this sector).  
The largest contribution comes from the standard gluino box
diagrams that give (in the ``vacuum insertion'' approximation)
\begin{equation}
\Delta m_D \sim (7.0 \times 10^{-10} \; \mathrm{MeV})
\times Z^{1i\,\ast}_{u,LL} Z^{2i}_{u,LL} Z^{1j\,\ast}_{u,LL} Z^{2j}_{u,LL}
\times \left( \frac{\alpha_s}{0.1} \right)^2 
\times \left( \frac{f_D}{200 \; \mathrm{MeV}} \right)^2
\end{equation}
for first and second generation masses that are $10$~TeV and a gluino mass 
that is 1~TeV\@.  For typical values of the above constants, the
Cabibbo-like mixing gives $x_D = \frac{\Delta m_D}{\Gamma} \sim 0.021$, 
which nearly saturates the current bound \cite{CLEO}, 
and is two orders of magnitude above the SM estimate \cite{DDbar}.

The $B$ system will also receive significant contributions from 
superpartners.  The signals are similar to those outlined in \cite{CKLN}.
In the flavor basis, the mixing comes from the quark mass matrix.  
The squark mass matrix is nearly diagonal in this basis
with respect to the third generation.  The phase in the gluino box
diagrams is different from the phase in the SM box diagrams ($W$-exchange)
because equal portions of $V_{td}$ come from the up and down sectors.
Here, quark-squark mixing is only due to the down matrix, and thus
has a different phase than $V_{td}$.

At tree-level, left-right mixing matrices (the off-diagonal blocks of the
squark (mass)$^2$ matrices) are of the same form as the corresponding
quark mass matrices.  Thus, left-right mixing is in general small in this
model.  Since the Cabibbo angle does not originate from the down sector 
\cite{MM}, the model does not predict any significant supersymmetric
contributions to the recent measurement of $\epsilon'$ \cite{KTeV}.

\section{Modifying the model}
\label{modify-model-sec}

The allowed superpotential terms in the FN sector of the MFMM
were guided by several restrictions \cite{KLMNR}:
\begin{itemize}
\item[(i)]   U(1)$_F$ charge assignments must be compatible with 
             an SU(5) GUT.
\item[(ii)]  FN fields must be added in complete SU(5) multiplets.
\item[(iii)] The number of FN fields must be less than (the equivalent of) 
             five $5^{FN}$, $\overline{5}^{FN}$ pairs to allow for at 
             least semi-perturbative couplings at the GUT scale.
\item[(iv)]  The ratio of the smallest to the largest superpotential 
             couplings must not be less than $0.1$.
\end{itemize}
Constraints from electroweak symmetry breaking and FCNC imposed additional
restrictions on the model (e.g., $H^u$ could not couple to U(1)$_F$-charged 
fields in the superpotential, and the first two generations had to be charged).
In this section, we examine the consequences of lifting some of these
constraints.  In addition, we discuss two important extensions of the 
model:  dynamical generation of both a $\mu$-term and neutrino masses.

There are, however, a few variations of the MFMM that do not 
require violating any of the above restrictions.  One example
is to take the charge of $\bar{5}^{FN(1)}$ to vanish 
($q_{\overline{5}^{FN(1)}} = 0$), remove the superpotential coupling 
$H^d 10^1 \overline{5}^{FN(1)}$, and include the couplings 
$H^d 10^3 \overline{5}^{FN(1)}$ and $H^d 10^{FN} \overline{5}^{FN(1)}$.  
While this change does create a new parameter space, the low energy 
spectrum does not change dramatically in this new version, e.g., 
we still expect one light stau with all other squarks and sleptons
to be at or above about $1$~TeV\@.

\subsection{Relaxing the SU(5) ansatz}
\label{m-u-nonzero-sec}

Of the four criteria listed above, the first three involve unification,
with criteria (iii) most relevant to ensuring the gauge couplings 
approximately unify at the high scale.  Criteria (i) (charge assignments 
which commute with an SU(5) GUT) is required for standard SU(5) unification.
Criteria (ii) (FN fields in complete SU(5) multiplets) is not necessarily
required for an SU(5)-invariant high energy theory, however one would
be faced with highly split multiplets that represent an additional 
hierarchy problem [analogous to the doublet-triplet splitting of the 
Higgs field that must occur with SU(5)].  Criteria (iii) (maximum 
number of FN fields to preserve unification) is a more general
prediction of GUTs that need not be specific to SU(5), and could be 
imposed independent of the first two criteria.  For example, some 
free fermionic string models break directly to the SM gauge group, 
also with possibly a nonstandard hypercharge normalization \cite{DFMR}.  
In the following,
we explore models that do not satisfy criteria (i) and (ii),
instead satisfying merely criteria (iii) and (iv), or just
criteria (iv) only.  One important consequence of these model
variations is that the mass for the up quark is no longer necessarily 
zero, although we find that this occurs only for models that violate
criteria (i)--(iii).

Relaxing the SU(5) ansatz, FN fields now come in three varieties: 
$(Q,\overline{Q})$, $(u,\overline{u})$ and $(d,\overline{d})$ pairs.  
One possible variation is to leave the field content alone and to 
allow the U(1)$_F$ charges to deviate from the SU(5) requirements.  
The benefit of this choice is that it allows the leptons to take 
different charges than the quarks, thus more easily accommodating
their different masses.  Moreover, this variation could lead
to a simple mechanism for generating neutrino masses (see 
Sec.~\ref{neutrinos-sec}).  However, the deviation from SU(5) comes
at a cost:  $\tr Y_i m_i^2 \simeq 0$ is no longer automatically satisfied. 
This trace must approximately vanish to avoid a large contribution to
a FI hypercharge $D$-term \cite{CKN}.  The soft masses are 
quadratic functions of the U(1)$_F$ charges, and therefore this 
requirement is highly restrictive.  Regardless, it is possible to 
find charge assignments which violate SU(5), but in all of these 
variations the structure of the (full) quark mass matrices remains 
the same as in the MFMM case.

Variations in the particle content are perhaps more interesting.
There are three viable FN sectors other than that of the MFMM that have 
$\leq 5$ pairs of additional quark triplets.  All other possibilities
are ruled out since they predict a zero eigenvalue in the down mass matrix
or two zero eigenvalues in the up mass matrix.  If $n_i$ (with $i = Q,u,d$) 
is the number of FN quark pairs, then these model variations can be 
labeled by the triplet $\left( n_Q ,n_u ,n_d \right)$.  In this language, 
the four allowed models satisfying criteria (iii) are $(1,1,2)$ [the MFMM], 
$(1,0,2)$, $(1,0,3)$, and $(2,0,1)$.  The second of these is the most 
interesting as it is the only one with only four pairs of FN quarks.  
In this model, the gauge couplings remain completely perturbative up 
to the unification scale.  The full up mass matrix for this
(as well as the third) model is
\begin{eqnarray}
\mathbf{M}^u &=& \left( \begin{array}{cccc}
    0 & 0 & 0 & b_1^{(Q)} \vev{\phi^{(1)}} \\
    0 & 0 & 0 & b_2^{(Q)} \vev{\phi^{(2)}} \\
    0 & 0 & Y_t \vev{H^u} & 0 \\
    0 & \lambda^u_{1} \vev{H^u} & \lambda^u_{2} \vev{H^u} 
				& a_1^{(Q)} \vev{\chi^{(1)}} \\
                        \end{array} \right)\; .
\end{eqnarray}
Notice this matrix requires the right-handed charm quark to be uncharged
with respect to U(1)$_F$.  If the right-handed up quark is also uncharged,
then the dominant contribution to the masses of the right-handed up-type
squarks comes from the standard two-loop gauge interactions.  For soft
masses near 1~TeV\@ (as is typical in the MFMM), these squarks are
very nearly degenerate and contributions to $D^0-\overline{D}^0$ 
mixing from this sector are far below experimental bounds.  In a sense, 
this is a hybrid model between gauge-mediated and flavor-mediated 
supersymmetry breaking.  However, with this field content, 
$\tr Y_i m_i^2 \gg 0$ because there is not a sufficient number
of U(1)$_F$ charged fields that have negative hypercharge
to compensate the fields with positive hypercharge.
The consequence of this large positive hypercharge $D$-term
is typically that a scalar (mass)$^2$, such as the stau, goes negative.
In fact, all three of these variant field contents, (1,0,2), (1,0,3), (2,0,1), 
do not work for this reason.  Thus, it is remarkable that the FN sector 
is completely fixed and unique simply by requiring $n_Q + n_u + n_d \leq 5$.

In this framework, it is impossible to satisfy criteria (iii) and
(iv) and have $m_u \not= 0$.  Giving up criteria (iv) and permitting 
$H^u$ to couple to charged fields does allow the up quark to gain a mass, 
but radiative EWSB is very difficult (if not impossible) to achieve.
Giving up criteria (iii) allows a somewhat reasonable model with a 
non-zero up mass (though it is still difficult to satisfy (iv) in 
such models).  The minimal field content for this type of model is (2,0,2), 
although models with a larger field content need fewer 
small couplings.  In these models the gauge couplings encounter a
Landau pole prior to the unification scale, and therefore apparently
require new physics to appear at some lower scale.  Finally,
while all of these models can produce $m_u \not= 0$, the
natural solution to the strong CP problem is lost.

\subsection{The $\mu$-term}
\label{q-chi1-nonzero-sec}

The one dimensionful parameter whose dynamical generation has not been 
accounted for is the superpotential mass parameter of the Higgs
superfields.  The difficulty of naturally generating the superpotential
coupling $\mu H^u H^d$ (the so-called ``$\mu$-problem''), exists in this 
model in the same way as it does in ordinary gauge-mediated scenarios.  
The problem is not one of generating $\mu$ of order the weak scale 
(required by naturalness); instead the problem 
is that most mechanisms that generate 
$\mu$ also generate a soft supersymmetry breaking scalar bilinear 
mass term $B_{\mu} H^u H^d$ that is much too large \cite{DGP}.  We do not 
have anything significantly new to add to the discussion, except to 
describe the situation in the MFMM\@.

As described in the conclusion of Ref.~\cite{KLMNR}, the $\mu$-term is 
naturally generated at one-loop when the couplings 
$\lambda H^u 10^{FN} 10^{FN}$ and 
$\overline{\lambda} H^d \overline{10}^{FN} \overline{10}^{FN}$ are added to
the superpotential.  The leading contribution is the right size:
$\mu \sim (\lambda \overline{\lambda}/16\pi^2) 
(F_{\chi^{(1)}}/\vev{\chi^{(1)}})$
times a group theory factor.  However, there is a similar one loop 
contribution \cite{DvGP} to the scalar bilinear 
$B_{\mu} = \mu F_{\chi^{(1)}}/\vev{\chi^{(1)}}$ which is much too large
for EWSB to work naturally (if at all).  The authors of Ref.~\cite{DvGP}
did offer a solution that should work in our scenario, but it requires 
a new dimensionful parameter (although this could be dynamically 
generated as in Ref.~\cite{DDR}).  An alternative approach using
an extra U(1) (unrelated to our flavor U(1)$_F$) whose breaking 
generates an effective $\mu$-term was pursued in the context of a 
chiral supersymmetric standard model in Refs.~\cite{DPchiral,
CDMchiral}.

The simplest method to dynamically generate a $\mu$-term is to add a 
superpotential coupling $S H^u H^d$, where $S$ is a gauge singlet 
that may carry some global charge.  There are several candidates 
in the model for $S$, namely the flavons.  The MFMM offers $\chi^{(1)}$ 
as a gauge singlet, however the vev of this field is clearly too 
large to generate $\mu$.  This provides sufficient motivation to give 
$\chi^{(1)}$ a non-zero U(1)$_F$ charge, although we note that there
are phenomenological constraints on the size and sign of this charge
that we do not discuss further here.  A more natural possibility is 
$S \equiv \phi^{(3)}$.  This choice simply requires the constraint
$q_{\overline{5}^1} = q_{5^{FN(1)}}$, and is natural because 
$\vev{\phi^{(3)}}/\vev{\chi^{(2)}} \sim 10^{-2} - 10^{-3}$.
However, again we face the problem of an overly large $B_{\mu}$
since, in general, we expect the $F$-term of $\phi^{(3)}$ to be of order 
$\vev{\phi^{(3)}}\vev{\chi^{(2)}}$.  A priori this is not guaranteed,
and the flavon interactions might produce a small $F$-term, though
it seems unlikely.

Finally, Nilles and Polonski \cite{NP} produce a $\mu$-term via a
gauge singlet that is coupled gravitationally to the DSB sector
via the K\"{a}hler potential coupling $(S+S^{\dagger}) Q Q^{\dagger}$, 
where $Q$ is a DSB sector chiral superfield.  If $S$ has a self-cubic
coupling in the superpotential, then for $\sqrt{F_Q} \simeq 
\LambdaDSB \sim 10^7$~GeV, and $S$ has a vev of order the weak scale that
can dynamically produce $\mu$ without producing a large $B_{\mu}$.  
This mechanism can be implemented in the MFMM by adding an 
additional singlet or by again taking $\phi^{(3)}$ to be the singlet 
(making the generation of its vev distinct from the other flavons).

\subsection{Neutrino masses}
\label{neutrinos-sec}

The MFMM is apparently incomplete in one important sector of flavor:
it fails to predict viable neutrino masses.  Recent observations of 
zenith angle dependence in the number of atmospheric muon neutrino
events \cite{superK} together with the lower than expected measurements
of solar neutrino flux \cite{solar} suggest that (at least two) neutrinos
have small non-zero masses.  These experiments measure differences in 
(mass)$^2$ and find the relevant scales to be less than $1$~eV\@.  
The MFMM cannot
reproduce such a low scale in any reasonable way without invoking new
interactions at a large mass scale.  We find that it is possible
to account for the observed neutrino oscillations in a flavor-mediated
model by introducing the effects of heavy right-handed neutrinos \cite{seesaw},
or by introducing spontaneous $R$-parity breaking \cite{rpvneumass,
DEK-Nelson}
and including Planck scale contributions.  The latter solution
does not work with the SU(5) restriction while the former solution 
does if an assumption is made about the right-handed neutrino spectrum.

Given the current experimental data with the assumptions that only 
the known three neutrinos are light, and the mixing scales are 
hierarchical (i.e., $\Delta m_{\mathrm{atm}}^2 \gg 
\Delta m_{\mathrm{solar}}^2$)\footnote{The hierarchy 
$\Delta m_{\mathrm{atm}}^2 \gg \Delta m_{\mathrm{solar}}^2$ is actually 
required unless one ignores the data from one of the solar neutrino 
measurement techniques or ignores the solar models \cite{BHSSW}.}, 
one can establish significant constraints on neutrino mass and 
mixing parameters \cite{fogli}:
\begin{itemize}
\item The mass squared difference associated with atmospheric neutrinos
      is in the range $\Delta m_{\mathrm{atm}}^2 
      \sim 1 \times 10^{-3} - 6 \times 10^{-3}$ eV$^2$ ($90\%$ C.L.).
\item The mixing between $\nu_{\mu}$ and $\nu_{\tau}$ is large (i.e., 
      the mixing angle $\theta$ is bounded by $\sin{2\theta} > 0.8$ 
      ($90\%$ C.L.). 
\item The $\nu_e$ fraction of the heaviest mass eigenstate is small
      ($\lsim 0.1$ at $90\%$ C.L.)\footnote{This bound includes
      results from the CHOOZ Collaboration \cite{chooz}.}.
\end{itemize}
These constraints severely restrict the form of the neutrino mass matrix.  
If the neutrino mass matrices are to come from some symmetry, 
the leading contributions must be of the form \cite{BHS}:
\begin{eqnarray}
\mathbf{M}^{\nu 1} \;=\; \left( \begin{array}{ccc}
    0 & B & A \\
    B & 0 & 0 \\
    A & 0 & 0 \\
                        \end{array} \right)
                &
                \quad \mbox{or} \quad
                &
\mathbf{M}^{\nu 2} \;=\; \left( \begin{array}{ccc}
    0 & 0 & 0 \\
    0 & B^2/A & B \\
    0 & B & A \\
                        \end{array} \right)
\end{eqnarray}
where $A,B \sim \mathcal{O}(\Delta m_{\mathrm{atm}})$ and corrections 
to these forms are much smaller than $A$ and $B$\@.  Flavor-mediated 
models that satisfy FCNC constraints and are consistent with SU(5) 
do not contain symmetries that could produce either of these matrices.  
The first matrix, however, could be produced if assumptions are made 
about the high scale physics that produced it.  Either matrix can 
be produced once the SU(5) restriction is dropped.

There is an inherent difficulty producing viable neutrino masses in
the SU(5)-symmetric case.  Limits on FCNC require that the first two 
generations of right-handed down-type quarks 
$\overline{d}^1$, $\overline{d}^2$ must have non-zero U(1)$_F$
charges, while the third generation is constrained to be uncharged so
that all the down-type fermions have nonzero masses.  At the same time, 
the neutrino spectrum 
requires large mixing between the second and third generation lepton 
doublets.  To naturally produce this large mixing, 
$\overline{L}^2$ and $\overline{L}^3$ should have the same 
charge (preferably zero).  However, $\overline{L}^i$ and
$\overline{d}^i$ must have the same charge, as required by SU(5).
One can get around this problem (albeit in a slightly obtuse manner) 
by requiring that all leading 
order pieces to the neutrino mass matrix vanish.  Using the SU(5) variation 
where $\overline{5}^{FN(1)}$ is uncharged, and choosing U(1)$_F$ charges 
such that $q_{\overline{5}^1} = -q_{\overline{5}^{FN(2)}}$, one can 
produce the following (Majorana) mass matrix for the left-handed neutrinos:
\begin{eqnarray}
\mathbf{M}^{\nu} = \left( \begin{array}{ccccccc}
    0 & 0 & 0 & 0 & \lambda^{\nu}_1 m_{\mathrm{LL}} &
                                b_3^{(L)}\vev{\phi^{(3)}} & 0 \\
    0 & 0 & 0 & 0 & 0 &
                                0 & b_4^{(L)} \vev{\phi^{(4)}} \\
    0 & 0 & Y_{\nu_{\tau}} m_{\mathrm{LL}} & \lambda^{\nu}_2 m_{\rm LL} & 0 &
                                0 & b_5^{(L)}\vev{\phi^{(5)}} \\
    0 & 0 & \lambda^{\nu}_2 m_{\rm LL} & \lambda^{\nu}_3 m_{\rm LL} & 0 &
                                a_2^{(L)} \vev{\chi^{(2)}} & 0 \\
    \lambda^{\nu}_1 m_{\rm LL} & 0 & 0 & 0 & 0 &
                                0 & a_3^{(L)} \vev{\chi^{(3)}} \\
    b_3^{(L)}\vev{\phi^{(3)}} & 0 & 0 & a_2^{(L)} \vev{\chi^{(2)}} & 0 &
                                0 & 0 \\
    0 & b_4^{(L)} \vev{\phi^{(4)}} & b_5^{(L)}\vev{\phi^{(5)}} & 0 &
                a_3^{(L)} \vev{\chi^{(3)}} & 0 & 0 \\
                        \end{array} \right) & &
\label{nu-matrix-eq}
\end{eqnarray}
where the rows and columns represent 
($\overline{L}^1$, $\overline{L}^2$, $\overline{L}^3$, $\overline{L}^{FN(1)}$,
$\overline{L}^{FN(2)}$, $L^{FN(1)}$, $L^{FN(2)}$), 
$m_{\mbox{LL}} = \vev{H^u}^2/M$, and $M$ is the mass scale of right
handed neutrinos (assuming couplings of order unity)\footnote{Note
that the coupling and U(1)$_F$ charge of $\phi^{(5)}$ in this
scenario differ from those of the MFMM.}.
After integrating out the FN fields, the $3\times3$ neutrino
mass matrix becomes:
\begin{eqnarray}
\mathbf{M}^{\nu}_{\mathrm{eff}} &=& m_{\rm LL} \left( \begin{array}{ccc}
    0 & \lambda^{\nu}_1 \epsilon_{43}^{(L)} & 
	\lambda^{\nu}_1 \epsilon_{53}^{(L)} 
	+ \lambda^{\nu}_2 \epsilon_{32}^{(L)} \\
    \lambda^{\nu}_1 \epsilon_{43}^{(L)} & 0 & 0 \\
    \lambda^{\nu}_1 \epsilon_{53}^{(L)}        
        + \lambda^{\nu}_2 \epsilon_{32}^{(L)} & 0 & Y_{\nu_{\tau}}
    \end{array} \right) + \mathcal{O}(\epsilon^2)
\end{eqnarray}
This becomes $\mathbf{M}^{\nu_1}$ only if the (3,3) element vanishes.  
To see how this may be possible, consider a high energy theory containing
the following superpotential couplings:
\begin{equation}
W\supset N_b^1 \overline{5}_{-b}^{FN(2)} H^u + N_{-b}^2 \overline{5}_{b}^1 H^u 
	+ M N_{b}^1 N_{-b}^2 
	\; ,
\end{equation}
where the $N^i$ are right handed neutrinos, the subscripts are U(1)$_F$ 
charges, and all order one couplings have been suppressed.  Below the scale
$M$, the mass matrix Eq.~(\ref{nu-matrix-eq}) would be produced
without the couplings $\lambda^{\nu}_2$, $\lambda^{\nu}_3$, and 
$Y_{\nu_{\tau}}$.
Similarly, a global symmetry could forbid these couplings.
From atmospheric neutrino measurements and the other fermion 
mass matrices, the scale of right-handed neutrinos in this framework 
must be $M\sim 10^{13}$~GeV\@.  For solar neutrinos, 
either uncharged right-handed neutrinos should appear at the GUT scale 
(for a large angle MSW solution \cite{msw}) or the global symmetry should 
be broken by Planck scale physics (for a vacuum oscillation solution).

The extension to the non-SU(5) invariant models allows one more flexibility.
As we saw in Sec.~\ref{m-u-nonzero-sec}, the field content cannot be
varied from the MFMM in order to produce the correct phenomenology without
encountering a Landau pole below the unification scale.  However, allowing
different charges in the lepton sector permits both $\overline{L}^2$ 
and $\overline{L}^3$ to be uncharged.  In such a scenario, it is possible 
to produce either $\mathbf{M}^{\nu 1}$ or $\mathbf{M}^{\nu 2}$.  
The latter matrix could be produced via spontaneous violation 
of $R$-parity \cite{rpvneumass, DEK-Nelson} in a 
different sector.  It may be possible to use the DSB sector as the source 
of bilinear $R$-parity violation \cite{DEK-Nelson}, thus eliminating 
the need for an additional scale all together.  We will not explore 
this option here, saving it for possible future work.

\section{Conclusions}
\label{conclusions-sec}

In this paper we have presented the phenomenology of a new
economical model of flavor.  Within the context of gauge-mediation,
we utilize the hierarchy of scales, extra vector-like 
(messenger) matter, and additional gauge group structure to provide 
not only a means to communicate supersymmetry breaking to the MSSM 
fields, but also to generate the first and second generation fermion masses
using a modified Froggatt-Nielsen mechanism.  The identification
of the messenger U(1) that communicates supersymmetry breaking
from the DSB fields to the messenger-Froggatt-Nielsen fields is also 
the flavor symmetry in our model is essential.  In particular, 
all of the first and second generations are charged under the U(1)$_F$ 
and obtain flavor-dependent supersymmetry breaking contributions 
to their scalar masses.
Ordinarily this would be disastrous, however it is precisely 
because these fields are charged under the U(1)$_F$ that implies 
these contributions 
are rather large, roughly equivalent to the DSB scale suppressed by 
a two-loop factor.  Hence, squark-induced FCNCs are suppressed
by the heavy mass scale, following the ``more minimal 
supersymmetry'' approach \cite{CKN}.  
Perhaps the most fascinating phenomenological 
aspect of this model is that the fermion mass hierarchy and heavy 
first and second generation squarks are inextricably linked together.

There are several subtle issues to construct a successful supersymmetric
model of flavor along the lines we have discussed, and
we merely summarize the results of these model-building efforts here.  
In Sec.\ref{dynamics-sec} a simplified model was advocated as the 
``minimal flavor-mediated model'', that depends on four main scales:  
$\LambdaDSB$, $\MFN \simeq \vev{\chi}$, $\tilde{m}$, and $\sqrt{F_{\chi}}$.  
Several other parameters play an important role, including the 
normalized U(1)$_F$ FI term $q_\xi$, the U(1)$_F$ gauge coupling $g_F$, 
three superpotential parameters ($b_5$, $\lambda^d_1$, and $\lambda^d_2$) 
that determine the size of Yukawa-induced supersymmetry breaking
contributions, and also $\tan\beta$.  We discovered that there 
is a tight connection between the weak scale masses, including 
$m_{\tilde{\tau}_1}$ and the mass scale of the other third
generation scalars, and the heavy first and second generation
mass scale.  Increasing the (absolute) size of $\tilde{m}^2$
causes an increase in the first and second generation masses,
but also causes a decrease in the mass of the lightest stau.
This important result suggests a no-lose theorem for the model:  
Either the stau is light and should discovered in an upcoming collider
experiment, or the first and second generations are ``lighter'',
and should be manifest in future measurements of FCNC processes.

The signals of the model depend heavily on three factors:
The identity of the NLSP, the NLSP decay length, and the
mass of the lightest stau.  We showed that in the MFMM
the decay length is always much larger than the scale of 
the detector (hence the NLSP always \emph{escapes} the detector), 
and the mass of the stau is always less
than the mass of the lightest chargino and the second lightest
neutralino.  The NLSP is therefore either the lightest neutralino
or the lightest stau.  In both cases the production of 
the lightest chargino and/or the second lightest neutralino
is expected to result in about one tau per chargino, and two taus
per neutralino due to the two-body decays 
$\tilde{C}_1^\pm \ra \tilde{\tau}_1^\pm \nu_\tau$ and
$\tilde{N}_2 \ra \tilde{\tau}_1^\pm \tau^\mp$.
If the NLSP is the lightest neutralino, searching for these taus 
is the best method of discovering supersymmetry.
In this scenario the most promising production processes are 
$\tilde{C}_1^\pm \tilde{N}_2$ and $\tilde{N}_2 \tilde{N}_2$, 
that lead to the $3\tau$ and $4\tau$ plus missing energy final states.
If the NLSP is the lightest stau, then the stau itself should
manifest itself as a charged track in the detector.
Depending on the energy of the stau, this could mimic a
muon (if $\beta\gamma \gsim 0.85$) or appear as a unique 
highly-ionizing track (HIT).  Searching for signals with at
least one HIT is the best search strategy for collider
experiments.  Indeed, the reach of a high luminosity Tevatron 
extends up to about two hundred GeV for the lightest stau, and 
several hundred GeV for the lightest chargino and second lightest
neutralino \cite{FengMoroi, MartinWells}.

We should also remark that there are several other observables
characteristic of our flavor model that we have not mentioned.
The heavy first and second generation superfields are highly
split in mass between their fermionic components and the the
scalar components.  It was shown in Refs.~\cite{superoblique} that 
a consequence of nondegeneracy is an observable weak scale distinction 
between between the gauge coupling of fermions to gauge bosons, and the 
associated gauge coupling of a fermion and sfermion to a gaugino.
Unfortunately, measuring the difference between a ``gauge
coupling'' and a ``gaugino coupling'' is not easy,
especially for the electroweak couplings, since it requires plenty of 
observed sparticle production data.  Another important
class of signals that are perhaps much more promising for our
model are more precise FCNC measurements.  We have already
mentioned the important $D^0 \leftrightarrow \overline{D}^0$
mixing process that is expected, although several other
FCNC processes could provide a window on the (scalar) flavor 
structure of our model.

The two outstanding problems of the MFMM that are not definitely 
resolved are the dynamical origin of the $\mu$-term, 
and the generation of neutrino masses consistent with
the recent atmospheric and solar neutrino experiments.
The value of $\mu^2$ for any successful weak scale supersymmetric 
model must be matched to $M_Z$, and thus a dynamical solution to the 
$\mu$-problem is really a model-building issue.  Several possible
methods to generate $\mu$ were presented, although we remain
agnostic about the mechanism that is best for this framework.
Generating neutrino masses is more a glaring problem for the MFMM,
and we presented a few possible ways that this could be 
accomplished.  In all cases one must introduce new physics
at scales larger than $\LambdaDSB$ to accomplish the usual see-saw 
mechanism.  Combining this solution to flavor with a more natural 
mechanism for generating neutrino masses is an very interesting 
avenue of research shall be explored in future work.

\section*{Acknowledgments}
\indent

We thank K.~Agashe, D.~Demir, F.~Lepeintre, and A.~Nelson for 
useful discussions.
G.D.K. would like to thank the CERN theory group for hospitality 
where part of this work was completed.  This work was supported in part 
the U.S. Department of Energy under grant numbers 
and DOE-FG03-96ER-40956 and DOE-ER-40682-143.

\begin{appendix}
\refstepcounter{section}

\section*{Appendix~\thesection:~~Superpotential in components}
\label{superpot-component-app}

Here we present the full superpotential of the model in
component form.
\begin{eqnarray}
W &=&{} + a_1^{(Q)} \chi^{(1)} Q^{FN} \overline{Q}^{FN}
      + a_1^{(u)} \chi^{(1)} u^{FN} \overline{u}^{FN}
      + a_1^{(e)} \chi^{(1)} e^{FN} \overline{e}^{FN} \nonumber \\
& &{} + a_2^{(d)} \chi^{(2)} d^{FN(1)} \overline{d}^{FN(1)}
      + a_2^{(L)} \chi^{(2)} L^{FN(1)} \overline{L}^{FN(1)} \nonumber \\
& &{} + a_3^{(d)} \chi^{(3)} d^{FN(2)} \overline{d}^{FN(2)}
      + a_3^{(L)} \chi^{(3)} L^{FN(2)} \overline{L}^{FN(2)} \nonumber \\
& &{} + b_1^{(Q)} \phi^{(1)} Q^{1} \overline{Q}^{FN}
      + b_1^{(u)} \phi^{(1)} u^{1} \overline{u}^{FN}
      + b_1^{(e)} \phi^{(1)} e^{1} \overline{e}^{FN} \nonumber \\
& &{} + b_2^{(Q)} \phi^{(2)} Q^{2} \overline{Q}^{FN}
      + b_2^{(u)} \phi^{(2)} u^{2} \overline{u}^{FN}
      + b_2^{(e)} \phi^{(2)} e^{2} \overline{e}^{FN} \nonumber \\
& &{} + b_3^{(d)} \phi^{(3)} \overline{d}^{1} d^{FN(1)}
      + b_3^{(L)} \phi^{(3)} \overline{L}^{1} L^{FN(1)} \nonumber \\
& &{} + b_4^{(d)} \phi^{(4)} \overline{d}^{2} d^{FN(2)}
      + b_4^{(L)} \phi^{(4)} \overline{L}^{2} L^{FN(2)} \nonumber \\
& &{} + b_5^{(d)} \phi^{(5)} \overline{d}^{3} d^{FN(1)}
      + b_5^{(L)} \phi^{(5)} \overline{L}^{3} L^{FN(1)} \nonumber \\
& &{} + \lambda^{d}_{1, 1} H^{d} Q^{1} \overline{d}^{FN(1)}
      + \lambda^{d}_{1, 2} H^{d} e^{1} \overline{L}^{FN(1)} \nonumber \\
& &{} + \lambda^{d}_{2, 1} H^{d} Q^{2} \overline{d}^{FN(2)}
      + \lambda^{d}_{2, 2} H^{d} e^{2} \overline{L}^{FN(2)} \nonumber \\
& &{} + Y_b H^{d} Q^{3} \overline{d}^{3}
      + Y_\tau H^{d} e^{3} \overline{L}^{3} \nonumber \\
& &{} + \lambda^{d}_{4, 1} H^{d} Q^{FN} \overline{d}^{3}
      + \lambda^{d}_{4, 2} H^{d} e^{FN} \overline{L}^{3} \nonumber \\
& &{} + Y_t H^{u} Q^{3} u^{3}
      + \lambda^{u}_{2, 1} H^{u} Q^{3} u^{FN} 
      + \lambda^{u}_{2, 2} H^{u} u^{3} Q^{FN} \; .
\label{superpot-component-eq}
\end{eqnarray}
Since no Higgs triplets are present in the model, only some
component terms remain after expanding the SU(5) tensor products.
It is also possible to identify $\lambda^{u}_1$ as the top
Yukawa coupling $Y_t$, and $\lambda^{d}_{3,1}, \lambda^{d}_{3,2}$
as the bottom and tau Yukawa couplings $Y_b, Y_\tau$.

To define our notation, the quantum numbers of these fields 
are given in Table~\ref{quantum-numbers-table}.
\begin{table}
\renewcommand{\baselinestretch}{1.3}\small\normalsize
\begin{center}
\begin{tabular}{rc|cccc} \hline
\multicolumn{2}{c|}{field} & SU(3)$_c$ & SU(2)$_L$ & U(1)$_Y$ & U(1)$_F$ \\
SU(5) & component & & & & \\ \hline
& $Q^{FN}$ & $\three$    & $\two$ & $\textfrac{1}{6}$ & $0$ \\
$10^{FN}$ & $u^{FN}$ & $\threebar$ & $\one$ & $-\textfrac{2}{3}$ & $0$ \\
& $e^{FN}$ & $\one$      & $\one$ & $1$ & $0$ \\ \cline{1-2}
& $\overline{Q}^{FN}$ & $\threebar$ & $\two$ & $-\textfrac{1}{6}$ 
    & $q_{\overline{10}^{FN}}$ \\
$\overline{10}^{FN}$ & $\overline{u}^{FN}$ & $\three$ & $\one$ 
    & $\textfrac{2}{3}$ & $q_{\overline{10}^{FN}}$ \\
& $\overline{e}^{FN}$ & $\one$ & $\one$ & $-1$ 
    & $q_{\overline{10}^{FN}}$ \\ \cline{1-2}
$5^{FN(1)}$ & $d^{FN(1)}$ & $\three$ & $\one$ & $-\textfrac{1}{3}$ 
    & $q_{5^{FN(1)}}$ \\
& $L^{FN(1)}$ & $\one$ & $\two$ & $\textfrac{1}{2}$ 
    & $q_{5^{FN(1)}}$ \\ \cline{1-2}
$\overline{5}^{FN(1)}$ & $\overline{d}^{FN(1)}$ & $\threebar$ & $\one$ 
    & $\textfrac{1}{3}$ & $q_{\overline{5}^{FN(1)}}$ \\
& $\overline{L}^{FN(1)}$ & $\one$ & $\two$ & $-\textfrac{1}{2}$ 
    & $q_{\overline{5}^{FN(1)}}$ \\ \cline{1-2}
$5^{FN(2)}$ & $d^{FN(2)}$ & $\three$ & $\one$ & $-\textfrac{1}{3}$ 
    & $q_{5^{FN(2)}}$ \\
& $L^{FN(2)}$ & $\one$ & $\two$ & $\textfrac{1}{2}$ 
    & $q_{5^{FN(2)}}$ \\ \cline{1-2}
$\overline{5}^{FN(2)}$ & $\overline{d}^{FN(2)}$ & $\threebar$ & $\one$ 
    & $\textfrac{1}{3}$ & $q_{\overline{5}^{FN(2)}}$ \\
& $\overline{L}^{FN(2)}$ & $\one$ & $\two$ & $-\textfrac{1}{2}$ 
    & $q_{\overline{5}^{FN(2)}}$ \\ \cline{1-2}
& $Q^1$  & $\three$  & $\two$ & $\textfrac{1}{6}$ & $q_{10^1}$ \\
$10^1$ & $u^1$  & $\threebar$  & $\one$ & $-\textfrac{2}{3}$ & $q_{10^1}$ \\
& $e^1$  & $\one$    & $\one$ & $1$ & $q_{10^1}$ \\ \cline{1-2}
& $Q^2$  & $\three$  & $\two$ & $\textfrac{1}{6}$ & $q_{10^2}$ \\
$10^2$ & $u^2$  & $\threebar$  & $\one$ & $-\textfrac{2}{3}$ & $q_{10^2}$ \\
& $e^2$  & $\one$    & $\one$ & $1$ & $q_{10^2}$ \\ \cline{1-2}
$\overline{5}^1$ & $\overline{d}^1$ & $\threebar$ & $\one$ & $\textfrac{1}{3}$ 
    & $q_{\overline{5}^1}$ \\
& $\overline{L}^1$ & $\one$ & $\two$ & $-\textfrac{1}{2}$ 
    & $q_{\overline{5}^1}$ \\ \cline{1-2}
$\overline{5}^2$ & $\overline{d}^2$ & $\threebar$ & $\one$ & $\textfrac{1}{3}$ 
    & $q_{\overline{5}^2}$ \\
& $\overline{L}^2$ & $\one$ & $\two$ & $-\textfrac{1}{2}$ 
    & $q_{\overline{5}^2}$ \\ \cline{1-2}
& $Q^3$ & $\three$  & $\two$ & $\textfrac{1}{6}$ & $0$ \\
$10^3$ & $u^3$ & $\threebar$  & $\one$ & $-\textfrac{2}{3}$ & $0$ \\
& $e^3$ & $\one$    & $\one$ & $1$ & $0$ \\ \cline{1-2}
$\overline{5}^3$ & $\overline{d}^3$ & $\threebar$ & $\one$ 
    & $\textfrac{1}{3}$ & $0$ \\ 
& $\overline{L}^3$ & $\one$ & $\two$ & $-\textfrac{1}{2}$ & $0$ \\ \hline
\end{tabular}
\end{center}
\renewcommand{\baselinestretch}{1.0}\small\normalsize
\caption{Charge assignments for the matter considered in this paper.  
Note that the $U(1)_Y$ charges are in the GUT normalization.}
\label{quantum-numbers-table}
\end{table}
We did not list the the flavons since they are uncharged under the
SM gauge groups.  

\subsection{Renormalization group equations}
\label{DSB-FN-RGE-app}

The renormalization group equations of the (mass)$^2$ for the scalars
in the superpotential, Eq.~(\ref{superpot-component-eq}), have been
computed below using Ref.~\cite{MartinVaughn}.
These are given to one-loop, which is sufficient for our purposes.
(The two-loop gauge contributions to the uncharged scalars are 
incorporated via the boundary conditions at $\MFN$.)

In general, we write the $\beta$-function for the scalar (mass)$^2$ as
\begin{eqnarray}
\frac{d}{dt} m^2_i &=& \frac{1}{16 \pi^2} \left( \beta_{i} 
    - 8 q_i^2 g_F^2 |M_F|^2 + 2 q_i g_F^2 S' \right) \; .
    \label{DSB-FN-RGE-eq}
\end{eqnarray}
where $t \equiv \ln Q$, $q_i$ is the U(1)$_F$ charge of field $i$, 
$g_F$ is the U(1)$_F$ gauge coupling, $M_F$ is the U(1)$_F$ gaugino mass, 
and we define the scalar (mass)$^2$ function $S'$ by
\begin{eqnarray}
S' &=& \sum_j S(j) q_j m_j^2 \; ,
    \label{weighted-sum-eq}
\end{eqnarray}
where $S(j)$ is the Dynkin index for the field $j$.

The Yukawa coupling-dependent components are given by
\begin{eqnarray*}
\beta_{\chi^{(1)}} &=&{} 
    + 12 \Big(a_1^{(Q)}\Big)^2 \Big[ m_{\chi^{(1)}}^2 + m_{Q^{FN}}^2 
                              + m_{\overline{Q}^{FN}}^2 \Big]
    +  6 \Big(a_1^{(u)}\Big)^2 \Big[ m_{\chi^{(1)}}^2 + m_{u^{FN}}^2 
                              + m_{\overline{u}^{FN}}^2 \Big] \nonumber \\
& &{} +  2 \Big(a_1^{(e)}\Big)^2 \Big[ m_{\chi^{(1)}}^2 + m_{e^{FN}}^2 
                              + m_{\overline{e}^{FN}}^2 \Big] \\
\beta_{\chi^{(2)}} &=&{} 
    + 6 \Big(a_2^{(d)}\Big)^2 \Big[ m_{\chi^{(2)}}^2 + m_{d^{FN(1)}}^2 
                              + m_{\overline{d}^{FN(1)}}^2 \Big]
    + 4 \Big(a_2^{(L)}\Big)^2 \Big[ m_{\chi^{(2)}}^2 + m_{L^{FN(1)}}^2 
                              + m_{\overline{L}^{FN(1)}}^2 \Big] \\
\beta_{\chi^{(3)}} &=&{} 
    + 6 \Big(a_3^{(d)}\Big)^2 \Big[ m_{\chi^{(3)}}^2 + m_{d^{FN(2)}}^2 
                              + m_{\overline{d}^{FN(2)}}^2 \Big]
    + 4 \Big(a_3^{(L)}\Big)^2 \Big[ m_{\chi^{(3)}}^2 + m_{L^{FN(2)}}^2 
                              + m_{\overline{L}^{FN(2)}}^2 \Big] \\
\beta_{Q^{FN}} &=&{} 
    + 2 \Big(a_1^{(Q)}\Big)^2 \Big[ m_{Q^{FN}}^2 + m_{\chi^{(1)}}^2 
                              + m_{\overline{Q}^{FN}}^2 \Big]
    + 2 \Big(\lambda^d_{4,1}\Big)^2 \Big[ m_{Q^{FN}}^2 + m_{H^d}^2 
                              + m_{\overline{d}^{3}}^2 \Big] \nonumber \\
& &{} + 2 \Big(\lambda^u_{2,2}\Big)^2 \Big[ m_{Q^{FN}}^2 + m_{H^u}^2 
                              + m_{u^3}^2 \Big] \\
\beta_{u^{FN}} &=&{} 
    + 2 \Big(a_1^{(u)}\Big)^2 \Big[ m_{u^{FN}}^2 + m_{\chi^{(1)}}^2 
                              + m_{\overline{u}^{FN}}^2 \Big]
    + 4 \Big(\lambda^u_{2,1}\Big)^2 \Big[ m_{u^{FN}}^2 + m_{H^u}^2 
                              + m_{\overline{Q}^3}^2 \Big] \nonumber \\
\beta_{e^{FN}} &=&{} 
    + 2 \Big(a_1^{(e)}\Big)^2 \Big[ m_{e^{FN}}^2 + m_{\chi^{(1)}}^2 
                              + m_{\overline{e}^{FN}}^2 \Big]
    + 4 \Big(\lambda^d_{4,2}\Big)^2   \Big[ m_{e^{FN}}^2 + m_{H^d}^2 
                              + m_{\overline{L}^{3}}^2 \Big] \nonumber \\
\beta_{\overline{Q}^{FN}} &=&{} 
    + 2 \Big(a_1^{(Q)}\Big)^2 \Big[ m_{\overline{Q}^{FN}}^2 + m_{\chi^{(1)}}^2 
                              + m_{Q^{FN}}^2 \Big]
    + 2 \Big(b_1^{(Q)}\Big)^2 \Big[ m_{\overline{Q}^{FN}}^2 + m_{\phi^{(1)}}^2 
                              + m_{Q^{1}}^2 \Big] \nonumber \\
& &{} + 2 \Big(b_2^{(Q)}\Big)^2 \Big[ m_{\overline{Q}^{FN}}^2 
                              + m_{\phi^{(2)}}^2 + m_{Q^{2}}^2 \Big] \\
\beta_{\overline{u}^{FN}} &=&{} 
    + 2 \Big(a_1^{(u)}\Big)^2 \Big[ m_{\overline{u}^{FN}}^2 + m_{\chi^{(1)}}^2 
                              + m_{u^{FN}}^2 \Big]
    + 2 \Big(b_1^{(u)}\Big)^2 \Big[ m_{\overline{u}^{FN}}^2 + m_{\phi^{(1)}}^2 
                              + m_{u^{1}}^2 \Big] \nonumber \\
& &{} + 2 \Big(b_2^{(u)}\Big)^2 \Big[ m_{\overline{u}^{FN}}^2 
                              + m_{\phi^{(2)}}^2 + m_{u^{2}}^2 \Big] \\
\beta_{\overline{e}^{FN}} &=&{} 
    + 2 \Big(a_1^{(e)}\Big)^2 \Big[ m_{\overline{e}^{FN}}^2 + m_{\chi^{(1)}}^2 
                              + m_{e^{FN}}^2 \Big]
    + 2 \Big(b_1^{(e)}\Big)^2 \Big[ m_{\overline{e}^{FN}}^2 + m_{\phi^{(1)}}^2 
                              + m_{e^{1}}^2 \Big] \nonumber \\
& &{} + 2 \Big(b_2^{(e)}\Big)^2 \Big[ m_{\overline{e}^{FN}}^2 
                              + m_{\phi^{(2)}}^2 + m_{e^{2}}^2 \Big] \\
\beta_{d^{FN(1)}} &=&{} 
    + 2 \Big(a_2^{(d)}\Big)^2 \Big[ m_{d^{FN(1)}}^2 + m_{\chi^{(2)}}^2 
                              + m_{\overline{d}^{FN(1)}}^2 \Big]
    + 2 \Big(b_3^{(d)}\Big)^2 \Big[ m_{d^{FN(1)}}^2 + m_{\phi^{(3)}}^2 
                              + m_{\overline{d}^1}^2 \Big] \nonumber \\
& &{} + 2 \Big(b_5^{(d)}\Big)^2 \Big[ m_{d^{FN(1)}}^2 + m_{\phi^{(5)}}^2 
                              + m_{\overline{d}^2}^2 \Big] \\
\beta_{L^{FN(1)}} &=&{} 
    + 2 \Big(a_2^{(L)}\Big)^2 \Big[ m_{L^{FN(1)}}^2 + m_{\chi^{(2)}}^2 
                              + m_{\overline{L}^{FN(1)}}^2 \Big]
    + 2 \Big(b_3^{(L)}\Big)^2 \Big[ m_{L^{FN(1)}}^2 + m_{\phi^{(3)}}^2 
                              + m_{\overline{L}^1}^2 \Big] \nonumber \\
& &{} + 2 \Big(b_5^{(L)}\Big)^2 \Big[ m_{L^{FN(1)}}^2 + m_{\phi^{(5)}}^2 
                              + m_{\overline{L}^2}^2 \Big] \\
\beta_{\overline{d}^{FN(1)}} &=&{} 
    + 2 \Big(a_2^{(d)}\Big)^2 \Big[ m_{\overline{d}^{FN(1)}}^2 
                              + m_{\chi^{(2)}}^2 + m_{d^{FN(1)}}^2 \Big]
    + 4 \Big(\lambda^d_{1,1}\Big)^2 \Big[ m_{\overline{d}^{FN(1)}}^2 
                              + m_{H^d}^2 + m_{Q^{1}}^2 \Big] \nonumber \\
\beta_{\overline{L}^{FN(1)}} &=&{} 
    + 2 \Big(a_2^{(L)}\Big)^2 \Big[ m_{\overline{L}^{FN(1)}}^2 
                              + m_{\chi^{(2)}}^2 + m_{L^{FN(1)}}^2 \Big]
    + 2 \Big(\lambda^d_{1,2}\Big)^2 \Big[ m_{\overline{L}^{FN(1)}}^2 
                              + m_{H^d}^2 + m_{e^{1}}^2 \Big] \nonumber \\
\beta_{d^{FN(2)}} &=&{} 
    + 2 \Big(a_3^{(d)}\Big)^2 \Big[ m_{d^{FN(2)}}^2 + m_{\chi^{(3)}}^2 
                              + m_{\overline{d}^{FN(2)}}^2 \Big]
    + 2 \Big(b_4^{(d)}\Big)^2 \Big[ m_{d^{FN(2)}}^2 + m_{\phi^{(4)}}^2 
                              + m_{\overline{d}^2}^2 \Big] \nonumber \\
\beta_{L^{FN(2)}} &=&{} 
    + 2 \Big(a_3^{(L)}\Big)^2 \Big[ m_{L^{FN(2)}}^2 + m_{\chi^{(3)}}^2 
                              + m_{\overline{L}^{FN(2)}}^2 \Big]
    + 2 \Big(b_4^{(L)}\Big)^2 \Big[ m_{L^{FN(2)}}^2 + m_{\phi^{(4)}}^2 
                              + m_{\overline{L}^2}^2 \Big] \nonumber \\
\beta_{\overline{d}^{FN(2)}} &=&{} 
    + 2 \Big(a_3^{(d)}\Big)^2 \Big[ m_{\overline{d}^{FN(2)}}^2 
                              + m_{\chi^{(3)}}^2 + m_{d^{FN(2)}}^2 \Big]
    + 4 \Big(\lambda^d_{2,1}\Big)^2 \Big[ m_{\overline{d}^{FN(2)}}^2 
                              + m_{H^d}^2 + m_{Q^{2}}^2 \Big] \nonumber \\
\beta_{\overline{L}^{FN(2)}} &=&{} 
    + 2 \Big(a_3^{(L)}\Big)^2 \Big[ m_{\overline{L}^{FN(2)}}^2 
                              + m_{\chi^{(3)}}^2 + m_{L^{FN(2)}}^2 \Big]
    + 2 \Big(\lambda^d_{2,2}\Big)^2 \Big[ m_{\overline{L}^{FN(2)}}^2 
                              + m_{H^d}^2 + m_{e^{2}}^2 \Big] \nonumber \\
\beta_{\phi^{(1)}} &=&{} 
    + 12 \Big(b_1^{(Q)}\Big)^2 \Big[ m_{\phi^{(1)}}^2 + m_{Q^1}^2 
                              + m_{\overline{Q}^{FN}}^2 \Big]
    + 6 \Big(b_1^{(u)}\Big)^2 \Big[ m_{\phi^{(1)}}^2 + m_{u^1}^2 
                              + m_{\overline{u}^{FN}}^2 \Big] \nonumber \\
& &{} + 2 \Big(b_1^{(e)}\Big)^2 \Big[ m_{\phi^{(1)}}^2 + m_{e^1}^2 
                              + m_{\overline{e}^{FN}}^2 \Big] \\
\beta_{\phi^{(2)}} &=&{} 
    + 12 \Big(b_2^{(Q)}\Big)^2 \Big[ m_{\phi^{(2)}}^2 + m_{Q^2}^2 
                              + m_{\overline{Q}^{FN}}^2 \Big]
    + 6 \Big(b_2^{(u)}\Big)^2 \Big[ m_{\phi^{(2)}}^2 + m_{u^2}^2 
                              + m_{\overline{u}^{FN}}^2 \Big] \nonumber \\
& &{} + 2 \Big(b_2^{(e)}\Big)^2 \Big[ m_{\phi^{(2)}}^2 + m_{e^2}^2 
                              + m_{\overline{e}^{FN}}^2 \Big] \\
\beta_{\phi^{(3)}} &=&{} 
    + 6 \Big(b_3^{(d)}\Big)^2 \Big[ m_{\phi^{(3)}}^2 + m_{\overline{d}^1}^2 
                              + m_{d^{FN(1)}}^2 \Big]
    + 4 \Big(b_3^{(L)}\Big)^2 \Big[ m_{\phi^{(3)}}^2 + m_{\overline{L}^1}^2 
                              + m_{L^{FN(1)}}^2 \Big] \nonumber \\
\beta_{\phi^{(4)}} &=&{} 
    + 6 \Big(b_4^{(d)}\Big)^2 \Big[ m_{\phi^{(4)}}^2 + m_{\overline{d}^2}^2 
                              + m_{d^{FN(2)}}^2 \Big]
    + 4 \Big(b_4^{(L)}\Big)^2 \Big[ m_{\phi^{(4)}}^2 + m_{\overline{L}^2}^2 
                              + m_{L^{FN(2)}}^2 \Big] \nonumber \\
\beta_{\phi^{(5)}} &=&{} 
    + 6 \Big(b_5^{(d)}\Big)^2 \Big[ m_{\phi^{(5)}}^2 + m_{\overline{d}^3}^2 
                              + m_{d^{FN(1)}}^2 \Big]
    + 4 \Big(b_5^{(L)}\Big)^2 \Big[ m_{\phi^{(5)}}^2 + m_{\overline{L}^3}^2 
                              + m_{L^{FN(1)}}^2 \Big] \nonumber \\
\beta_{Q^1} &=&{} 
    + 2 \Big(b_1^{(Q)}\Big)^2 \Big[ m_{Q^1}^2 + m_{\phi^{(1)}}^2 
                              + m_{\overline{Q}^{FN}}^2 \Big]
    + 2 \Big(\lambda^d_{1,1}\Big)^2 \Big[ m_{Q^1}^2 + m_{H^d}^2 
                              + m_{\overline{d}^{FN(1)}}^2 \Big] \nonumber \\
\beta_{Q^2} &=&{} 
    + 2 \Big(b_2^{(Q)}\Big)^2 \Big[ m_{Q^2}^2 + m_{\phi^{(2)}}^2 
                              + m_{\overline{Q}^{FN}}^2 \Big]
    + 2 \Big(\lambda^d_{2,1}\Big)^2 \Big[ m_{Q^2}^2 + m_{H^d}^2 
                              + m_{\overline{d}^{FN(2)}}^2 \Big] \nonumber \\
\beta_{u^1} &=&{} 
    + 2 \Big(b_1^{(u)}\Big)^2 \Big[ m_{u^1}^2 + m_{\phi^{(1)}}^2 
                              + m_{\overline{u}^{FN}}^2 \Big] \nonumber \\
\beta_{u^2} &=&{} 
    + 2 \Big(b_2^{(u)}\Big)^2 \Big[ m_{u^2}^2 + m_{\phi^{(2)}}^2 
                              + m_{\overline{u}^{FN}}^2 \Big] \nonumber \\
\beta_{e^1} &=&{} 
    + 2 \Big(b_1^{(e)}\Big)^2 \Big[ m_{e^1}^2 + m_{\phi^{(1)}}^2 
                              + m_{\overline{e}^{FN}}^2 \Big]
    + 4 \Big(\lambda^d_{1,2}\Big)^2 \Big[ m_{e^1}^2 + m_{H^d}^2 
                              + m_{\overline{L}^{FN(1)}}^2 \Big] \nonumber \\
\beta_{e^2} &=&{} 
    + 2 \Big(b_2^{(e)}\Big)^2 \Big[ m_{e^2}^2 + m_{\phi^{(2)}}^2 
                              + m_{\overline{e}^{FN}}^2 \Big]
    + 4 \Big(\lambda^d_{2,2}\Big)^2 \Big[ m_{e^2}^2 + m_{H^d}^2 
                              + m_{\overline{L}^{FN(2)}}^2 \Big] \nonumber \\
\beta_{\overline{d}^1} &=&{} 
    + 2 \Big(b_3^{(d)}\Big)^2 \Big[ m_{\overline{d}^1}^2 + m_{\phi^{(3)}}^2 
                              + m_{d^{FN(1)}}^2 \Big] \nonumber \\
\beta_{\overline{d}^2} &=&{} 
    + 2 \Big(b_4^{(d)}\Big)^2 \Big[ m_{\overline{d}^2}^2 + m_{\phi^{(4)}}^2 
                              + m_{d^{FN(2)}}^2 \Big] \nonumber \\
\beta_{\overline{L}^1} &=&{} 
    + 2 \Big(b_3^{(L)}\Big)^2 \Big[ m_{\overline{L}^1}^2 + m_{\phi^{(3)}}^2 
                              + m_{L^{FN(1)}}^2 \Big] \nonumber \\
\beta_{\overline{L}^2} &=&{} 
    + 2 \Big(b_4^{(L)}\Big)^2 \Big[ m_{\overline{L}^2}^2 + m_{\phi^{(4)}}^2 
                              + m_{L^{FN(2)}}^2 \Big] \nonumber \\
\beta_{Q^3} &=&{} 
    + 2 \Big(Y_t\Big)^2 \Big[ m_{Q^3}^2 + m_{H^u}^2 
                              + m_{u^3}^2 \Big]
    + 2 \Big(\lambda^u_{2,1}\Big)^2 \Big[ m_{Q^3}^2 + m_{H^u}^2 
                              + m_{u^{FN}}^2 \Big] \nonumber \\
& &{} + 2 \Big(Y_b\Big)^2 \Big[ m_{Q^3}^2 + m_{H^d}^2 
                              + m_{\overline{d}^{3}}^2 \Big] \\
\beta_{u^3} &=&{} 
    + 4 \Big(Y_t\Big)^2 \Big[ m_{u^3}^2 + m_{H^u}^2 
                              + m_{Q^3}^2 \Big]
    + 4 \Big(\lambda^u_{2,2}\Big)^2 \Big[ m_{u^3}^2 + m_{H^u}^2 
                              + m_{Q^{FN}}^2 \Big] \nonumber \\
\beta_{e^3} &=&{} 
    + 4 \Big(Y_\tau\Big)^2 \Big[ m_{e^3}^2 + m_{H^d}^2 
                              + m_{\overline{L}^3}^2 \Big] \nonumber \\
\beta_{\overline{d}^3} &=&{} 
    + 2 \Big(b_5^{(d)}\Big)^2 \Big[ m_{\overline{d}^3}^2 + m_{\phi^{(5)}}^2 
                              + m_{d^{FN(1)}}^2 \Big]
    + 4 \Big(Y_b\Big)^2 \Big[ m_{\overline{d}^3}^2 + m_{H^d}^2 
                              + m_{Q^3}^2 \Big] \nonumber \\
& &{} + 4 \Big(\lambda^d_{4,1}\Big)^2 \Big[ m_{\overline{d}^3}^2 + m_{H^d}^2 
                              + m_{Q^{FN}}^2 \Big] \\
\beta_{\overline{L}^3} &=&{} 
    + 2 \Big(b_5^{(L)}\Big)^2 \Big[ m_{\overline{L}^3}^2 + m_{\phi^{(5)}}^2 
                              + m_{L^{FN(1)}}^2 \Big]
    + 2 \Big(Y_\tau\Big)^2 \Big[ m_{\overline{L}^3}^2 + m_{H^d}^2 
                              + m_{e^3}^2 \Big] \nonumber \\
& &{} + 2 \Big(\lambda^d_{4,2}\Big)^2 \Big[ m_{\overline{L}^3}^2 + m_{H^d}^2 
                              + m_{e^{FN}}^2 \Big] \\
\beta_{H^u} &=&{} 
    + 6 \Big(Y_t\Big)^2 \Big[ m_{H^u}^2 + m_{Q^3}^2 
                              + m_{u^3}^2 \Big]
    + 6 \Big(\lambda^u_{2,1}\Big)^2 \Big[ m_{H^u}^2 + m_{Q^3}^2 
                              + m_{u^{FN}}^2 \Big] \nonumber \\
& &{} + 6 \Big(\lambda^u_{2,2}\Big)^2 \Big[ m_{H^u}^2 + m_{Q^{FN}}^2 
                              + m_{u^3}^2 \Big] \\
\beta_{H^d} &=&{} 
    + 6 \Big(\lambda^d_{1,1}\Big)^2 \Big[ m_{H^d}^2 + m_{Q^1}^2 
                              + m_{\overline{d}^{FN(1)}}^2 \Big]
    + 2 \Big(\lambda^d_{1,2}\Big)^2 \Big[ m_{H^d}^2 + m_{e^1}^2 
                              + m_{\overline{L}^{FN(1)}}^2 \Big] \nonumber \\
& &{} + 6 \Big(\lambda^d_{2,1}\Big)^2 \Big[ m_{H^d}^2 + m_{Q^2}^2 
                              + m_{\overline{d}^{FN(2)}}^2 \Big]
    + 2 \Big(\lambda^d_{2,2}\Big)^2 \Big[ m_{H^d}^2 + m_{e^2}^2 
                              + m_{\overline{L}^{FN(2)}}^2 \Big] \nonumber \\
& &{} + 6 \Big(Y_b\Big)^2 \Big[ m_{H^d}^2 + m_{Q^3}^2 
                              + m_{\overline{d}^3}^2 \Big] 
    + 2 \Big(Y_\tau\Big)^2 \Big[ m_{H^d}^2 + m_{e^3}^2 
                              + m_{\overline{L}^3}^2 \Big] \nonumber \\
& &{} + 6 \Big(\lambda^d_{4,1}\Big)^2 \Big[ m_{H^d}^2 + m_{Q^{FN}}^2 
                              + m_{\overline{d}^3}^2 \Big]
    + 2 \Big(\lambda^d_{4,2}\Big)^2 \Big[ m_{H^d}^2 + m_{e^{FN}}^2 
                              + m_{\overline{L}^3}^2 \Big] \\
\end{eqnarray*}

\refstepcounter{section}
\section*{Appendix~\thesection:~~Example parameter set and sparticle spectrum}
\label{example-app}

\begin{figure}[!t]
\centering
\epsfxsize=5.0in
\hspace*{0in}
\epsffile{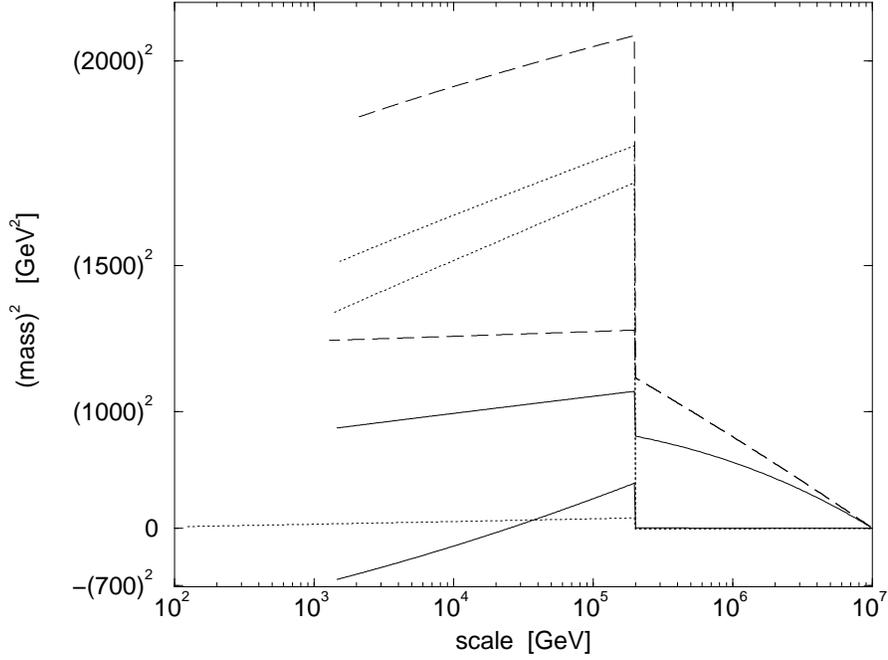}
\caption{Using the example set of parameters given in 
Table~\ref{example-table}, the evolution of several soft scalar 
(mass)$^2$ is shown as a function of the renormalization scale.  
The (top, bottom) solid lines correspond to ($m_{H^d}^2$, $m_{H^u}^2$),
the (top, middle, bottom) dotted lines correspond to ($m_{Q^3}^2$, 
$m_{u^3}^2$, $m_{e^3}^2$), and the (top, bottom) dashed lines 
correspond to ($m_{\overline{d}^3}^2$, $m_{\overline{L}^3}^2$).
The ``break'' in the evolution at $\MFN = 2 \times 10^5$ GeV
corresponds to where the FN sector is integrated out, and 
thus additional supersymmetry breaking contributions are
induced for the light MSSM fields.}
\label{example-fig}
\end{figure}

We present one example set of parameters of the MFMM, and
the resulting weak scale spectra.  The input parameters
and calculated weak scale quantities are shown in
Table~\ref{example-table}.  The importance of the RG
\begin{table}
\renewcommand{\baselinestretch}{1.3}\small\normalsize
\begin{center}
\begin{tabular}{r|c||r|c} \hline\hline
\multicolumn{1}{c|}{input parameters} & 
  value & 
  \multicolumn{1}{c|}{calculated quantities} & 
  value \\ \hline
$\vev{\chi^{(1)}} = \vev{\chi^{(2)}} = \vev{\chi^{(3)}}$ &
  $200$~TeV &
  $m_{\tilde{N}_1} \simeq M_1(M_1)$ 
  & $170$ GeV \\
$(-\tilde{m}^2)^{1/2} = (F_{\chi}/\vev{\chi})$ &
  $24$ TeV &
  $m_{\tilde{N}_2} \simeq m_{\tilde{C}_1} \simeq M_2(M_2)$ & 
  $325$ GeV \\
$\LambdaDSB$ &
  $10^7$ GeV &
  $m_{\tilde{N}_3} \simeq m_{\tilde{N}_4} \simeq m_{\tilde{C}_2} \simeq |\mu|$ 
  & $700$ GeV \\
$q \equiv q_{10^1} \sim q_{10^2} \sim q_{\overline{5}^1} \sim 
q_{\overline{5}^2}$ & 
  $-0.5$ & 
  $m_{\mathrm{gluino}} \sim M_3(M_3)$ & 
  $900$ GeV \\
$q_r \equiv q_{\chi^{(2)}}/q_{\chi^{(3)}}$ & 
  $1$ &
  $m_{1st,2nd}$ & 
  $9.8 \ra 10.8$ TeV \\
$-q_\xi/q$ & 
  $2$ &
  $m_{\tilde{t}_1}$, $m_{\tilde{t}_2}$ & 
  $1365$, $1520$ GeV \\ 
$g_F$ &
  $0.3$ &
  $m_{\tilde{b}_1}$, $m_{\tilde{b}_2}$ & 
  $1510$, $1870$ GeV \\ 
$b_5 = \lambda^d_1 = \lambda^d_2$ &
  $0.1$ &
  $m_{\tilde{\tau}_1}$, $m_{\tilde{\tau}_2}$, $m_{\tilde{\nu}_\tau}$ & 
  $125$, $1270$, $1265$ GeV \\
$\tan\beta$ &
  $26$ &
  $\theta_{\tilde{\tau}}$ & 
  $0.987 \times \pi/2$ \\
$\mathrm{sign}(\mu)$ &
  $+$ &
  $m_h$ & 
  $120$ GeV \\
&
&
  $m_A \simeq m_H \simeq m_{H^\pm}$ & 
  $1150$ GeV \\
&
&
  $B_\mu$ & 
  $-(230 \; \mathrm{GeV})^2$ \\
&
&
  $g_1$, $g_2 \simeq g_3$ at $\Munif$ & 
  $1.80$, $2.13$ \\ 
&
&
  $\Munif$ &
  $1.8 \times 10^{16}$ GeV \\ \hline\hline
\end{tabular}
\end{center}
\renewcommand{\baselinestretch}{1.0}\small\normalsize
\caption{Example set of input parameters and some of the
calculated weak scale masses and couplings for an example MFMM\@.}
\label{example-table}
\end{table}
evolution in this model was emphasized in Sec.~\ref{RGE-sec},
and thus we also show in Fig.~\ref{example-fig} the evolution 
of several soft (mass)$^2$ as a function of the renormalization scale.
Notice that there is significant evolution between both
the DSB scale and the FN scale, as well as between the FN scale
and the weak scale.  Since all of the fields shown in the figure 
are uncharged under the U(1)$_F$, their mass at the DSB scale
is assumed to vanish.

\end{appendix}



\begin{thebibliography}{99}
\singlespaced

\bibitem{Nimaframework}
    N. Arkani-Hamed, C.D. Carone, L.J. Hall, and H. Murayama,
    \PRD{54}{7032}{1996}.

\bibitem{GGMS}
    See for example F. Gabbiani, E. Gabrielli, A. Masiero, 
    and L. Silvestrini, \NPB{477}{321}{1996}, and references
    therein.

\bibitem{NS}
    Y. Nir and N. Seiberg, \PLB{309}{337}{1993}.

\bibitem{DG}
    M. Dine, A. Kagan, and S. Samuel, \PLB{243}{250}{1990}; \\
    S. Dimopoulos and G.F. Giudice, \PLB{357}{573}{1995}; \\ 
    A. Pomarol and D. Thommasini, \NPB{466}{3}{1996}.

\bibitem{CKN}
    A.G. Cohen, D.B. Kaplan, and A.E. Nelson, \PLB{388}{588}{1996}.

\bibitem{nilles}
    For a review, see H.P. Nilles, \PREP{110}{1}{1984}.

\bibitem{GM82} 
    M. Dine and W. Fischler, \PLB{110}{227}{1982};\\
    M. Dine and W. Fischler, \NPB{204}{346}{1982};\\
    L. Alvarez-Gaum\'e, M. Claudson, and M. Wise, \NPB{207}{96}{1982};\\
    C.R. Nappi and B.A. Ovrut, \PLB{113}{175}{1982};\\
    S. Dimopoulos and S. Raby, \NPB{219}{479}{1983}.

\bibitem{DNS}
    M. Dine and A.E. Nelson, \PRD{48}{1277}{1993};\\
    M. Dine, A.E. Nelson, and Y. Shirman, \PRD{51}{1362}{1995};\\
    M. Dine, A.E. Nelson, Y. Nir, and Y. Shirman, \PRD{53}{2658}{1996}.

\bibitem{gmsbrev}
    For a thorough review, see
    G.F. Giudice and R. Rattazzi, \xxx{hep-ph/9801271}.

\bibitem{sugra}
    S. Dimopoulos and H. Georgi, \NPB{193}{150}{1981}; \\
    N. Sakai, \ZPC{11}{153}{1981}.

\bibitem{BFP}
    J.L. Feng, C. Kolda, and N. Polonsky, \NPB{546}{3}{1999}; \\
    J. Bagger, J.L. Feng, and N. Polonsky, \xxx{hep-ph/9905292}.

\bibitem{LNS} 
    M. Leurer, Y. Nir, and N. Seiberg, \NPB{398}{319}{1993};
    \emph{ibid.}, \NPB{420}{468}{1994}; \\ 
    P. Ramond, R.G. Roberts, and G.G. Ross, \NPB{406}{19}{1993}; \\
    L. Ib\'{a}\~{n}ez and G.G. Ross, \PLB{332}{100}{1994};\\
    P. Bin\'{e}truy and P. Ramond, \PLB{350}{49}{1995}; \\
    V. Jain and R. Shrock, \PLB{352}{83}{1995}; \\
    E. Dudas, S. Pokorski, and C.A. Savoy, \PLB{356}{45}{1995};
    P. Bin\'{e}truy, S. Lavignac, and P. Ramond, \NPB{477}{353}{1996}; \\
    E. Dudas, C. Grojean, S. Pokorski, and C.A. Savoy, \NPB{481}{85}{1996}; \\
    E.J. Chun and A. Lukas, \PLB{387}{99}{1996}; \\
    Z. Berezhiani and Z. Tavartkiladze, \PLB{396}{150}{1997}; \\
    K. Choi, E.J. Chun, and H. Kim, \PLB{394}{89}{1997}; \\
    P. Bin\'{e}truy, N. Irges, S. Lavignac, and P. Ramond, \PLB{403}{38}{1997}.

\bibitem{hall}
    R. Barbieri, G. Dvali, and L.J. Hall, \PLB{377}{76}{1996}.

\bibitem{GS}
    M.B. Green and J.H. Schwarz, \PLB{149}{117}{1984}.

\bibitem{BDDP}
    P. Bin\'{e}truy and E. Dudas, \PLB{389}{503}{1996}; \\
    G. Dvali and A. Pomarol, \PRL{77}{3728}{1996}.

\bibitem{MRNW}
    R.N. Mohapatra and A. Riotto, \PRD{55}{1138}{1997}; 
        \emph{ibid.}, \PRD{55}{4262}{1997}; \\
    A.E. Nelson and D. Wright, \PRD{56}{1598}{1997}; \\
    G. Eyal, \xxx{hep-ph/9903423}.

\bibitem{FN}
    C.D. Froggatt and H.B. Nielsen, \NPB{147}{277}{1979}.

\bibitem{preons}
    N. Arkani-Hamed, M.A. Luty, and J. Terning, \PRD{58}{015004}{1998}; \\
    M.A. Luty and J. Terning, \xxx{hep-ph/9812290}.

\bibitem{AHCH}
    N. Arkani-Hamed, H.-C. Cheng, and L.J. Hall, 
    \NPB{472}{95}{1996}; \emph{ibid.}, \PRD{54}{2242}{1996} 

\bibitem{BFPT}
    F. Borzumati, G.R. Farrar, N. Polonsky, and S. Thomas, 
    \xxx{hep-ph/9902443}.

\bibitem{KLMNR}
    D.E. Kaplan, F. Lepeintre, A. Masiero, A.E. Nelson, and A. Riotto,
    \xxx{hep-ph/9806430}.

\bibitem{superK} 
    Super-Kamiokande Collaboration, Y. Fukuda et al.,
    \PLB{433}{9}{98}; \emph{ibid.}, \PLB{436}{33}{98}; 
    \emph{ibid.}, \PRL{81}{1562}{1998}.

\bibitem{solar} 
    B.T. Cleveland et al., \NPB{38}{47}{95} (proc. suppl); \\
    Kamiokande collaboration, Y. Fukuda et al., \PRL{77}{1683}{96}; \\
    Gallex Collaboration, W. Hampel et al., \PLB{388}{384}{96}; \\
    SAGE Collaboration, J.N. Abdurashitov et al., \PRL{77}{4708}{96}; \\
    J.N. Bahcall and M.H. Pinsonneault, \RMP{67}{781}{95}; \\
    J.N. Bahcall, S. Basu, and M.H. Pinsonneault, \PLB{433}{1}{98}.

\bibitem{AHM}
    N. Arkani-Hamed and H. Murayama, \PRD{56}{6733}{1997}.

\bibitem{AG}
    K. Agashe and M. Graesser, \PRD{59}{015007}{1999}. 

\bibitem{CKLN}
    A.G. Cohen, D.B. Kaplan, F. Lepeintre, and A.E. Nelson, 
    \PRL{78}{2300}{1997}.

\bibitem{other-effective-SUSY}
    S. Ambrosanio and A.E. Nelson, \PLB{411}{283}{1997}; \\
    J. Hisano, K. Kurosawa, and Yasunori Nomura, \PLB{445}{316}{1999}; \\
    S. Ambrosanio and J.D. Wells, talk given at The Physics at Run II: 
    Workshop on Supersymmetry/Higgs, \xxx{hep-ph/9902242}. 

\bibitem{baggertalk}
    J. Bagger, Talk given at Higgs \& SUSY: Search and Discovery,
    Gainesville, FL, March 1999.

\bibitem{m-u-zero}
    H. Georgi and I.N. McAuthur, Harvard University Report, HUTP-81/A011 
        (1981); \\
    D.B. Kaplan and A.V. Manohar, \PRL{56}{2004}{1986}; \\
    K. Choi, C.W. Kim, and W.K. Sze, \PRL{61}{794}{1988}.

\bibitem{Leut}  
    For two opposing views, see H. Leutwyler, Talk given at 
    Conference on Fundamental Interactions of Elementary particles, 
    Moscow, Russia, 23-26 Oct. 1995, \xxx{hep-ph/9602255}, 
    and T. Banks, Y. Nir and N. Seiberg, Presented at 2nd IFT 
    Workshop on Yukawa Couplings and the Origins of Mass, 
    Gainesville, FL, 11-13 Feb. 1994, \xxx{hep-ph/9403203}.

\bibitem{MartinGMSB}
    S.P. Martin, \PRD{55}{3177}{1997}. 

\bibitem{GDK-gaugino}
    G.D. Kribs, \NPB{535}{41}{1998}.

\bibitem{KTeV} 
    KTeV Collaboration, \xxx{hep-ex/9905060}.

\bibitem{PT}
    E. Poppitz and S.P. Trivedi, \PLB{401}{38}{1997}.

\bibitem{JJMVY}
    I. Jack, D.R.T. Jones, S.P. Martin, M.T. Vaughn, and Y. Yamada,
    \PRD{50}{5481}{1994}. 

\bibitem{DGP}
    S. Dimopoulos, G.F. Giudice, and A. Pomarol, \PLB{389}{37}{1996}. 

\bibitem{MartinVaughn}
    S.P. Martin and M.T. Vaughn, \PRD{50}{2282}{1994}. 

\bibitem{Yamada}
    Y. Yamada, \PRD{50}{3537}{1994}. 

\bibitem{RandallStr}
    L. Randall, \NPB{495}{37}{1997}.

\bibitem{AHMRM}
    N. Arkani-Hamed, J. March-Russell, and H. Murayama, 
    \NPB{509}{3}{1998}. 

\bibitem{GiudiceRattazziwf}
    G.F. Giudice and R. Rattazzi, \NPB{511}{25}{1998}. 

\bibitem{DTW}
    S. Dimopoulos, S. Thomas, and J.D. Wells, \NPB{488}{39}{1997}. 

\bibitem{BMPZ}
    J.A. Bagger, K. Matchev, D.M. Pierce, and R. Zhang, \PRD{55}{3188}{1997}.

\bibitem{AKM1}
    S. Ambrosanio, G.D. Kribs, and S.P. Martin, \PRD{56}{1761}{1997}. 

\bibitem{KaneKing}
    G.L. Kane and S.F. King, \PLB{451}{113}{1999}. 

\bibitem{PBMZ}
    D.M. Pierce, J.A. Bagger, K. Matchev, and R. Zhang,
    \NPB{491}{3}{1997}.

\bibitem{KMR}
    C. Kolda and J. March-Russell, \PRD{55}{4252}{1997}. 

\bibitem{DFMR}
    K.R. Dienes, A.E. Faraggi, and J. March-Russell,
    \NPB{467}{44}{1996}. 

\bibitem{Wells-taus}
    J.D. Wells, \MPL{A13}{1923}{1998}.

\bibitem{CarenaWagnerhiggs}
    M. Carena, J.R. Espinosa, M. Quir\'{o}s, and C.E.M. Wagner 
    \PLB{355}{209}{1995}; \\
    M. Carena, M. Quir\'{o}s, and C.E.M. Wagner, \NPB{461}{407}{1996}.

\bibitem{higgsbounds}  
    H.E. Haber, R. Hempfling, and A.H. Hoang, \ZPC{75}{539}{1997}; \\
    S. Heinemeyer, W. Hollik, and G. Weiglein \PLB{440}{296}{1998}; 
    \emph{ibid.}, \xxx{hep-ph/9812472}; \\
    H.E. Haber, \xxx{hep-ph/9901365}.

\bibitem{LEPhiggsbound}
    See e.g.\ T. Greening (for the LEP Collaborations),
    \xxx{hep-ex/9903013}.

\bibitem{DDRT}
    S. Dimopoulos, M. Dine, S. Raby, and S. Thomas, \PRL{76}{3494}{1996}. 

\bibitem{BaerGMSB}
    H. Baer, M. Brhlik, C.-h. Chen, and X. Tata, \PRD{55}{4463}{1997}.

\bibitem{Nandipapers}
    D.A. Dicus, B. Dutta, and S. Nandi, \PRL{78}{3055}{1997}; 
    \emph{ibid.}, \PRD{56}{5748}{1997}; \\
    B. Dutta, D.J. Muller, and S. Nandi, \NPB{544}{451}{1999}; \\
    D.J. Muller and S. Nandi, \xxx{hep-ph/9811248}.

\bibitem{AKM2}
    S. Ambrosanio, G.D. Kribs, and S.P. Martin, \NPB{516}{55}{1998}.

\bibitem{Baerandcompany}
    H. Baer, C.-H. Chen, M. Drees, F. Paige, and X. Tata, \PRL{79}{986}{1997};
    \emph{ibid.}, \PRD{58}{075008}{1998}.
  
\bibitem{MatchevLykken}
    J.D. Lykken and K.T. Matchev, \xxx{hep-ph/9903238}.

\bibitem{MatchevPierce}
    K.T. Matchev and D.M. Pierce, \xxx{hep-ph/9904282}.

\bibitem{FengMoroi}
    J.L. Feng and T. Moroi, \PRD{58}{035001}{1998}. 

\bibitem{MartinWells}
    S.P. Martin and J.D. Wells, \PRD{59}{035008}{1999}.

\bibitem{CDFslepton}
    A. Connolly, The CDF Collaboration, FERMILAB-CONF-99/092-E,
    talk given at DPF 99, UCLA, Los Angeles, CA (1999)
    \xxx{hep-ex/9904010}.

\bibitem{CLEO}
    J. Gronberg, \xxx{hep-ph/9903368}; \\
    E791 Collaboration, \xxx{hep-ex/9903012}; \\
    D.M. Asner, \xxx{hep-ph/9905223}.

\bibitem{DDbar}
    L. Wolfenstein, \PLB{164}{170}{1985}; \\
    J.F. Donoghue et al., \PRD{33}{179}{1986}; \\
    S. Pakvasa, Chin. J. Phys. \textbf{32}, 1163 (1994); \\
    H. Georgi, \PLB{297}{353}{1992}; \\
    T. Ohl, G. Ricciardi, and E.H. Simmons, \NPB{403}{605}{1993}.

\bibitem{MM} 
    A. Masiero and H. Murayama, \xxx{hep-ph/9903363}.

\bibitem{DvGP} 
    G. Dvali, G.F. Giudice, and A. Pomarol, \NPB{478}{31}{1996}.

\bibitem{DDR} 
    S. Dimopoulos, G. Dvali, and R. Rattazzi, \PLB{413}{336}{1997}.

\bibitem{DPchiral}
    G. Dvali and A. Pomarol, \NPB{522}{3}{1998}. 

\bibitem{CDMchiral}
    H.-C. Cheng, B.A. Dobrescu, and K.T. Matchev, \PLB{439}{301}{1998}; 
    \emph{ibid.}, \NPB{543}{47}{1999}.

\bibitem{NP} 
    H.P. Nilles and N. Polonsky, \PLB{412}{69}{1997}.

\bibitem{seesaw}
    T. Yanagida, \PRL{45}{71}{1980}.

\bibitem{rpvneumass} 
    L.J. Hall and M. Suzuki, /NPB{231}{419}{1984}; \\
    S. Dawson, \NPB{261}{297}{1985}; \\
    T. Banks, Y. Grossman, E. Nardi, and Y. Nir, \PRD{52}{5319}{1995}; \\
    Z. Berezhiani and A. Rossi, \PLB{367}{219}{1996}; \\
    M. Drees, S. Pakvasa, X. Tata, and T. ter~Veldhuis, 
    \PRD{57}{5335}{1998}; \\
    S. King, \xxx{hep-ph/9806440}.

\bibitem{DEK-Nelson}
    D.E. Kaplan and A.E. Nelson, \xxx{hep-ph/9901254}.

\bibitem{BHSSW} 
    R. Barbieri, L. Hall, D. Smith, A. Strumia, and N. Weiner,
    JHEP 9812, 017 (1998).

\bibitem{fogli} 
    For atmospheric neutrinos, see G.L. Fogli, E. Lisi, A. Marrone, 
    and G. Scioscia, \PRD{59}{033001}{1999}; \emph{ibid.}, 
    \xxx{hep-ph/9904465}; and for solar neutrinos, see 
    G.L. Fogli, E. Lisi, and D. Montanino, \PRD{54}{2048}{1996}.

\bibitem{chooz} 
    CHOOZ Collaboration, M. Apollonio et al., \PLB{420}{397}{1998}.

\bibitem{BHS} 
    R. Barbieri, L. Hall, and A. Strumia, \PLB{445}{407}{1999}.

\bibitem{msw} 
    L. Wolfenstein, \PRD{17}{2369}{78}; \\
    S. P. Mikheyev and A. Yu. Smirnov, Yad. Fiz. \textbf{42}, 1441 (1985); 
    Nuovo Cimento \textbf{C9}, 17 (1986).

\bibitem{superoblique}
    Extensive discussion of this can be found in: \\
    H.-C. Cheng, J.L. Feng, and N. Polonsky, \PRD{56}{6875}{1997}; 
    \emph{ibid.}, \PRD{57}{152}{1998}; \\
    L. Randall, E. Katz, and S. Su, \NPPS{62}{299}{1998}; \\
    M.M. Nojiri, D.M. Pierce, and Y. Yamada, \PRD{57}{1539}{1998}; \\
    E. Katz, L. Randall, and S. Su, \NPB{536}{3}{1998}.


\end{thebibliography}
\end{document}